\documentclass{elsarticle}

\usepackage{amssymb}
\usepackage{wasysym}
\usepackage{color}
\usepackage{lineno,hyperref}
\usepackage{fixltx2e}
\modulolinenumbers[5]

\journal{Advances in Atomic Molecular and Optical Physics, Vol. 66, }

\biboptions{authoryear,sort}

\begin{document}

\begin{frontmatter}

\title{Nonclassical Light Generation from III-V and Group-IV Solid-State Cavity Quantum Systems}

\author{Marina Radulaski\corref{mycorrespondingauthor}}

\author{Kevin A. Fischer\corref{mycorrespondingauthor}}

\author{Jelena Vu\v ckovi\' c\corref{mycorr}}
\ead{jela@stanford.edu}

\cortext[mycorrespondingauthor]{These authors contributed equally}

\address{E. L. Ginzton Laboratory, Stanford University, Stanford CA 94305, USA}

\begin{abstract}
In this chapter, we present the state-of-the-art in the generation of nonclassical states of light using semiconductor cavity quantum electrodynamics (QED) platforms. Our focus is on the photon blockade effects that enable the generation of indistinguishable photon streams with high purity and efficiency. Starting with the leading platform of InGaAs quantum dots in optical nanocavities, we review the physics of a single quantum emitter strongly coupled to a cavity. Furthermore, we propose a complete model for photon blockade and tunneling in III-V quantum dot cavity QED systems. Turning toward quantum emitters with small inhomogeneous broadening, we propose a direction for novel experiments for nonclassical light generation based on group-IV color-center systems. We present a model of a multi-emitter cavity QED platform, which features richer dressed-states ladder structures, and show how it can offer opportunities for studying new regimes of high-quality photon blockade.

\end{abstract}

\end{frontmatter}


\newpage

\tableofcontents

%%%%%%%%%%%%%%%%%%%%%%%%%%%%%%%%%%%%%%%%%%%%%%%%%%%%%%%%%%%%%%%%
% Section 1
%%%%%%%%%%%%%%%%%%%%%%%%%%%%%%%%%%%%%%%%%%%%%%%%%%%%%%%%%%%%%%%%

\newpage

%\linenumbers

\section{Introduction}

Many quantum technologies, including quantum key distribution and photonic-qubit based quantum computation, require on-demand sources of light that produce pulses containing a well-defined number of photons \citep{OBrien2009-dq}. Such sources are expected to have high efficiencies, rarely emit a wrong number of photons, and produce indistinguishable photons that interfere with each other.
The majority of scientific efforts toward creating sources of nonclassical light have so far been focused on engineering single-photon sources \citep{buckley2012engineered}, and the basic idea used to generate single photons on demand in state-of-the-art approaches is very simple: a single quantum emitter is excited with a pulsed source and its emission is filtered in order to isolate a single-photon with the desired properties \citep{sanders2005single}. For example, an optical or electrical pulse can generate carriers--electrons and holes--inside a quantum dot that recombine to produce several photons at different frequencies. Subsequent spectral filtering can be used to isolate a single-photon \citep{santori2002indistinguishable, yuan2002electrically}. Although these systems are already characterized by high multi-photon probability suppression, both the efficiency and indistinguishability of such a source can be further improved by embedding the quantum emitter into a cavity that has a high quality factor and a small mode volume, or by resonant excitation approaches such as photon blockade \citep{Birnbaum2005-yk,Faraon2008-zh}.

Epitaxial, III-V semiconductor quantum dots in cavities have been a leading platform for nonclassical light generation experiments, but they suffer from large inhomogeneous broadening, which impedes scaling of the systems to the substatial number of quantum emitters necessary for implementation of large entangled photon states, quantum networks or quantum simulators \citep{greentree2006quantum, hartmann2008quantum, kimble2008quantum, carusotto2013quantum}. Namely, a stream of indistinguishable single photons from one or more quantum emitters could be employed to generate larger photon number states or entangled photon states by quantum interference, but this requires photons to be indistinguishable and thus the emitters must be  indistinguishable. Additionally, these quantum dots require operation at cryogenic temperatures. Recently, color-centers in group-IV semiconductors have emerged as alternative quantum emitters, with significantly smaller inhomogenous broadening and the ability to operate at room temperature \citep{beveratos2002room, neu2011single, rogers2014multiple, widmann2015coherent}.

In this chapter, we present state-of-the-art demonstrations of nonclassical light generation in these semiconductor systems, and we propose a model of a system that improves the quality of the sources (efficiency, indistinguishability) through the multi-emitter cavity QED. We believe that photon blockade effects will continue to play an important role in high-throughput generation of nonclassical states of light. These effects, which have been initially developed in atomic systems \citep{Birnbaum2005-yk}, have been demonstrated only in quantum-dot-based systems among solid-state optical platforms \citep{Faraon2008-zh, volz2012ultrafast}. However, ensembles of color-centers in group-IV semiconductor cavities present a very promising platform for the implementation of new regimes of photon blockade and new quantum light sources.

\subsection{InAs Quantum Dots in GaAs}

\begin{figure}
\centering
\textsf{
\includegraphics[width=5cm]{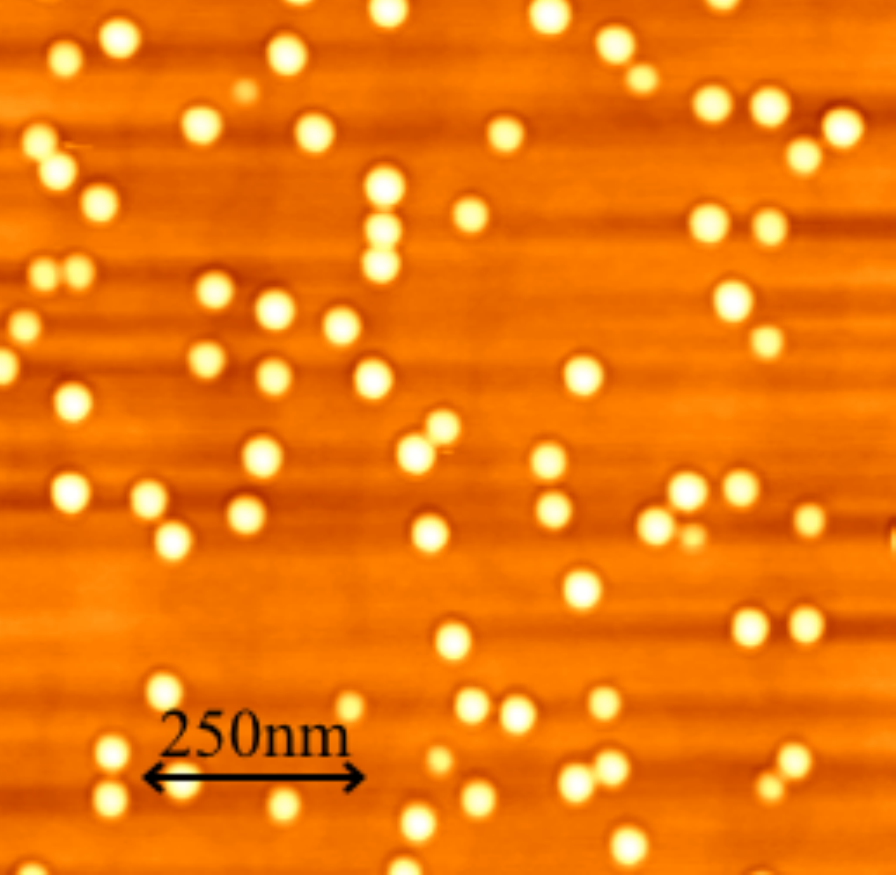}
\caption{\label{figure:1-1} AFM image of a 1x1 micron square array of uncapped, self-assembled InGaAs quantum dots.}
}
\end{figure}

\begin{figure}
\centering
\textsf{
\includegraphics[width=11.43cm]{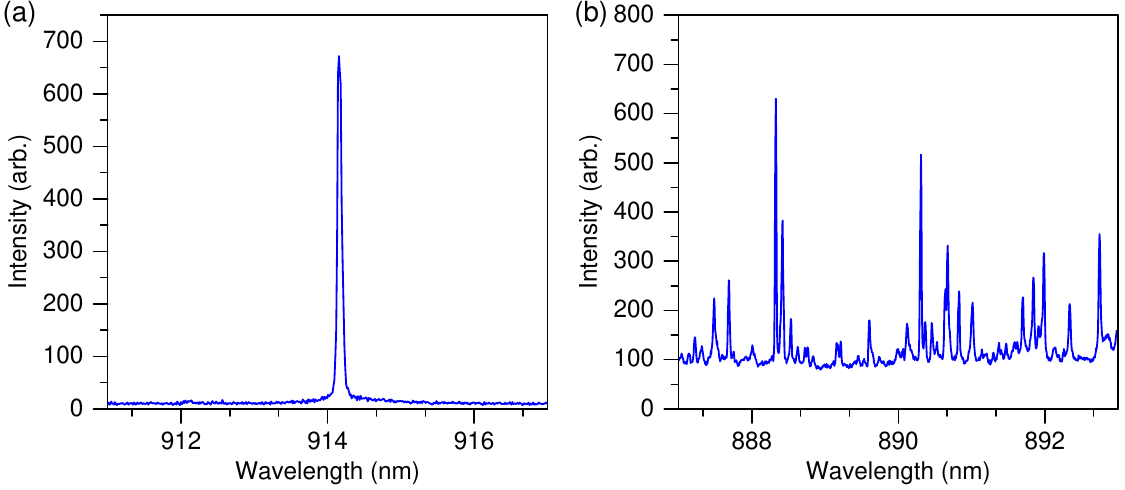}
\caption{\label{figure:1-2} Spectra of quantum dots, (a) of a neutral excitonic transition and (b) of an ensemble of excitonic transitions from many quantum dots that feature a large inhomogeneous broadening.}
}
\end{figure}

InAs quantum dots in GaAs have served as a leading platform for solid-state quantum optics and cavity QED experiments for the past 20 years \citep{michler2009single,dietrich2016gaas}. Such quantum dots are formed by self-assembly during the growth process called molecular beam epitaxy (MBE), as a result of lattice mismatch between InAs and GaAs. Since quantum dot islands are formed by self-assembly, their locations are random and they have a distribution of sizes and shapes (Fig. \ref{figure:1-1}), leading to variation in the transition energies of different quantum dots on the same wafer (inhomogeneous broadening) (Fig. \ref{figure:1-2}). Because of a shallow confining potential, these quantum  dots require operation at cryogenic temperatures, lower than 70 K and typically around 4 K. 

\subsection{Color-Centers in Group-IV Materials}

Color-centers are emerging solid-state photonics systems, that hold great promise for nonclassical light generation. They can serve both as bright single-photon emitters \citep{aharonovich2009two,castelletto2014room} with fast\,GHz generation rates and as spin qubits \citep{childress2006coherent, widmann2015coherent} with demonstrated lifetimes of close to a second \citep{bar2013solid}. Structurally, color-centers are point-defects in the semiconductor lattice whose localized electronic orbitals have optical transitions \citep{weber2010quantum}. They can occur naturally or be generated through electron, neutron or ion irradiation \citep{widmann2015coherent, fuchs2015engineering, kurtsiefer2000stable}, or doping during growth \citep{zhang2015hybrid}, where masking, dose and beam energy can be used to control defect density and position. In addition, the angstrom-scale size of defect centers opens opportunities to interface an ensemble of emitters with a nanocavity \citep{Radulaski2016-rb, cui2015hybrid, Sipahigil847}. Moreover, the inhomogeneous broadening of color-center optical transitions can be as narrow as four lifetime-limited linewidths \citep{sipahigil2014indistinguishable}, making these emitters nearly identical. This is an advantageous property compared to other semiconductor emitters which can be utilized to access novel many-body studies in the solid-state. Additional flexibility in designing photonic systems is granted by the variety of operating wavelengths, as well as the difference in optical, mechanical and electronic properties between the host materials. 

Diamond and silicon carbide have prominently served as hosts of various vacancy-related defects \citep{kurtsiefer2000stable, sipahigil2014indistinguishable, castelletto2014room, falk2013polytype, widmann2015coherent}. These substrates are chemically inert which makes them biocompatible for interfacing with or within living cells as dyes and nanoscale sensors \citep{kucsko2013nanometre, saddow2012silicon}. Color-centers also have a unique set of properties as nearly identical solid-state emitters with strong dipole moment, which we will explore for novel opportunities in cavity QED.

%%%%%%%%%%%%%%%%%%%%%%%%%%%%%%%%%%%%%%%%%%%%%%%%%%%%%%%%%%%%%%%%
% Section 2
%%%%%%%%%%%%%%%%%%%%%%%%%%%%%%%%%%%%%%%%%%%%%%%%%%%%%%%%%%%%%%%%

\section{Overview of Cavity QED With a Single Quantum Emitter}

Cavity quantum electrodynamical systems hold great promise for exploring fundamental light-matter interactions, generating novel quantum states of light, and playing integral roles in quantum information networks \citep{OBrien2009-dq}. Originally, such investigations were centered on atomic systems where both the quantum matter and light confining components of the cavity QED system possessed intrinsically narrow linewidths \citep{Birnbaum2005-yk}. As a result, the hallmark signature of strong light-matter interaction with the emergence of new hybridized states of light and matter or \emph{strong coupling}, was more easily achievable. Such hybridized states are anharmonic, resulting in giant nonlinearities at the single-photon level. However, these atomic systems will unlikely serve as practical elements in communication networks due to their slow interaction rates and lack of scalability. In the 2000's, major advances towards the study of light-matter interaction in systems based on InGaAs quantum dots and photonic crystal nanocavities enabled the observation of strong coupling in an optical solid-state system \citep{Englund2007-wy,Yoshie2004-yd}. Since then, the boundaries of quantum-dot-based cavity QED work have been pushed from ultra-low threshold lasers \citep{Ellis2011-xb} and ultra-fast single-photon sources \citep{Muller2015-il}, to photon blockade and single-photon phase gates \citep{Fushman2008-jt}. More recently, group-IV solid-state quantum emitters, in the form of the popular nitrogren \citep{Wolters2010-rz,Englund2010-vk} and silicon \citep{Radulaski2016-rb,Riedrich-Moller2014-tb} vacancy-related complexes, have generated strong interest for future cavity QED experiments with their demonstrations of Purcell enhancement, single-photon emission, and exceptionally small inhomogeneous broadening. In this section, we provide an introduction to the basic concepts relevant to nonclassical light generation with a single-emitter cavity QED system.

%%%%%%%%%%%%%%%%%%%%%%%%%%%%%%%%%%%%%%%%%%%%%%%%%%%%%%%%%%%%%%%%

\subsection{Single-Photon Emission With Color-Centers}

For emission purposes, color-centers can be described as two-level quasi-atoms whose single-photon emission can be employed for optical networks and quantum cryptography. Similarly to quantum dots, the excited state lifetimes are in the nanosecond range. Due to the presence of vibronic sublevels of the ground and excited states, the emission spectra consist of a zero-phonon line (ZPL), a narrow feature corresponding to the energetically highest transition, and a phonon sideband, which comprises phonon-assisted transitions \citep{fitchen1968physics}. The ratio of the emission into the ZPL varies between color-centers, often based on the symmetry of the defect system, and is higher at low temperatures.

The negatively charged nitrogen-vacancy (NV) center in diamond has, until recently, been the dominant color-center system in photonics \citep{aharonovich2011diamond}. Its ZPL at 637\,nm contains 3--5\% of the emission. Single-photon generation rates are in the tens of~kHz and have been enhanced by the use of nanopillar structures to 168 kHz \citep{babinec2010diamond}. The NV center in diamond has been utilized for cryptography \citep{beveratos2002single}, and integrated systems with silicon nitride waveguides have been demonstrated \citep{mouradian2015scalable}. One of the challenges in working with the NV center is its spectral instability \citep{siyushev2013optically}, which has motivated a search for novel color-center systems \citep{weber2010quantum}.

\begin{figure}
\textsf{
\includegraphics[width=11.43cm]{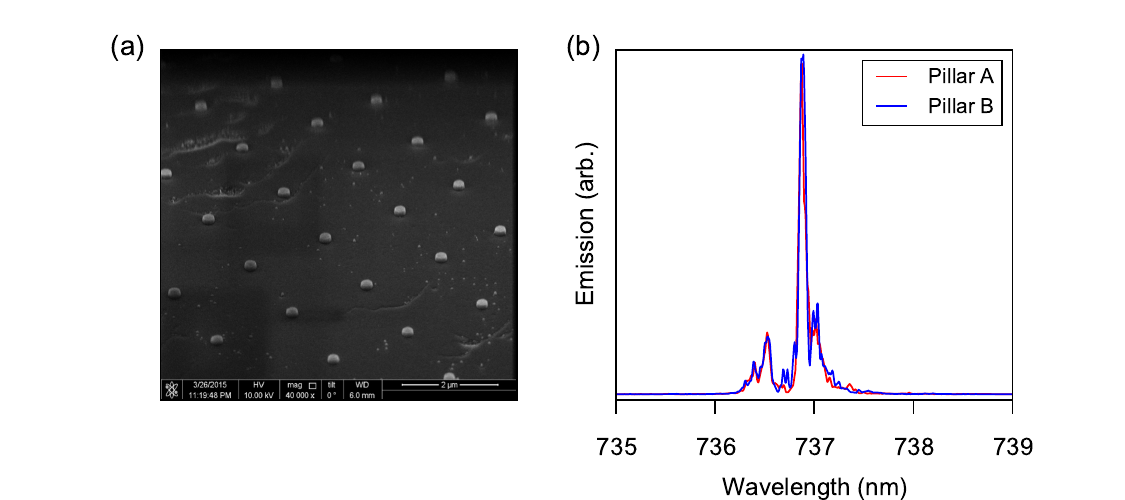}
\caption{\label{figure:2-1} (a) Etched nanopillars in color-center rich diamond substrate. (b) Silicon-vacancy emission from two different pillars indicating low strain in the fabricated structures. \textit{Data from \citet{zhang2015hybrid}}.}
}
\end{figure}

The negatively charged silicon vacancy (SiV) in diamond has emerged as an emitter with a large (70\%) emission ratio into its ZPL at 738\,nm. In addition, the ensemble linewidths are as narrow as 400\,MHz at cryogenic temperatures \citep{sipahigil2014indistinguishable}, which qualifies these systems as nearly identical solid state quasi-atoms. An order of magnitude increase in the ZPL extraction and two to three orders of magnitude reduction in the ZPL linewidth compared to the NV center makes SiV a more attractive center for scalable systems. Additionally, reactive ion etching can be performed while maintaining the SiV's optical properties (Fig. \ref{figure:2-1}), which has been demonstrated with an array of nanopillars \citep{zhang2015hybrid} and in nanobeam cavities \citep{Sipahigil847}. These recent results pave a promising path for integrating nearly identical quantum emitters in the suspended photonic structures needed for cavity QED.

Silicon carbide (SiC) has also been discovered as an alternative host of color-centers to diamond. Across its many polytypes (such as 3C, 4H, or 6H), SiC offers a variety of quantum emitters. Emitters in 3C-SiC, such as the carbon-anti-site-vacancy (Si\textsubscript{C}V\textsubscript{Si}) \citep{castelletto2014room} and oxidation-induced centers \citep{lohrmann2016activation} have exceptional brightness of around 1\,MHz count rates. The divacancy (V\textsubscript{C}V\textsubscript{Si}) in 4H-SiC gives rise to six ZPL transitions in the range of 1,100--1,200\,nm, two of which are active at room temperatures \citep{koehl2011room}. The silicon vacancy (V\textsubscript{Si}) in 4H-SiC and 6H-SiC emits into 25--30\,GHz narrow ZPLs capturing several percent of emission at cryogenic temperatures in their two and three lines \citep{sorman2000silicon}, respectively, in the wavelength region 860--920\,nm. In particular, the 916\,nm ZPL from V\textsubscript{Si} in 4H-SiC has been incorporated in nanopillars where nanofabrication processing left the optical properties unaltered but increased the collection efficiency of single photons by several times \citep{radulaski2016scalable}.

An increase in single-photon emission rate can be achieved through Purcell enhancement where a color-center is coupled to a cavity with small mode volume and high quality factor. This has been achieved with Purcell factor as high as $F=26$ in a hybrid approach where NV and SiV centers in diamond were coupled to GaP \citep{gould2016efficient, Wolters2010-rz} and SiC \citep{Radulaski2016-rb} microresonators, as well as in monolithic approaches with V\textsubscript{Si} in 4H-SiC \citep{bracher2016selective}.

For an overview of the established field surrounding single-photon emission from quantum dots we refer the reader to \citet{buckley2012engineered}.

%%%%%%%%%%%%%%%%%%%%%%%%%%%%%%%%%%%%%%%%%%%%%%%%%%%%%%%%%%%%%%%%

\subsection{Strong Coupling With Quantum Dots}

\begin{figure}
\textsf{
  \includegraphics[width=11.43cm]{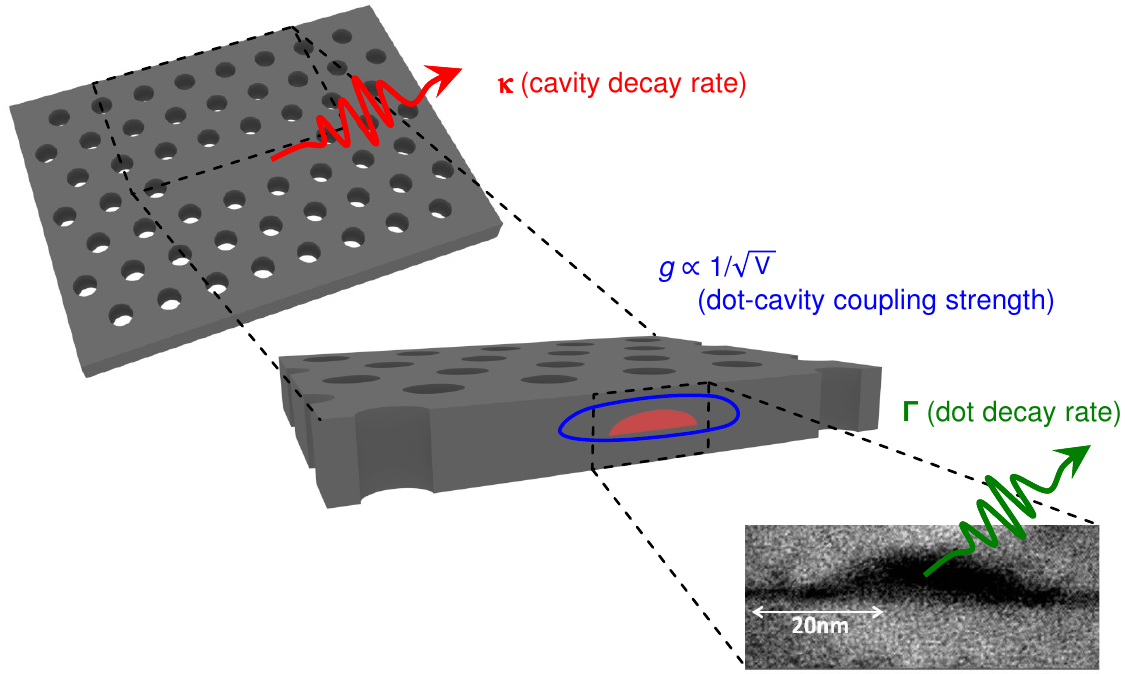}
  \caption{Schematic depiction of a strongly-coupled system based on a single InGaAs quantum dot embedded within an L3 photonic crystal cavity. The red partial-dome in the cross-section (middle) represents the quantum dot and a TEM image is shown of a typical dot in the bottom right. \textit{TEM adapted with permission from \citet{Krenner2005-mj}.}
}
\label{figure:3-1}}
\end{figure}

One of the most successful and established platforms for single-emitter cavity QED is the InGaAs quantum dot embedded within an L3 photonic crystal cavity \citep{Akahane2003-ma,lodahl2015interfacing}. Self-assembed InGaAs quantum dots are widely used as quantum emitters due to their excellent optical quality, and are particularly useful for strong coupling experiments due to their narrowband optical transitions and ability to be integrated into planar photonic crystal cavities \citep{Aharonovich2016-pq}. Planar photonic crystal resonators provide extremely small mode volumes ($V$) that enable a large enhancement of the light-matter interaction strength with any embedded quantum emitters, $g\propto 1/\sqrt{V}$ \citep{Vahala2003-xx}. A schematic depiction of such a system is shown in Fig. \ref{figure:3-1}, with the L3 photonic crystal cavity shown in the upper left (characterized by energy decay rate $\kappa$) and the quantum dot shown in the bottom right (characterized by energy decay rate $\Gamma$). The quantum dot, illustrated as the red half-dome, is either probabilistically or deliberately positioned at the center of the crystal where it can maximally couple to the fundamental mode of the cavity for optimal interaction rate $g$ \citep{Kuruma2016-jc}. Ideally, the quantum dot behaves like a quantum two-level system, and it can be placed inside of a diode structure to control which specific excitonic transition forms the two-level system \citep{Warburton2000-rf,Carter2013-bm}. However, most of the work on strongly-coupled systems in the solid-state has been performed without such a structure, and usually either the neutral or singly charged exciton couples to the cavity mode.

%%%%%%%%%%%%%%%%%%%%%%%%%%%%%%%%%%%%%%%%%%%%%%%%%%%%%%%%%%%%%%%%

\subsection{The Jaynes--Cummings Model} \label{sec:diag}

At its simplest, a strongly-coupled system is modeled by the Jaynes--Cummings (JC) Hamiltonian \citep{Shore1993-zy}
\begin{equation}
H_\textrm{\scriptsize JC}=\omega_a a^\dagger a + \left(\omega_a + \Delta_\textrm{\scriptsize e}\right)\sigma^\dagger \sigma + g \left(a^\dagger \sigma  + a \sigma^\dagger\right)\textrm{,}
\end{equation}
which represents a single cavity mode coupled to a single quantum two-level system. Here, $a$ is the operator which represents the cavity's mode, $\omega_a$ is the mode's frequency, $\sigma$ is the dipole operator representing a single two-level subsystem of a quantum emitter, and $\Delta_\textrm{\scriptsize e}$ is the detuning of the quantum emitter's transition frequency from that of the cavity. As we will discuss in subsequent sections, specific solid-state considerations will have important implications in exactly modeling such a system, but for now we will consider the idealized JC model.

Because of the coherent interaction between the bosonic cavity mode and fermionic emitter mode, a new set of eigenstates arise \citep{Shore1993-zy}. The optimal basis is no longer the \emph{bare} states of the system, the direct product of individual emitter (ground $| g \rangle$ or excited $| e \rangle$) and cavity (photon number $| n \rangle$) states, but rather a so-called \emph{dressed state} basis that comprises states known as polaritons. These polaritons are entangled states of light and matter, with eigenstates
\begin{equation}
| n,+ \rangle = \textrm{cos}\left( \alpha_n/2 \right)| n-1 \rangle| e \rangle + \textrm{sin}\left( \alpha_n/2 \right)| n \rangle| g \rangle
\end{equation}
and
\begin{equation}
| n,- \rangle = -\textrm{sin}\left( \alpha_n/2 \right)| n-1 \rangle| e \rangle + \textrm{cos}\left( \alpha_n/2 \right)| n \rangle| g \rangle\textrm{,}
\end{equation}
where $\alpha_n=\textrm{tan}^{-1}\left( 2g\sqrt{n+1}/\Delta_\textrm{\scriptsize e} \right)$. These states have eigenenergies
\begin{equation}
E^n_\pm = n\omega_a+\Delta_\textrm{\scriptsize e}/2\pm\sqrt{\left(\sqrt{n}g\right)^2 + \left(\Delta_\textrm{\scriptsize e}/2\right)^2}\textrm{.}
\end{equation}
Consider the first two polaritons of non-zero energy ($n=1$), called the first upper and lower polaritons: when the cavity and emitter are tuned on resonance ($\Delta_\textrm{\scriptsize e}=0$), then the polaritons each contain equal contributions of a single electronic or a single photonic excitation. However, as the emitter is detuned from the cavity (by increasing $\Delta_\textrm{\scriptsize e}$) then the polaritons trend towards either the electronic or photonic characters.

%%%%%%%%%%%%%%%%%%%%%%%%%%%%%%%%%%%%%%%%%%%%%%%%%%%%%%%%%%%%%%%%

\subsection{Observing Strong Coupling} \label{sec:osc}

The evolution of the system dynamics is governed by a Liouville equation $\partial_t\rho(t)=\mathcal{L}\rho(t)$, which accounts for the non-unitary evolution induced by the cavity and emitter dissipation \citep{Laussy2012-hr}. Specifically,
\begin{equation}
\mathcal{L}_\textrm{\scriptsize JC}\rho(t) =\textrm{i}\left[ \rho(t), H_\textrm{\scriptsize JC}\right] + \frac{\kappa}{2}\mathcal{D}[a]\rho(t) + \frac{\Gamma}{2}\mathcal{D}[\sigma]\rho(t) \textrm{,}
\end{equation}
where $\mathcal{D}[c]\rho(t)=2c\rho(t) c^\dagger-c^\dagger c\rho(t) - \rho(t) c^\dagger c$ is a super-operator called the Lindblad dissipator and $c$ is an arbitrary system operator.

\begin{figure}
\textsf{
  \includegraphics[width=11.43cm]{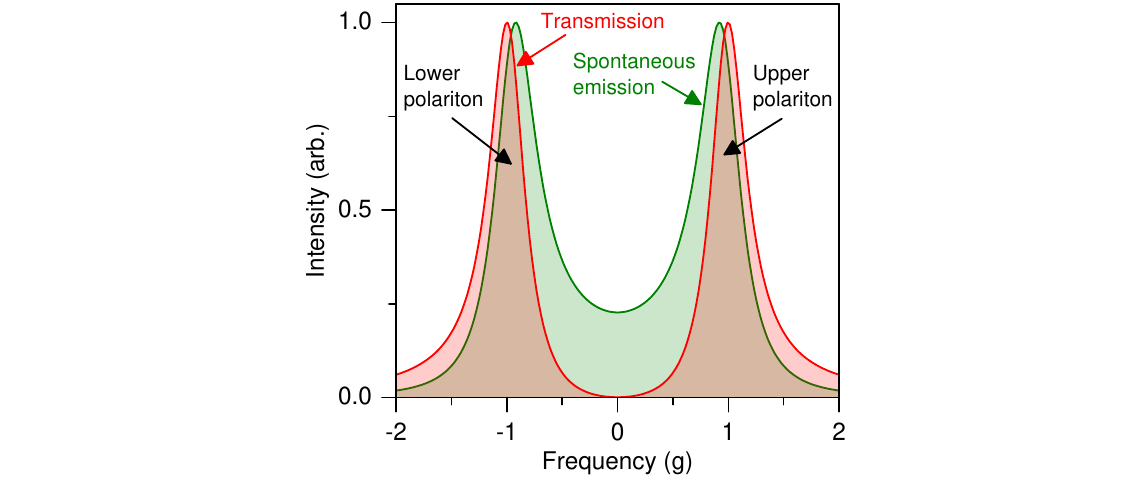}
  \caption{Observation of strong coupling by measuring either transmission or spontaneous emission spectra for zero emitter detuning ($\Delta_\textrm{\scriptsize e}=0$). Typical solid-state cavity QED parameters were used, where $\kappa\approx 3/4 g$ and $\Gamma$ is small enough to be insignificant ($\Gamma\ll g,\kappa$).
}
\label{figure:3-2}}
\end{figure}

Now, we can investigate how strong coupling behavior can be experimentally explored. The most common way to initially observe the effects of strong coupling is to directly measure the splitting of the two lowest energy polaritons either in a transmission experiment \citep{Englund2007-wy} or spontaneous emission experiment \citep{Reithmaier2004-fu}, as shown in Fig. \ref{figure:3-2}. First, we discuss a spontaneous emission experiment. Here, the emitter is incoherently pumped by exciting charge carriers in the semiconductor to higher energy than the emitter's transition [e.g. in a quantum dot's multi-excitonic states, quasi-resonant levels, or above the band-gap \citep{Santori2003-dk}], where they relax into the transition coupled to the cavity. Subsequently, the system relaxes through cavity emission. This process can easily be modeled by adding the term $\frac{P_\sigma}{2}\mathcal{D}[\sigma^\dagger]\rho(t)$ to $\mathcal{L}_\textrm{\scriptsize JC}\rho(t)$, where $P_\sigma$ is the incoherent pumping rate of the quantum emitter \citep{Laussy2008-nr}.

To get an idea of whether the emission occurs through the cavity or the emitter decay channel, consider their typical emission rates. Typical rates for a well-performing cavity QED system are approximately $\Gamma/2\pi\approx0.2$\,GHz (which is often further suppressed by a photonic bandgap) and $\kappa/2\pi\approx10$\,GHz \citep{Muller2015-il}. Because solid-state systems work in the bad-cavity limit where $\kappa\gg\Gamma$, the photons are almost exclusively emitted by the cavity. Therefore, to obtain the measured spectrum of spontaneous emission (e.g. through an ideal spectrometer or scanning Fabry-Perot cavity), we compute the spectrum of the cavity mode operator~$a$ \citep{Laussy2008-nr}
\begin{equation}
S_a(\omega)= \lim_{t\rightarrow\infty} \int_{-\infty}^\infty \mathop{\textrm{d} \tau}\,\langle a^\dagger(t+\tau) a(t) \rangle \textrm{e}^{-\textrm{\footnotesize i}\omega \tau} \textrm{.}
\end{equation}

The astute reader may also notice that it would be possible to incoherently excite the cavity mode using the term $\frac{P_a}{2}\mathcal{D}[a^\dagger]\rho(t)$, where $P_a$ is the cavity's incoherent pumping rate. While this configuration is difficult to realize experimentally, we note that the resulting incoherent spectrum of emission is formally equivalent to a transmission spectrum for arbitrarily small excitation powers. When the incoherent excitation rate is much slower than the decay rates of the first-rung polaritons, $\mathcal{D}[a^\dagger]$ randomly initializes the system with a maximum of one photon in the cavity mode. Thus, calculating a spectrum under incoherent cavity excitation is equivalent to calculating a one-photon spectra \citep{diniz2011strongly}, which represents the linear impulse response of the system and hence its transmission spectrum. Additionally, the computational complexity of this approach for calculating a transmission spectrum is much lower, especially for multi-emitter cavity QED systems, so we use this approach for modeling transmission spectra in Section~\ref{sec:meCQED}.

On the other hand in a transmission \textit{experiment}, a weak continuous-wave (CW) laser incident on the cavity is scanned in frequency and the transmitted light is measured. Now, as opposed to adding the excitation source as a dissipator, the input light is modeled as a coherent state coupled to the cavity \citep{Laussy2012-hr}. This is represented by a Hamiltonian driving term $H_\textrm{\scriptsize drive} = \mathcal{E}\left( a \,\textrm{e}^{\textrm{\footnotesize i}\omega_\textrm{\tiny L} t} + a^\dagger \textrm{e}^{-\textrm{\footnotesize i}\omega_\textrm{\tiny L} t}\right)$, where $\mathcal{E}$ is the real-valued driving strength (proportional to the incident field) and $\omega_\textrm{\scriptsize L}$ is the coherent state frequency. For simplicity, a rotating-frame transformation $\tilde{H}=U H U^\dagger + \textrm{i} \left(\partial_t U\right) U^\dagger$ with $U=\textrm{e}^{\textrm{\footnotesize i}\omega_\textrm{\tiny L} ( a^\dagger a + \sigma^\dagger \sigma)t}$ is used to remove the time-dependence of the excitation term \citep{Majumdar2013-bv} so that $\tilde{H}_\textrm{\scriptsize drive} = \mathcal{E}\left( a + a^\dagger\right)$, and written as a super-operator
\begin{equation}
\tilde{\mathcal{L}}_\textrm{\scriptsize drive}\tilde{\rho}(t) =\textrm{i}\left[ \tilde{\rho}(t), \tilde{H}_\textrm{\scriptsize drive}\right] \textrm{.}
\end{equation}
Similarly, the Jaynes--Cummings Liouvillian transforms according to
\begin{equation}
\tilde{\mathcal{L}}_\textrm{\scriptsize JC}\tilde{\rho}(t) =\textrm{i}\left[ \tilde{\rho}(t), \tilde{H}_\textrm{\scriptsize JC}\right] + \frac{\kappa}{2}\mathcal{D}[a]\tilde{\rho}(t) + \frac{\Gamma}{2}\mathcal{D}[\sigma]\tilde{\rho}(t)
\end{equation}
with
\begin{equation}
\tilde{H}_\textrm{\scriptsize JC}=\left(\omega_a-\omega_\textrm{\tiny L}\right) a^\dagger a + \left(\omega_a + \Delta_\textrm{\scriptsize e} - \omega_\textrm{\tiny L}\right)\sigma^\dagger \sigma + g\left(a^\dagger \sigma  + a \sigma^\dagger\right)\textrm{.}
\end{equation}

Calculating the transmission spectra is then performed first by calculating the steady-state density matrix through solving
\begin{equation}
\left(\tilde{\mathcal{L}}_\textrm{\scriptsize JC}(\omega_\textrm{\tiny L}) + \tilde{\mathcal{L}}_\textrm{\scriptsize drive}\right)\tilde{\rho}_\textrm{\scriptsize ss}(\omega_\textrm{\tiny L}) = 0
\end{equation}
and then using $\tilde{\rho}_\textrm{\scriptsize ss}(\omega_\textrm{\tiny L})$ to obtain the transmission spectrum $S_\textrm{\scriptsize T}(\omega_\textrm{\tiny L})=\langle a^\dagger a \rangle(\omega_\textrm{\tiny L})=\textrm{Tr}\{ a^\dagger a \,\tilde{\rho}_\textrm{\scriptsize ss}(\omega_\textrm{\tiny L}) \}$. In this way, we only simulated light transmitted through the cavity rather than including interference effects of the reflected light. Although we focus on numerical techniques in this section, we note that analytic solutions to the spectra under incoherent and coherent pumping may be found in \citet{Tian1992-ly} and \citet{Waks2006-cc}, respectively. From these analytic forms with dissipation, one can definitively establish a border for the strong coupling regime: when are the upper and lower branches of the polaritons non-degenerate ($g>|\kappa-\Gamma|/4$) \citep{Andreani1999-td}. In the incoherently excited spectra, this is slightly different than the Rayleigh criterion for when the two emission peaks are resolvable ($g>|\kappa-\Gamma|/2$).

Returning to the traces in Fig. \ref{figure:3-2}, one can immediately identify that both the transmission and emission spectra clearly identify the correct splitting of the first two polaritons as $2g$. However, both the correct linewidths of the polaritons and their precise locations (which are slightly shifted by the effects of loss) are only captured in the spontaneous emission spectrum; we will derive this effect in Section~\ref{sec:dissipative}. In transmission, because the two polaritons have cavity components that are shifted by $\pi$ phase, they interfere destructively at frequencies between the polaritons and artificially decrease the apparent polariton linewidths. In fact, this interference occurs for systems not even in strong coupling and is referred to as dipole-induced transparency \citep{Waks2006-cc}. Nevertheless, transmission experiments are more popular for identifying strong coupling due to the fact that incoherently exciting the quantum emitters causes experimental non-idealities. For instance, it generates excess carriers that can induce effects such as field noise \citep{Kuhlmann2015-hh}, and specifically in quantum dot samples with randomly positioned emitters, the carriers can cause nearby dots to indirectly pump the cavity \citep{Majumdar2012-eb}.

\begin{figure}
\textsf{
  \includegraphics[width=11.43cm]{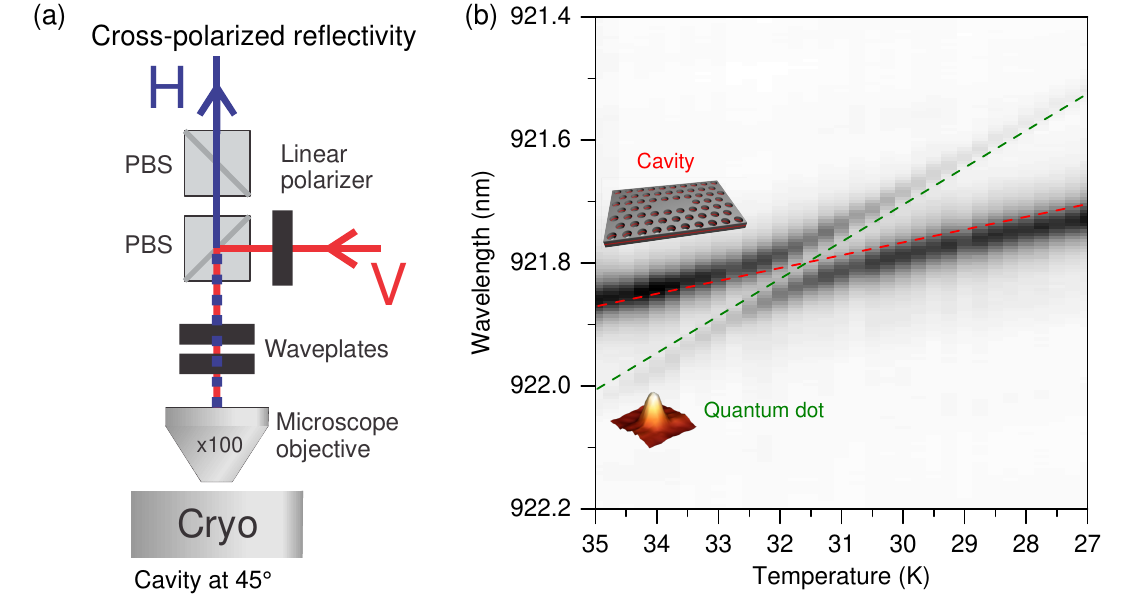}
  \caption{A typical reflectivity measurement used to observe strong coupling. (a) Experimental setup of a cross-polarized reflectivity measurement that is equivalent to transmission. The linear polarizer and two polarizing beamsplitter cubes perform the polarization selection, while the waveplates allow for polarization rotation and correction for bi-refringence in the optical elements \citep{Kuhlmann2013-mj}. (b) Typical reflectivity experiment on a strongly-coupled system, where the emitter detuning relative to the cavity frequency is controlled by tuning the temperature of the sample. An anticrossing between the bare cavity (red) and dot (green) states is observed. \textit{Data from \citet{Muller2015-il}.}
}
\label{figure:3-3}}
\end{figure}

Now that we have discussed transmission experiments, we compare with experiments performed in a configuration called cross-polarized reflectivity \citep{Englund2007-wy} (Fig.~\ref{figure:3-3}a). Here, both the excitation and detection light traverse along (approximately) the same physical path into and out-of the cavity. However, this setup is still mathematically equivalent to a transmission experiment due to the polarization degree of freedom and the configuration of polarizers. The linearly polarized cavity is rotated 45$^{\circ}$ relative to the linear polarization of the incident light. This way, the incident light projects onto the cavity polarization with $1/2$ efficiency. Meanwhile, the orthogonal polarization of the output light is filtered via the polarizing beamsplitters. Therefore, the input and output channels are strictly orthogonal and the reflectivity experiment is equivalent to transmission, to within a scaling factor of $1/2$. This scaling factor does not matter in realistic experiments because the collection losses are typically at least an order of magnitude larger \citep{Englund2007-wy}. As a result, normalized quantities are typically used to study the systems.

A typical cross-polarized reflectivity measurement revealing the strong coupling of an InGaAs-based system is shown in Fig. \ref{figure:3-3}b. The emitter detuning $\Delta_\textrm{\scriptsize e}$ is controlled relative to the cavity frequency $\omega_a$ by changing the lattice temperature. The quantum dot has a stronger dependence on temperature than the cavity due to the bandgap's quadratic temperature dependence compared to the cavity's linear change in permittivity \citep{Faraon2007-wf}. A clear avoided crossing can be seen between the two polaritons near 32.5\,K, as compared to the bare dot (green) and cavity (red) states. Here, the character of each of the polaritons switches between electronic and photonic, providing clear evidence for strong coupling between the cavity and dot. This crossing matches the change in character present in the eigenstate equations for $| n,\pm \rangle(\Delta_\textrm{\scriptsize e})$.

%%%%%%%%%%%%%%%%%%%%%%%%%%%%%%%%%%%%%%%%%%%%%%%%%%%%%%%%%%%%%%%%

\subsection{Basics of Nonclassical Light Generation}

\label{sec:4}

\begin{figure}
\textsf{
  \includegraphics[width=11.43cm]{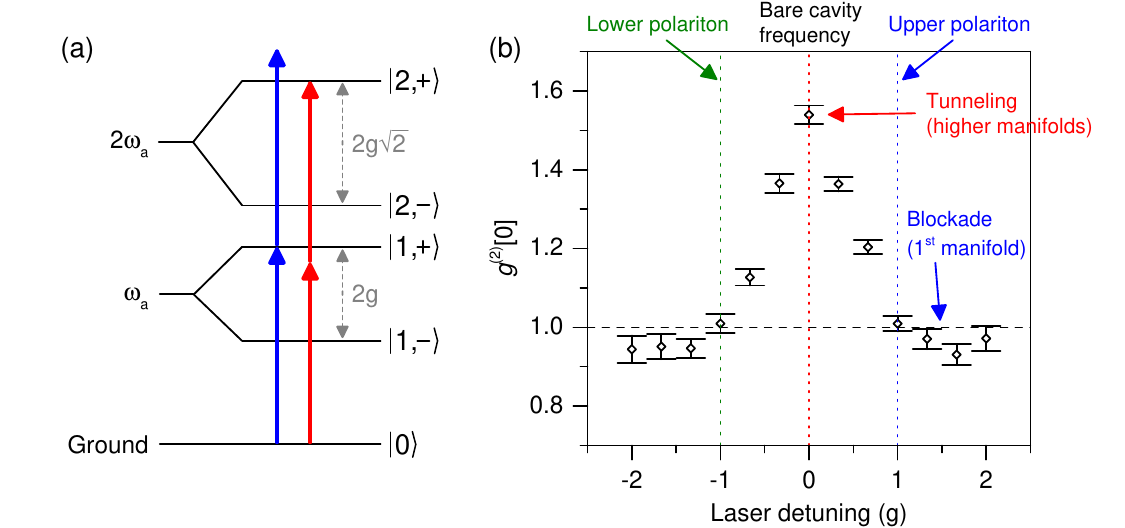}
  \caption{Photon blockade and photon tunneling. (a) Schematic depiction of the Jaynes--Cummings ladder in the dressed basis $|n,\pm\rangle$ when the emitter and cavity are resonant. The upward blue arrows (on the left side) depict photon blockade, while the upward red arrows (on the right side) depict photon tunneling. (b) Measured degree of second-order coherence $g^{(2)}[0]$ for an InGaAs-based resonant system, illustrating tunable nonclassical light generation from a strongly-coupled system (with fitted system parameters $g/2\pi=10.9$\,GHz and $\kappa/2\pi=10$\,GHz). \textit{Data from \citet{Muller2015-om}.}
}
\label{figure:3-4}}
\end{figure}

All of the above dynamics in weak excitation regimes can be fully captured by linear theories, however, the generation of nonclassical light is inherently non-linear. In a cavity QED system, nonclassical light is generated by filtering a stream of incident coherent light through a single strongly-coupled system \citep{Muller2015-il,Faraon2008-zh,Reinhard2011-ye}. Owing to the highly nonlinear character of the interaction between the input light and a strongly-coupled system, the admission of a single-photon into the cavity may enhance (photon tunneling) or diminish (photon blockade) the probability for a second photon to enter the cavity. Ideal photon blockade is depicted schematically in Fig. \ref{figure:3-4}a, where the upward blue arrows represent the laser tuned for photon blockade (left side). Absorption of a photon into state $|1,+\rangle$ blocks the admission of a second photon because no state is present to absorb the second photon. Meanwhile, in photon tunneling (depicted by the upward red arrows on the right side), two photons are absorbed together in a multiphoton transition to directly excite the state $|2,+\rangle$. Following these interactions, a light beam exits the cavity with the nonlinear action imprinted on its quantum character.

The quality of the nonclassical light is typically characterized by the measured degree of second-order coherence $g^{(2)}[0]$, as discussed in \citet{Fischer2016-pl}. If a given pulse results in a photocount distribution $P_n$, then
\begin{equation}
g^{(2)}[0]=\langle n (n-1) \rangle/\langle n \rangle^2,
\end{equation}
which is a normalized second-order factorial moment. Notably, $g^{(2)}[0]=1$ for a coherent pulse with Poissonian counting statistics, while $g^{(2)}[0]<1$ for sub-Poissonian or $g^{(2)}[0]>1$ for super-Poissonian counting statistics; for ideal photon-blockade $g^{(2)}[0]=0$. Importantly, this statistic is completely independent of collection losses so it gives an accurate representation of the internal system dynamics in almost any experiment. Although complete characterization of the photocount distribution would require measurement of all factorial moments of the photocount distribution, $g^{(2)}[0]$ is an important measure in determining whether the source is acting as a single- or multi-photon source, and is the most readily accessible in experiments \citep{Rundquist2014-kf}. Therefore, we will focus on the measured degree of second-order coherence in this section.

In Fig. \ref{figure:3-4}b, we present experimental evidence of photon blockade and photon tunneling in a series of typical experiments measuring $g^{(2)}[0]$ on emission from an InGaAs-based strongly-coupled system. When the laser detuning is near the upper and lower polaritons (states $|1,\pm\rangle$), $g^{(2)}[0]<1$ anti-bunches for photon blockade. The precise location of photon blockade occurs slightly detuned from the polaritons due to the strong dissipation in III-V systems \citep{Muller2015-il}, which will be thoroughly discussed in the next section. Meanwhile, when the laser is tuned in-between the polaritonic rungs, multi-photon transitions are emphasized such that photon tunneling occurs and the light bunches causing $g^{(2)}[0]>1$. The bunching occurs because a large component of the vacuum state is present in conjunction with super-Poissonian statistics \citep{Rundquist2014-kf}. However, because of the strong dissipation in the system (since $g\approx\kappa$), the second-order coherence statistics do not deviate much from the laser statistics of $g^{(2)}[0]=1$, as discussed in \citet{Muller2015-il}. In the next section, we will explore several nonidealities that result in this relatively poor performance and ways to leverage these nonidealities for interesting physics and better nonclassical light generation.

%%%%%%%%%%%%%%%%%%%%%%%%%%%%%%%%%%%%%%%%%%%%%%%%%%%%%%%%%%%%%%%%
% Section 3
%%%%%%%%%%%%%%%%%%%%%%%%%%%%%%%%%%%%%%%%%%%%%%%%%%%%%%%%%%%%%%%%

\section{Single-Emitter Cavity QED Beyond the Jaynes--Cummings Model} \label{sec:seCQED}

There are several key considerations that alter the performance of a solid-state cavity QED system from that of an ideal Jaynes--Cummings system:

\begin{enumerate}

\item Strong dissipation caused by cavity loss rates comparable to the coherent interaction strength; dissipation decreases the fidelity of nearly all cavity QED effects \citep{Muller2015-om,Muller2015-il}.
\item Detuning of the emitter from the cavity; this is an important tool in improving the nonclassical light generated from a dissipative JC system \citep{Muller2015-il}.
\item Pulse-wise experiments, which are potentially more interesting for applications in quantum networks and are regardless required by the timing resolution of most single-photon detectors \citep{Fischer2016-pl}; the pulse length and pulse shape have a strong influence on the fidelity of nonclassical light generation \citep{Muller2015-om}.
\item The solid-state environment results in an important interaction with phonons \citep{Roy2011-pb}; this dissipation can both be a detrimental and a positive influence towards nonclassical light generation \citep{Muller2015-om}.
\item Temporal variation of the ground state of a quantum emitter, a phenomenon known as blinking \citep{santori2004-su}; when the emitter blinks, it either decouples from the cavity or is so far spectrally detuned that the system is effectively no longer strongly coupled \citep{Reinhard2011-ye}.
\item Interferometric effects owing to the photonic crystal's background density of states; the light transmitted through the cavity cannot simply be modeled as a single cavity mode, but also requires the modeling of a continuum scattering channel \citep{Fischer2016-dj,Muller2016-fs,Fischer2016-nk}.

\end{enumerate}

In carefully studying these effects, we have learned to significantly improve both the modeling of a solid-state strongly-coupled system and its ability to generate high-purity states of nonclassical light. Although this section details these nonidealities using investigations with III-V quantum emitters due to their technological maturity, the physics and the theory will be equally applicable to future experiments with group-IV systems.

%%%%%%%%%%%%%%%%%%%%%%%%%%%%%%%%%%%%%%%%%%%%%%%%%%%%%%%%%%%%%%%%

\subsection{Dissipative Structure of a Jaynes--Cummings System} \label{sec:dissipative}

The first nonideality that we explore in detail is the large dissipation present in solid-state cavity QED systems. Due to limitations in current material fabrication technologies for III-V and group-IV systems, the cavity dissipation rates are comparable to the coherent coupling rates. We explore this point through the language of non-Hermitian effective Hamiltonians \citep{Garraway1997-ii,Garraway1997-ap}. By using a non-Hermitian Hamiltonian, the effects of decay may be incorporated in a linear manner, but at the cost of generating an evolution equation that leaks probability. This non-Hermitian evolution plays an important role in the Monte-Carlo wavefunction or trajectory approach to quantum simulation \citep{Steck2007-qk,Carmichael2009-uc}: it governs the evolution of the wavefunction in-between photon emission events. Thus, the non-Hermitian Hamiltonian is useful because after a state has been prepared, its initial evolution and hence decay rate is set by the effective Hamiltonian. For a given system, such a Hamiltonian is
\begin{equation}
H_\textrm{\scriptsize EFF} = H_0 - \textrm{i} \sum_k\frac{\gamma_k}{2} c^\dagger_k c_k
\end{equation}
where $c_k$ is an arbitrary system operator and $\gamma_k$ is its decay rate. For the Jaynes--Cummings system specifically, this takes the form
\begin{equation}
H_\textrm{\scriptsize EFF} = H_\textrm{\scriptsize JC} - \textrm{i} \frac{\kappa}{2}a^\dagger a  - \textrm{i} \frac{\Gamma}{2}\sigma^\dagger \sigma\textrm{.}
\end{equation}

\begin{figure}
\textsf{
  \includegraphics[width=11.43cm]{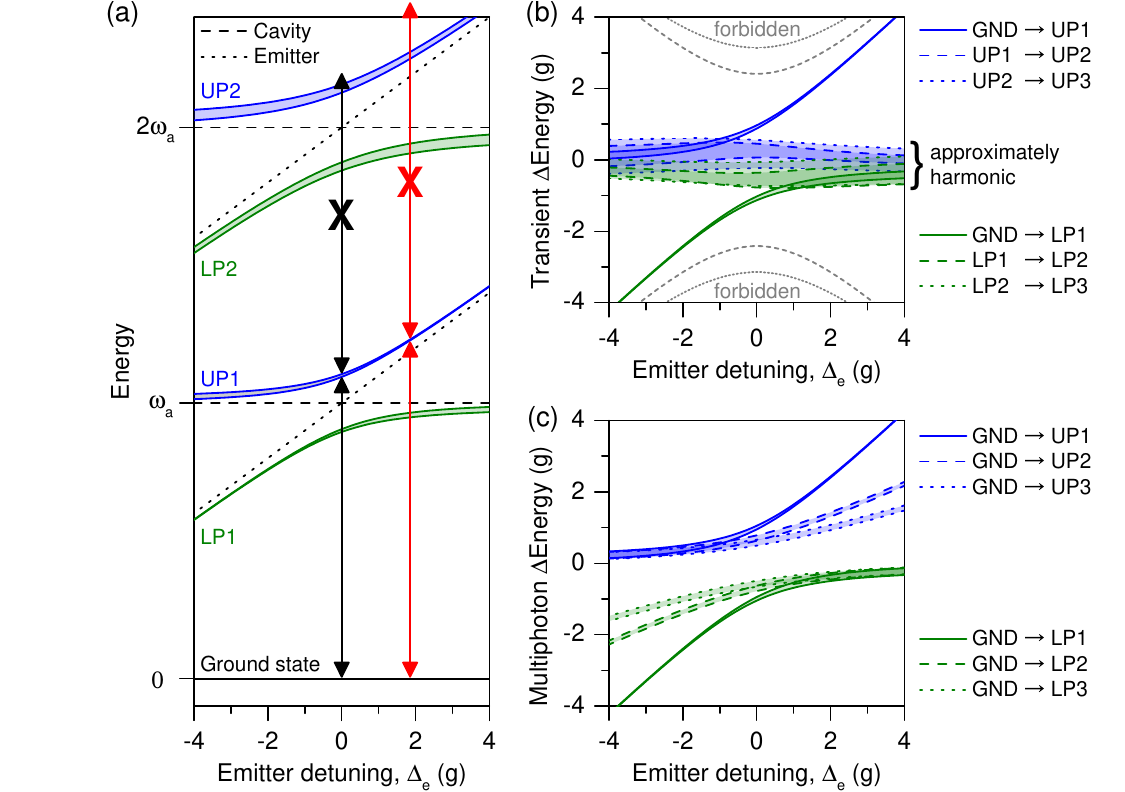}
  \caption{Energetic structure of a strongly-coupled system. (a) Dressed states of the Jaynes--Cummings ladder and their linewidths. Note: the separation between the rungs of the dressed-states ladder is not to scale. (b) Transient energies to climb the Jaynes--Cummings ladder rung-by-rung. (c) Resonant laser frequencies for multi-photon transitions. (a-c) The heights of the bounding regions represent the full-width half-maxes of the linewidths for each transition. The $g/\kappa=5$ ratio was used in order to cleanly illustrate the trends in each plot (which is now becoming achievable with state-of-the-art parameters for solid-state systems). UPn and LPn label the upper and lower polaritons, respectively.
}
\label{figure:4-1}}
\end{figure}

In a similar manner to diagonalizing $H_\textrm{\scriptsize JC}$ to obtain the dressed states in Section~\ref{sec:diag}, $H_\textrm{\scriptsize EFF}$ can be diagonalized to additionally obtain the loss rates of the dressed states \citep{Laussy2012-hr}. Diagonalization yields the complex eigenenergies for the $n$-th rung of the system to be
\begin{equation}E^n_\pm= n\omega_a + \frac{\Delta_\textrm{\scriptsize e}}{2} - \textrm{i} \frac{(2n-1)\kappa + \Gamma}{4} \pm \sqrt{\left( \sqrt{n}g\right)^2 - \left(\frac{\kappa-\Gamma}{4}+\textrm{i}\frac{\Delta_\textrm{\scriptsize e}}{2} \right)^2}\textrm{,}
\end{equation}
where the full-width half-maxes of the linewidths are given by the imaginary parts $\pm2\,\textrm{Im}\{E^n_\pm\}$. These complex eigenenergies are depicted in Fig. \ref{figure:4-1}a, with the bounding lines showing $\textrm{Re}\{E^n_+\}\pm\,\textrm{Im}\{E^n_+\}$ and $\textrm{Re}\{E^n_-\}\pm\,\textrm{Im}\{E^n_-\}$. One can derive the bounds for strong coupling by solving for when $\Delta_\textrm{\scriptsize e}=0$ and $\textrm{Re}\{E^n_+\}\ne\textrm{Re}\{E^n_-\}$, which occurs when $g>|\kappa-\Gamma|/4$.

For large emitter detunings~$\Delta_\textrm{\scriptsize e}$, the statement made in Section~\ref{sec:diag} that the polaritons take on primarily an electronic or photonic character can be understood even more intuitively: here, one polariton has approximately the emitter lifetime $\Gamma$ and the other the cavity lifetime~$\kappa$. At zero detuning, the upper (UP) polaritons have equivalent character to the lower (LP) polaritons. From this plot alone, we can already suspect a way to improve photon blockade in highly dissipative systems. Considering the black arrows for photon blockade at zero detuning where the laser is tuned in resonance with UP1, then the second photon is nearly resonant with UP2 due to the finite linewidths. Notably, this problem arises because the anharmonicity of the Jaynes--Cummings ladder scales as $\sqrt{n}$, while the decay rates scale as $n$. However, by introducing a small detuning, the effective non-linearity of the system is much higher. Now consider the red arrows for detuned photon blockade, where UP2 is much further off resonance in comparison. Thus, we would expect better quality photon blockade to occur for the detuned system.

This information can also be visualized in an alternative manner, as shown in Fig. \ref{figure:4-1}b. Here, the energies to climb the dressed-states ladder one-by-one \citep{Laussy2012-hr} are shown, with their linewidths, by plotting
\begin{equation}
\Delta E^n_{++} = \textrm{Re}\{E^{n}_+\}-\textrm{Re}\{E^{n-1}_+\}\pm\left(\textrm{Im}\{E^n_+\}+\textrm{Im}\{E^{n-1}_+\}\right)
\end{equation}
and
\begin{equation}
\Delta E^n_{--} = \textrm{Re}\{E^{n}_-\}-\textrm{Re}\{E^{n-1}_-\}\pm\left(\textrm{Im}\{E^n_-\}+\textrm{Im}\{E^{n-1}_-\}\right)\textrm{.}
\end{equation}
(Note, we define $E^0_\pm=0$). The linewidths add for the eigenstates involved in $E^n_{++}$ and $E^n_{--}$ because the transient energies are used to consider a resonant process that excites the the system rung-by-rung. For small detunings, the jumps to the second (dashed) and third (dotted) lines are almost on top of the first jump from the ground state. Thus, we would not expect good quality nonclassical light generation to come from the system. On the other hand, for non-trivial detunings the effective anharmonicity increases linearly because the jump to the first rung differentially increases relative to the second-rung with detuning. After the first jump, however, the ladder is relatively harmonic since the higher jumps are centered around ${\Delta E=0}$. Importantly, these jumps occur along either the upper or lower branches, but not between branches. Jumps between branches are disallowed for the same reason the transmission dips to zero in-between the upper and lower polaritons (as seen in Fig. \ref{figure:3-2}). Specifically, the transitions between branches, $\Delta E^n_{+-} = \textrm{Re}\{E^{n}_+\}-\textrm{Re}\{E^{n-1}_-\}$ and $\Delta E^n_{-+} = \textrm{Re}\{E^{n}_-\}-\textrm{Re}\{E^{n-1}_+\}$, are forbidden because the states have cavity components that are shifted by $\pi$ phase and hence zero dipolar overlap between ${\langle n-1,\mp |a| n,\pm\rangle=0}$. We have plotted the forbidden transitions as dashed/dotted lines with no bounding regions. These transitions occur through emission by the quantum emitter, i.e. $\langle n-1,\mp |\sigma| n,\pm\rangle\ne0$, but at a negligible rate due to typical solid-state cavity QED parameters.

Finally, we discuss the multi-photon structure of the dissipative Jaynes--Cummings ladder. Just like for the photon blockade argument where the detuning increased the effective nonlinearity between the first and second rungs of the dressed-states ladder, increasing the emitter detuning has the effect to separate the multi-photon transitions. This can be seen in Fig. \ref{figure:4-1}c, where the multiphoton transitions fan out with detuning. Interestingly, the absorption linewidths for multi-photon emission are all $\kappa$ after the first blockaded rung, because the multi-photon transitions occur between the ground state and an upper dressed state in an idealized model that assumes the intermediate levels remain unpopulated. (Certainly, this approximation breaks down when the multi-photon transitions strongly overlap, but then it's difficult to identify linewidths for distinctive multi-photon processes anyway.) In the idealized model, the target dressed state determines the linewidth since the ground state has approximately zero dephasing. Although the dephasing rate of the $n$th dressed state scales with $n$, the $n$-photon absorption linewidth scales with $1/n$ because the laser detuning is compounded by the number of photons involved in the process. Thus, the multi-photon absorption linewidths are left at $\kappa$, so we plot
\begin{equation}
\textrm{Re}\{E^{n}_+\}/n-\omega_a\pm \kappa/2
\end{equation}
and
\begin{equation}
\textrm{Re}\{E^{n}_-\}/n-\omega_a\pm \kappa/2\textrm{.}
\end{equation}
We note that it's possible to derive an effective multi-photon absorption Hamiltonian based on adiabatic elimination of the unpopulated states which rigorously supports this analysis \citep{Linskens1996-yh}. 

%%%%%%%%%%%%%%%%%%%%%%%%%%%%%%%%%%%%%%%%%%%%%%%%%%%%%%%%%%%%%%%%

\subsection{Emitter-Cavity Detuning} \label{sec:det}

\begin{figure}
\textsf{
  \includegraphics[width=11.43cm]{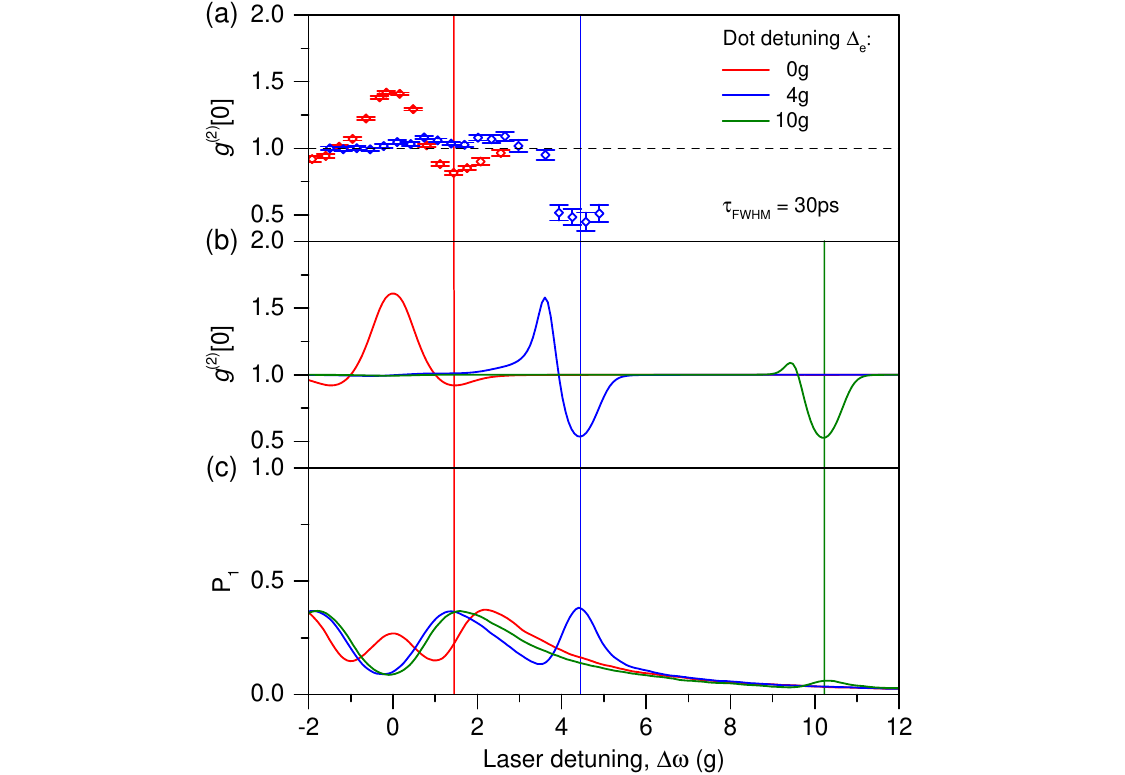}
  \caption{Detuned photon blockade, experiment and theory. (a) Measured second-order coherence $g^{(2)}[0]$ as a function of laser detuning for both a tuned and detuned strongly-coupled system based on an InGaAs quantum dot, \textit{data from \citet{Muller2015-il}.} Horizontal black dashed line represents statistics of the incident laser pulses. (b) Theoretical $g^{(2)}[0]$ as a function of laser detuning for one resonant and two detuned strongly-coupled systems. (c) Single-photon detection probability $P_1$ of the photocount distribution $P_n$, calculated using a quantum trajectory approach. The theoretical model for the simulations is a pure Jaynes--Cummings system with a time-dependent driving term.
}
\label{figure:4-2}}
\end{figure}

In this section, we experimentally and numerically consider the effects of photon blockade by measuring $g^{(2)}[0]$ as a function of the excitation laser frequency, for multiple emitter-cavity detunings. We excited an InGaAs-based strongly-coupled system with short Gaussian pulses and experimentally measured the degree of pulse-wise second-order coherence, as shown in Fig. \ref{figure:4-2}a. [For the tuned system (red), the correlation statistics are similar to the ones presented in Fig. \ref{figure:3-4}b.] However, when the emitter detuning $\Delta_\textrm{\scriptsize e}$ is increased to $4g$, the quality of the photon blockade is increased dramatically. This can be seen in the significantly decreased value of $g^{(2)}[0]$ at the vertical red line compared to the vertical blue line. This provides experimental evidence for our discussion in the Section 3.1 of how the emitter detuning modulates the effective anharmonicity of the system. [We note that in this set of experiments the cross-polarized suppression was not optimized for large detunings so in the tunneling region, where less light is emitted, the statistics were dominated by unwanted coherently scattered light \citep{Muller2015-il}].

To ensure that we fully understand the experimental behaviors, we discuss a numerical model for capturing the observed trend in photon blockade. As briefly mentioned in Section~\ref{sec:4}, the measurements of the degree of second-order coherence are performed in a pulsed manner, and hence we must adjust our master equation accordingly. Specifically, the driving Hamiltonian changes to $H_\textrm{\scriptsize drive} = \mathcal{E}(t)\left( a \,\textrm{e}^{\textrm{\footnotesize i}\omega_\textrm{\tiny L} t} + a^\dagger \textrm{e}^{-\textrm{\footnotesize i}\omega_\textrm{\tiny L} t}\right)$, where $\mathcal{E}(t)$ is the time-dependent driving strength. As before, we use the same rotating frame transformation to remove the time-dependence of the excitation term so that $\tilde{H}_\textrm{\scriptsize drive}(t) = \mathcal{E}(t)\left( a + a^\dagger\right)$ and written as a super-operator
\begin{equation}
\tilde{\mathcal{L}}_\textrm{\scriptsize drive}(t)\tilde{\rho}(t) =\textrm{i}\left[ \tilde{\rho}(t), \tilde{H}_\textrm{\scriptsize drive}(t)\right] \textrm{.}
\end{equation}
Typically, a Gaussian pulse shape represents experiment, where $\mathcal{E}(t)=\mathcal{E}_0\textrm{e}^{-t/2\tau_p^2}$ and $\tau_p=\tau_\textrm{\tiny FWHM}/2\sqrt{\textrm{ln}\,2}$ is the Gaussian pulse parameter. The overall system evolution is governed by the Liouvillian $\tilde{\mathcal{L}}(t)=\tilde{\mathcal{L}}_\textrm{\scriptsize JC}+ \tilde{\mathcal{L}}_\textrm{\scriptsize drive}(t)$.

This Liouvillian can be used to calculate the measured degree of second-order coherence in two ways. The first is to unravel $\tilde{\mathcal{L}}(t)\tilde{\rho}(t)$ into a quantum trajectory equation, approximate the expected photocount distribution $P_n$ over the entire pulsed emission from an ensemble of trajectories, and estimate $g^{(2)}[0]=\sum_n P_n n (n-1)/\left(\sum_n P_n n \right)^2$ \citep{Carmichael2009-uc}. The second way is to use a time-dependent form of the quantum regression theorem to calculate
\begin{equation} \label{eq:g2}
g^{(2)}[0] = \frac{\int_0^T \int_0^T \mathop{\textrm{d} t} \mathop{\textrm{d} t'} \, \langle \mathcal{T}_-[a^{\dagger}(t) a^{\dagger}(t')] \mathcal{T}_+[a(t') a(t)] \rangle}{\left(\int_0^T \mathop{\textrm{d} t} \, \langle a^{\dagger}(t) a(t) \rangle\right)^2}\equiv\frac{G^{(2)}[0]}{\langle N \rangle^2}\textrm{,}
\end{equation}
where the time range $0\rightarrow T$ encompasses the entire emission pulse and the operators $\mathcal{T}_\pm$ indicate the time-ordering required of a physical measurement (operators with higher time indices towards the center of the expression) \citep{Fischer2016-pl}. The second method is typically more computationally efficient and has an intellectually satisfying connection to instantaneous two-time correlations.

Using this method and a best fit to the Jaynes--Cummings model with no dephasing (yielding $\{g/2\pi, \kappa/2\pi\} = \{10.9\,\textrm{GHz}, 10\,\textrm{GHz} \}$), we use the driving strength $\mathcal{E}_0$ as a fitting parameter for the observed photon blockade regions (Fig.~\ref{figure:4-2}b). Just as experimentally measured, the blockade dip grows with increasing detuning from $\Delta_\textrm{\scriptsize e}=0g$ to $\Delta_\textrm{\scriptsize e}=4g$, which further supports the pictorial description of detuned blockade from the Section 3.1. The simulations do show a difference compared to this intuition for the very large detuning of $\Delta_\textrm{\scriptsize e}=10g$: the blockade dip saturates. With increasing detuning, the oscillator strength of the emitter-like polariton decreases until its emission strength is comparable to off-resonant transmission through the cavity-like polariton. The light from the cavity-like polariton then begins to destroy the photon blockade.

Because the emitter-like polariton has a smaller oscillator strength with increasing detuning, one might be concerned with how the efficiency of single-photon generation is affected by changing the emitter detuning \citep{Muller2015-il}. To answer this question, we simulated the probabilities of single-photon generation $P_1$ using quantum Monte-Carlo techniques and present these data in Fig. \ref{figure:4-2}. Surprisingly, the probability for single-photon generation actually increases for small emitter detunings by approximately a factor of two over the case in resonant photon blockade (compare $P_1$ at the red and blue lines). This occurs because the highly dissipative nature of the strongly-coupled system spoils the blockade so completely on resonance, that with increasing detuning there are plenty of multi-photon photocounts that can be suppressed and converted to single-photon counts. Of course, this effect wears out with large enough detuning and the $P_1$ for $10g$ is noticeably less efficient. Thus, emitter detuning has been shown both intuitively, experimentally, and theoretically to be a valuable mechanism for enhancing nonclassical light generation.

We make two brief comments on the figure:
\begin{enumerate}
\item In the photon tunneling regions, the bunching values of $g^{(2)}[0]$ are largest when the minimum amount of light is transmitted. This can easily be seen by comparing the maximum simulated bunching in the blue curve to the local minimum value of $P_1$ near the blockade region. Such bunching behavior is consistent with having a photocount distribution that has multi-photon components that are emphasized over a coherent state, but additionally has a large vacuum component \citep{Rundquist2014-kf}. For instance, although the photocount distribution $P_2 = 1$ anti-bunches with $g^{(2)}[0]=0.5$, the distribution $\{P_0, P_2\} = \{ 0.75, 0.25\}$ bunches with $g^{(2)}[0]=2$. This discussion further suggests that a highly dissipative Jaynes--Cummings system by itself is not necessarily ideal for the generation of multi-photon states.
\item The modeled photon blockade is actually \textit{weaker} for the ideal Jaynes--Cummings system than experimentally measured, even for arbitrarily low powers. Thus, even though the theory and experiment do not perfectly match, we trust the strength of our general arguments in this section. This additionally suggests missing physics from the model of the solid-state strongly-coupled system, which will be thoroughly discussed in the latter sections of this section.
\end{enumerate}

%%%%%%%%%%%%%%%%%%%%%%%%%%%%%%%%%%%%%%%%%%%%%%%%%%%%%%%%%%%%%%%%

\subsection{Excitation Pulse Length}

\begin{figure}
\textsf{
  \includegraphics[width=11.43cm]{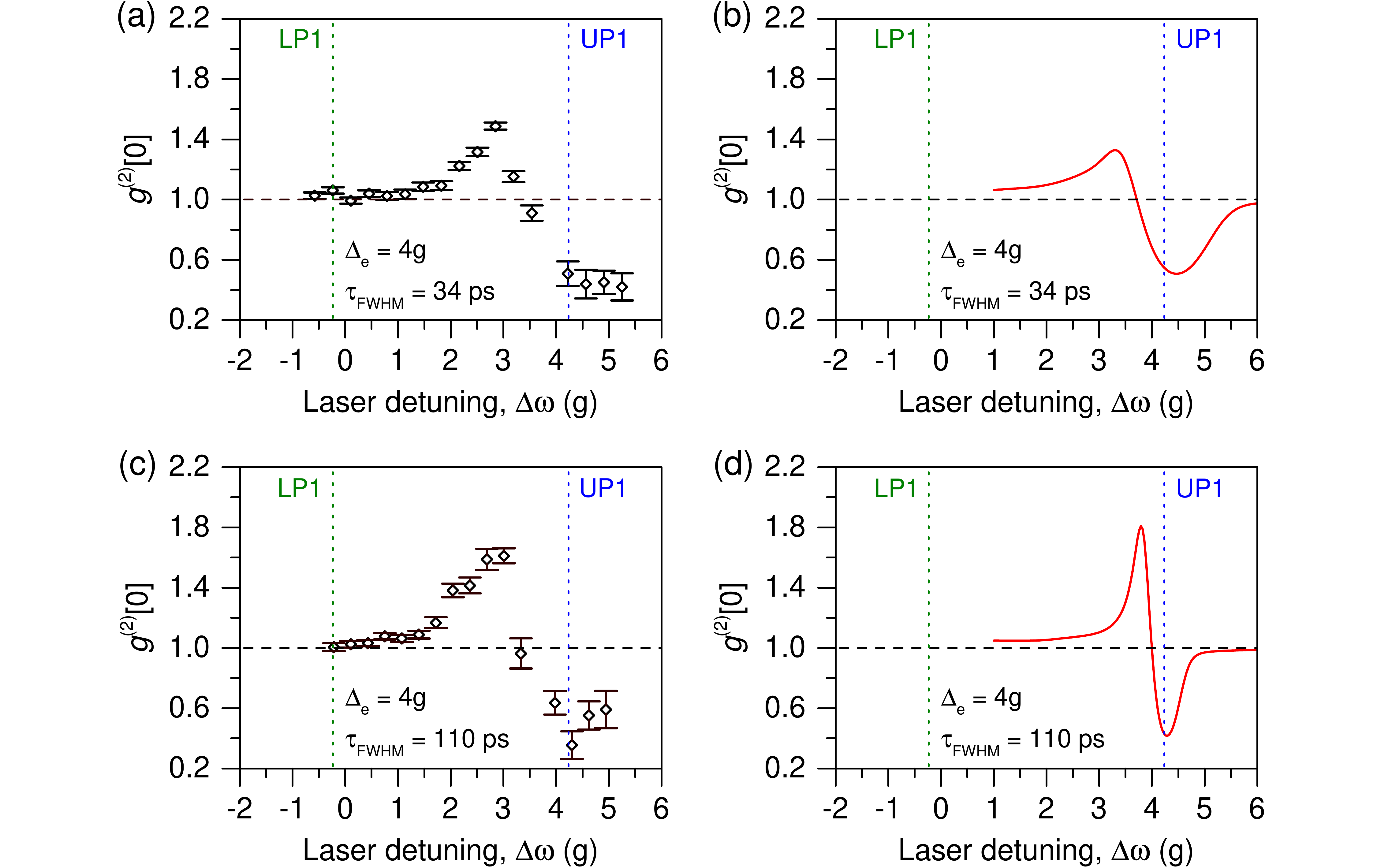}
  \caption{Pulse-length dependence of photon blockade. Experimentally measured degrees of second-order coherence $g^{(2)}[0]$ as a function of laser detuning for a detuned strongly-coupled system based on an InGaAs quantum dot under short (a) and long (c) pulses. Theoretical $g^{(2)}[0]$ as a function of laser detuning for a detuned strongly-coupled system under (b) short and (d) longer pulses. The theoretical model for the simulations is a pure Jaynes--Cummings system with a time-dependent driving term. Horizontal black dashed lines represent statistics of the incident laser pulses. \textit{Data from \citet{Muller2015-om}.}
}
\label{figure:4-3}}
\end{figure}

As has recently been shown in depth \citep{Muller2015-om}, the length of an excitation pulse is critically important to optimizing emission from photon blockade. Specifically, optimizing photon blockade in a detuned strongly-coupled system by changing the pulse length is a careful trade-off between avoiding excitation of the higher rungs for short pulses or re-excitation of the first polariton for long pulses. Here, we show experimental results from an InGaAs-based system in \citet{Muller2015-om}; Figs. \ref{figure:4-3}a and \ref{figure:4-3}c) and discuss their comparison with simulated photon blockade in a pulsed regime from a Jaynes--Cummings system (Figs.~\ref{figure:4-3}b and \ref{figure:4-3}d). The blockade dip has two important characteristics: its depth and its width. Here, the better pulse length of the two is seen to be 110\,ps since this pulse length better optimizes between higher rung excitation and lower rung re-excitation. Regarding the blockade dip, it is widest for short pulses but narrowest for long pulses; because the detuned polariton has a relatively long lifetime, the width of the blockade dip is roughly determined by the spectral width of the laser pulse. These trends are clearly observable in both experiment and theory.

Although the trends of the blockade dips are again reproduced well by the pulsed Jaynes--Cummings model, the agreement is only qualitative. For instance, the blockade dips are again smaller in simulation than in experiment, but now shown for multiple pulse lengths. This suggests that the disparity is not simply an error in the pulse length but rather additional physics. Especially in this set of experimental data where the cross-polarized setup was better optimized for suppression than in Fig. \ref{figure:4-2}a, the simulated tunneling regions quite poorly agree with the experimental data. While we supposed initially that the experimental difference was potentially an imprecision in our ability to accurately determine the laser detuning or an experimental drift, further investigation has shown the difference to be the result of an unexpected interference effect. We will discuss this effect later.

%%%%%%%%%%%%%%%%%%%%%%%%%%%%%%%%%%%%%%%%%%%%%%%%%%%%%%%%%%%%%%%%

\subsection{Electron-Phonon Interaction}

In this section we discuss the first true non-ideality to our model for the strongly-coupled systems: the effect of phonons on solid-state cavity QED platforms. Any subsystem embedded within a solid-state environment, at any temperature, has the potential to feel the effects of interactions with the phonon bath. In particular, the electron-phonon interaction between solid-state emitters and their environment can lead to a variety of incoherent phenomena. Due to the differing sizes of the III-V quantum dots and group-IV color-centers, however, the types of phononic excitations they couple to are different. The physically large size of quantum dots lends well towards coupling of only acoustic phonons \citep{Roy2011-pb}, while the small size of color-centers lends well towards coupling of both acoustic and optical phonons \citep{davies1974vibronic}. In this section, we discuss how phonon coupling manifests in single-emitter cavity QED systems using an InGaAs-based device as an example.

\begin{figure}
\textsf{
  \includegraphics[width=11.43cm]{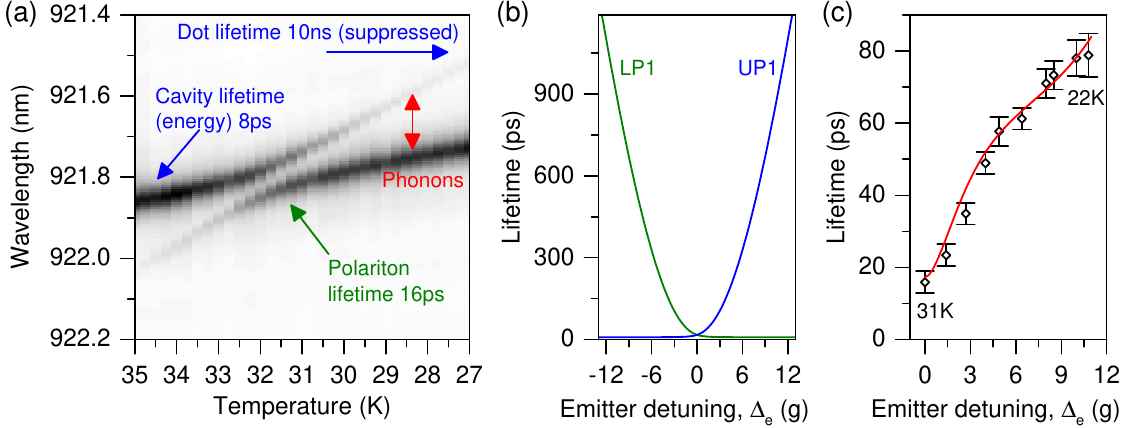}
  \caption{Effects of electron-phonon interaction on polariton lifetimes. (a) Annotated avoided-crossing plot that matches transmission spectra to measured lifetimes. (b) Lifetimes as calculated from the cavity and dot dissipation rates alone in the Jaynes--Cummings model. (c) Measured lifetimes from a typical solid-state cavity QED system, which are an order of magnitude shorter for large detunings than those predicted by a pure Jaynes--Cummings model. \textit{Data from \citet{Muller2015-om}.}
}
\label{figure:4-4}}
\end{figure}

First, consider the avoided crossing spectrum of a typical InGaAs-based system \citep{Muller2015-om}, reproduced in Fig. \ref{figure:4-4}a. Now, the spectra are annotated according to the state lifetimes. The measured cavity lifetime is $1/\kappa=8$\,ps, while the suppressed dot lifetime is expected to be approximately 10\,ns, and hence the maximally entangled polariton has a lifetime of 16\,ps since it is half electronic and half photonic character. As a function of the emitter detuning, the expected lifetimes of the LP1 and UP1 polaritons from the Jaynes--Cummings model are plotted in Fig. \ref{figure:4-4}b. However, from experimental measurements performed on a streak camera the lifetimes are an order of magnitude shorter for large detunings (Fig.~\ref{figure:4-4}c). This difference results from the electron-phonon interaction in solid-state cavity QED systems.

While the primary effect of electron-phonon interaction for quantum dots embedded in bulk GaAs is to generate a power-dependent damping, the effect in a cavity QED system is dramatically different. Because only acoustic phonons couple to the dots, there is an arbitrarily small density of phononic states to couple to under weak driving in bulk. However, in a cavity QED system the dressed ladder provides a constant energy difference between polaritons whereby the electron-phonon interaction samples the density of states, resulting in constant phonon emission and absorption. Three models have primarily been used to explore this effect in the solid-state:
\begin{enumerate}
\item Non-markovian models such as a path-integral form of the system dynamics \citep{Vagov2011-ew}. This formalism includes phonon effects to all orders.
\item The polaron master equation, which is a powerful formalism for including phonon effects to all orders in a Markovian model \citep{Roy2011-pb}.
\item The effective phonon master equation, which only includes first-order phonon effects \citep{Roy2012-gn,Muller2015-om}. This model is by far the simplest, but readily captures most experimental effects observed thus-far in solid-state cavity QED systems.
\end{enumerate}

For all of the nonclassical light generation phenomenon we have observed thus far, we have found the effective phonon master equation sufficient, and we will briefly review its findings here. For a cavity-driven system only (driven by $H_\textrm{\scriptsize drive}$), then the effects of phonons are captured by the addition of two incoherent dissipators $\frac{\Gamma_\textrm{\scriptsize f}}{2}\mathcal{D}[a^\dagger\sigma]$ and $\frac{\Gamma_\textrm{\scriptsize r}}{2}\mathcal{D}[\sigma^\dagger a]$, i.e. with the addition of the new Liouvillian term
\begin{equation} \label{eq:dissipators}
\tilde{\mathcal{L}}_\textrm{\scriptsize phonon}= \frac{\Gamma_\textrm{\scriptsize f}}{2}\mathcal{D}[a^\dagger\sigma] + \frac{\Gamma_\textrm{\scriptsize r}}{2}\mathcal{D}[\sigma^\dagger a]
\end{equation}
where $\Gamma_\textrm{\scriptsize f}$ and $\Gamma_\textrm{\scriptsize r}$ stand for forward and reverse phonon transfer rates, respectively. At the operating temperatures of 25\,K and small emitter detunings used in our experiments, these rates are each approximately $\Gamma_{f,r}\approx(80\,\textrm{ps})^{-1}$. Importantly, these rates vary with detuning and fall off rapidly at large detunings; full details for the extracted rates can be found in \citet{Muller2015-om}.

Because the dissipators in equation~(\ref{eq:dissipators}) transfer excitations between the cavity and the dot, at large detunings they result in an effective transfer of population between polaritons. The transfer rates then dominate the lifetime of the emitter-like polariton for large detunings because the system emits through a phonon-induced transfer (Fig.~\ref{figure:4-4}a red arrows) and cavity emission over dot emission. Specifically, the phonon-involved pathway has a lifetime of $\approx80\,\textrm{ps}+ 8$\,ps, which is much shorter than the dot's spontaneous lifetime of $\approx$10\,ns.

%%%%%%%%%%%%%%%%%%%%%%%%%%%%%%%%%%%%%%%%%%%%%%%%%%%%%%%%%%%%%%%%

\subsection{Blinking of the Quantum Emitter} \label{sec:blink}

While we hope the quantum emitter behaves as an ideal two-level system and is described by a single dipole operator $\sigma$, in reality it possesses a very complicated level structure. Consider InGaAs quantum dots: when placed in a high-quality electrical diode to control their precise ground states through controlling the local charge environment, they have been shown to behave as nearly ideal two-level systems \citep{Kuhlmann2015-hh}. However, this technology was later introduced into devices with planar photonic crystal cavities \citep{Laucht2008-ce,Warburton2000-rf,Carter2013-bm}, and therefore, much of the work on solid-state strongly-coupled systems still shows residual effects of a slightly unstable charge environment. Because the different charge configurations of the quantum dot have different binding energies, and hence emission frequencies, when the quantum dot sequentially absorbs a charge from the environment \citep{santori2004-su} its resonant frequency is highly detuned from the cavity and the strong coupling is no longer visible \citep{Reinhard2011-ye,Rundquist2014-kf}. During this time, the system behaves like the bare cavity by itself and coherently scatters the incident light with a Lorentzian lineshape. When this process periodically occurs, slowly modulating the transmission spectrum between that of a resonant strongly-coupled system and a bare cavity, the phenomenon is referred to as blinking. Although the experiments we discuss are with III-V quantum dots, group-IV color-centers have also been shown to blink \citep{Bradac2010-zl,Castelletto2013-qf}. Thus, the general physics of blinking in strongly-coupled systems, which we will elaborate in this section, applies to future realizations of cavity QED systems with group-IV quantum emitters.

\subsubsection{Effects on Transmission Spectra} \label{sec:blink}

While it's certainly possible to provide a complete description of the charge states of a quantum emitter, in many situations it is sufficient to model the system as behaving as the bare cavity for some fraction of the time $f_\textrm{\scriptsize blink}$ and like a strongly-coupled system for the rest. The statistical independence of these two situations holds because the blinking timescales are at least 10's of nanoseconds \citep{Davanco2014-ri}, while the emission timescales are 10's of picoseconds. When the system blinks, we calculate the steady-state density matrix by solving
\begin{equation}
\left(\tilde{\mathcal{L}}_\textrm{\scriptsize blink}(\omega_\textrm{\tiny L}) + \tilde{\mathcal{L}}_\textrm{\scriptsize drive}\right)\tilde{\rho}'_\textrm{\scriptsize ss}(\omega_\textrm{\tiny L}) = 0\textrm{,}
\end{equation}
where the bare cavity dynamics are represented by
\begin{equation}
\tilde{\mathcal{L}}_\textrm{\scriptsize blink}\tilde{\rho}'_{ss}=\textrm{i}\left[\tilde{\rho}'_{ss} ,\left( \omega_a-\omega_\textrm{\tiny L} \right)a^\dagger a\right]\textrm{.}
\end{equation}
Then, the transmission spectrum while blinking is $S_\textrm{\scriptsize blink}(\omega_\textrm{\tiny L})=\langle a^\dagger a \rangle(\omega_\textrm{\tiny L})$, using $\tilde{\rho}'_{ss}(\omega_\textrm{\tiny L})$.

\begin{figure}
\textsf{
  \includegraphics[width=11.43cm]{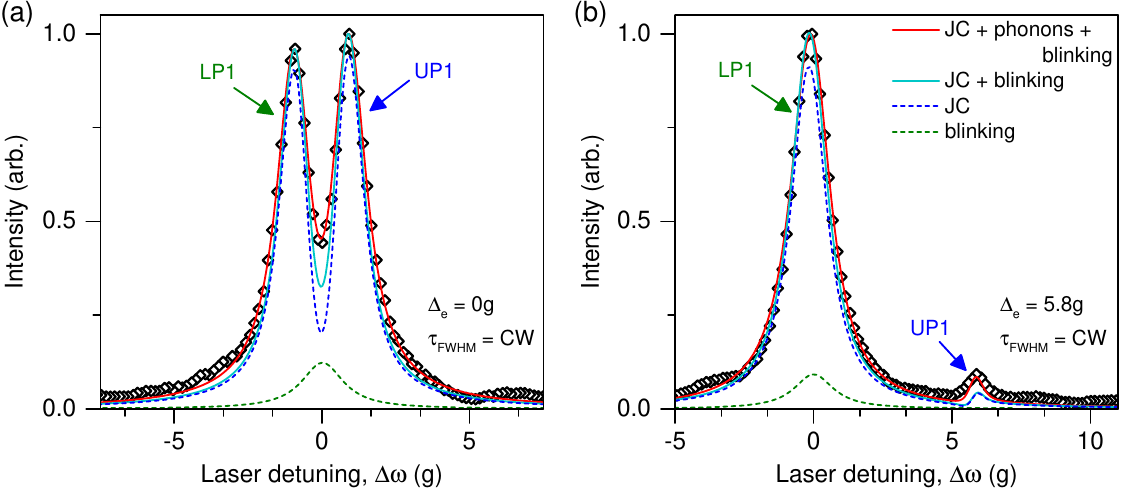}
  \caption{Transmisison spectra from an InGaAs quantum dot strongly coupled to an L3 photonic crystal cavity mode, taken in cross-polarized reflectivity. (a) Spectrum at nearly zero dot detuning. (b) Spectrum at a large dot detuning. The spectra were taken using weak broad-band diode, which is equivalent to transmission in the linear regime of cavity QED. Simulated spectra were convolved with the response function of the spectrometer (linewidth $\Gamma_\textrm{\tiny FWHM}/2\pi = 4.5$\,GHz), which is why the spectra for pure Jaynes--Cummings transmission do not dip to zero between LP1 and UP1.
}
\label{figure:4-5}}
\end{figure}

To incorporate the experimental strongly-coupled system that is often modeled as a Jaynes--Cummings system with phonons, we calculate the steady-state density matrix by solving
\begin{equation}
\left(\tilde{\mathcal{L}}_\textrm{\scriptsize JC}(\omega_\textrm{\tiny L}) + \tilde{\mathcal{L}}_\textrm{\scriptsize phonon} + \tilde{\mathcal{L}}_\textrm{\scriptsize drive}\right)\tilde{\rho}_\textrm{\scriptsize ss}(\omega_\textrm{\tiny L}) = 0\textrm{.}
\end{equation}
Using $\tilde{\rho}_\textrm{\scriptsize ss}(\omega_\textrm{\tiny L})$, we calculate the transmission spectrum of the strongly-coupled system $S_\textrm{\scriptsize sc}(\omega_\textrm{\tiny L})=\langle a^\dagger a \rangle(\omega_\textrm{\tiny L})$.

Finally, we weight and combine the two transmission profiles to obtain
\begin{equation}
S_\textrm{\scriptsize T}(\omega_\textrm{\tiny L}) = \left(1-f_\textrm{\scriptsize blink}\right)S_\textrm{\scriptsize sc}(\omega_\textrm{\tiny L}) + f_\textrm{\scriptsize blink} S_\textrm{\scriptsize blink}(\omega_\textrm{\tiny L})\textrm{.}
\end{equation}
With this quantum-optical model, we are ready to fit realistic transmission spectra with $S_\textrm{\scriptsize T}(\omega_\textrm{\tiny L})$. Experimental transmission spectra from an InGaAs quantum dot strongly coupled to an L3 photonic crystal cavity mode are shown in Fig. \ref{figure:4-5}; in Fig. \ref{figure:4-5}a the system is tuned nearly in resonance, and in Fig. \ref{figure:4-5}b the system is tuned significantly off resonance. In this model, we used the phonon-induced dot-cavity transfer rates extracted from the data in Fig. \ref{figure:4-4}, while $g$, $\kappa$, and $f_\textrm{\scriptsize blink}$ were taken as fitting parameters. From the fits, values of $\{g/2\pi, \kappa/2\pi\}=\{9.2\,\textrm{GHz}, 12.3\,\textrm{GHz}\}$ and $f_\textrm{\scriptsize blink}=0.09$ were extracted; by fitting one spectra on- and one spectra off-resonance $\kappa$ and $g$ can be extracted almost independently of one another.

We discuss several interesting features in these transmission spectra. Building the spectra component by component, we first show the blinking spectra and pure Jaynes--Cummings spectra as the dashed green and blue lines, respectively. The simulated spectra were convolved with the response function of the spectrometer (linewidth $\Gamma_\textrm{\tiny FWHM}/2\pi = 4.5$\,GHz), which is why the spectra for pure Jaynes--Cummings transmission do not dip to zero between LP1 and UP1. Then, the two spectra were added together (cyan lines) to show the effects of the blinking term: blinking decreases the visibility of the strong coupling dip in the resonant case and increases the height of the cavity-like polariton in the off-resonant case. Finally, we added  the effects of electron-phonon interaction to fully model our strongly-coupled system. The effect of phonons is to decrease the depth of the transmission dip in the resonant case due to the additional incoherent dephasing. With this dephasing also comes a small increase in the linewidths of the polaritons. Meanwhile in the off-resonant case, the effect of the electron-phonon interaction is to reduce the lifetime of the emitter-like polariton (small peak) so that the polariton can emit at a faster rate, resulting in a higher count rate than without the interaction.

%%%%%%%%%%%%%%%%%%%%%%%%%%%%%%%%%%%%%%%%%%%%%%%%%%%%%%%%%%%%%%%%

\subsubsection{Effects on Nonclassical Light Generation}

Although the effects of blinking resulted in simply adding the strongly-coupled and blinking spectra together, this approach is not sufficient for photon statistics. To understand the effect of blinking on nonclassical photon statistics, consider the standard setup for measuring pulsed $g^{(2)}[0]$, as thoroughly discussed in \citet{Fischer2016-pl}. Using a Hanbury--Brown and Twiss interferometer operated in a pulsed manner a binned temporal coincidence histogram is built up, with peaks separated by the pulse reptition rate $\tau_r$, i.e. $h_\textrm{\scriptsize HBT}[n\tau_r]$ where $n\in\{\mathbb{Z}\ge0\}$. The measured degree of second-order coherence is estimated by taking the ratio
\begin{equation}
\hat{g}^{(2)}[0]=\frac{h_\textrm{\scriptsize HBT}[0]}{h_\textrm{\scriptsize HBT}[\tau_r]}\textrm{,}
\end{equation}
where each histogrammed time-bin $h_\textrm{\scriptsize HBT}[n\tau_r]$ is an independent and binomially-distributed random variable. Here, we additionally consider the effects of blinking. Often the blinking time-scale is very long compared to~$\tau_r$, where the counts due to transmission through the strongly-coupled system or the blinking system are modeled by adding their individual histograms together, i.e. $h_\textrm{\scriptsize HBT}[n\tau_r]=h_\textrm{\scriptsize SC}[n\tau_r]+h_\textrm{\scriptsize blink}[n\tau_r]$. Hence,
\begin{equation}
\hat{g}^{(2)}[0]=\frac{h_\textrm{\scriptsize SC}[0]+h_\textrm{\scriptsize blink}[0]}{h_\textrm{\scriptsize SC}[\tau_r]+h_\textrm{\scriptsize blink}[\tau_r]}\textrm{.}
\end{equation}
Because the statistics of the transmitted light are directly inherited from the laser when the system blinks, $h_\textrm{\scriptsize blink}[n\tau_r]\propto\langle N_\textrm{\scriptsize blink}  \rangle^2 = \left( \int_0^T \mathop{\textrm{d} t} \, \langle a^{\dagger}(t) a(t) \rangle \right)^2$ and is calculated using the Liouvillian $\mathcal{L}(t)=\tilde{\mathcal{L}}_\textrm{\scriptsize blink}+ \tilde{\mathcal{L}}_\textrm{\scriptsize drive}(t)$. Then, in terms of the instantaneous correlations of the system and by extension of equation~(\ref{eq:g2})
\begin{equation} \label{eq:22}
g^{(2)}[0] = \frac{\left(1-f_\textrm{\scriptsize blink}\right)g^{(2)}_\textrm{\scriptsize sc}[0]\langle N_\textrm{\scriptsize sc}  \rangle^2 + f_\textrm{\scriptsize blink}\langle N_\textrm{\scriptsize blink} \rangle^2}{\left(1-f_\textrm{\scriptsize blink}\right)\langle N_\textrm{\scriptsize sc}  \rangle^2+f_\textrm{\scriptsize blink}\langle N_\textrm{\scriptsize blink}  \rangle^2}\textrm{,}
\end{equation}
where $ \langle N_\textrm{\scriptsize sc}\rangle$ and $g^{(2)}_\textrm{\scriptsize sc}[0]$ are calculated using the Liouvillian
\begin{equation}
\mathcal{L}_\textrm{\scriptsize sc}(t)=\tilde{\mathcal{L}}_\textrm{\scriptsize JC}+\tilde{\mathcal{L}}_\textrm{\scriptsize phonon}+ \tilde{\mathcal{L}}_\textrm{\scriptsize drive}(t)\textrm{.}
\end{equation}

\begin{figure}
\textsf{
  \includegraphics[width=11.43cm]{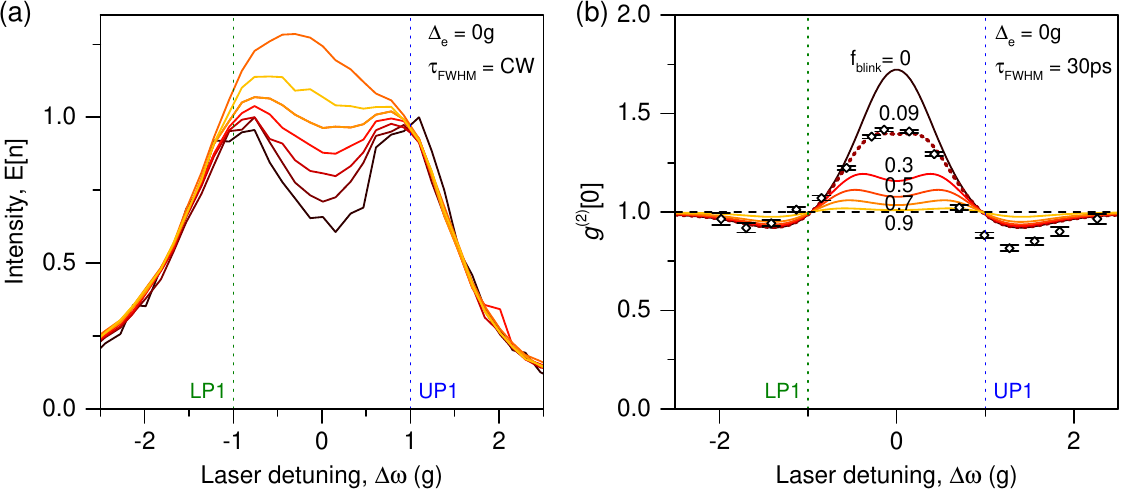}
  \caption{Effects of blinking on a resonant strongly-coupled system. (a) Normalized CW transmission spectra with increasing blinking rate (black to orange). (b) Pulse-wise second-order coherence ${g}^{(2)}[0]$ as a function of laser detuning with increasing blinking rate. Horizontal black dashed line represents statistics of the incident laser pulses. \textit{Experimental data for ${g}^{(2)}[0]$ from \citet{Muller2015-il}.}
}
\label{figure:4-6}}
\end{figure}

The strongly-coupled system examined in this section had a relatively low fraction of blinking time, at only $f_\textrm{\scriptsize blink}=9$\,\%. However, many quantum dots may have higher blinking fractions due to variation in their local charge environments. Here, we explore the effects of stronger blinking, both on the transmission spectrum and on the second-order coherence statistics, as shown in Fig. \ref{figure:4-6}.

Consider the transmission spectra (Fig.~\ref{figure:4-6}a): the doublet that signifies strong coupling disappears with increasing blinking. Such a result is easily controlled in theory via changing $f_\textrm{\scriptsize blink}$, however, here we modified the local charge environment with a broad-band diode. While collecting information about the transmission spectrum, the diode power was increased in order to encourage blinking in the system. Note: this effect is different than saturation of the strongly-coupled system, which also destroys the signature of strong coupling. As a Jaynes--Cummings system is driven into saturation, the separation between the doublet decreases until the spectra is comparable to that of the bare cavity, while no such linewidth narrowing is present here \citep{Fushman2008-jt}. Of course, if the $g/\kappa$ ratio were larger then the addition of blinking would be easy to identify as a triplet in the transmission spectrum \citep{ota2009investigation}.

Meanwhile in the second-order coherence statistics (Fig.~\ref{figure:4-6}b), blinking most strongly affects the tunneling region because this is where the blinking spectrum is maximized, i.e. at the bare cavity frequency. Here, the effect to push down the ${g}^{(2)}[0]$ near zero laser detuning manifests itself as a broader tunneling peak and even a slight dip at zero detuning. For the case of the resonant system, the agreement between experiment and theory is excellent. Blinking has a large influence due to the spectral proximity of the strongly-coupled system's emission to the bare cavity peak. By extension, it has negligible effect on the detuned spectra and cannot be used to explain the disparity between the detuned tunneling experiments and theory in Section~\ref{sec:det}. Finally, we note that the disparity in the blockade region of Fig. \ref{figure:4-6}b will be addressed in the next section.

%%%%%%%%%%%%%%%%%%%%%%%%%%%%%%%%%%%%%%%%%%%%%%%%%%%%%%%%%%%%%%%%

\subsection{Self-Homodyne Interference}

\begin{figure}
\textsf{
  \includegraphics[width=11.43cm]{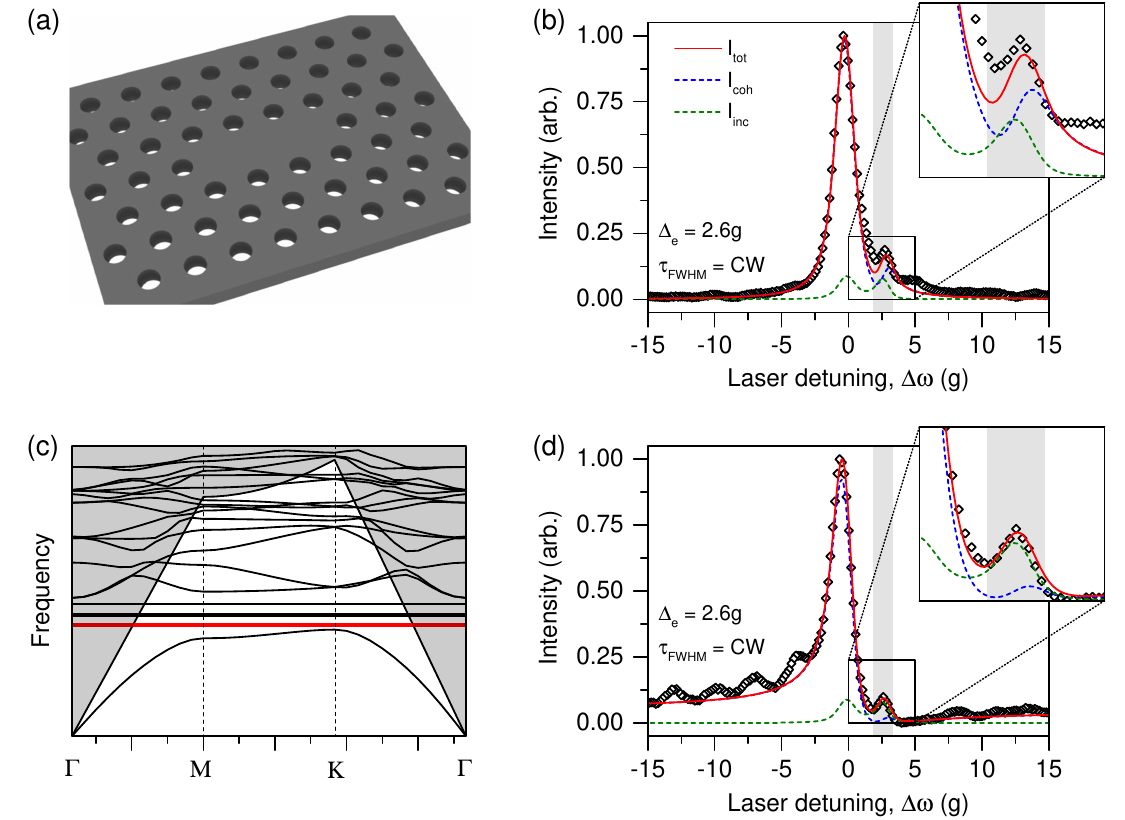}
  \caption{Origin of self-homodyne interference. (a) Schematic of a planar L3 photonic crystal cavity. (b) Transmission spectra under excitation conditions that produce a Lorentzian-like profile, fit to a quantum optical model. (c) Complicated mode structure of a planar L3 Photonic crystal cavity, calculated using the MIT Photonic-Bands (MPB) package. Red and black horizontal lines depict the cavity's fundamental and higher order modes, respectively. Curved lines represent photonic crystal guided modes. Grey region indicates leaky modes that are above the light line. (d) Transmission spectra under excitation conditions that produce a Fano-like profile, fit to a quantum optical model. \textit{Transmission data in (B) and (D) from \citet{Fischer2016-dj}.} Simulated spectra were convolved with the response function of the spectrometer (linewidth $\Gamma_\textrm{\tiny FWHM}/2\pi = 4.5$\,GHz).
}
\label{figure:4-7}}
\end{figure}

Previously, we discussed how both the experimental blockade and tunneling data seemed to reveal stronger nonclassical correlations than an ideal Jaynes--Cummings model would predict. In this section, we explain the missing physics that allows for better performance through enhancement of the nonclassical light emission \citep{Fischer2016-dj,Fischer2016-nk}. It is very tempting to model the physics of an L3 photonic crystal cavity (Fig.~\ref{figure:4-7}a) as a single mode of a harmonic oscillator, i.e. with a single Heisenberg mode operator $a(t)$ which we will refer to as the discrete scattering channel. Under certain cross-polarized reflectivity conditions, such as those used in Fig. \ref{figure:4-7}b, this model accurately captures the system dynamics. In this scenario, the detuned transmission profile nearly resembles the bare cavity's Lorentzian profile at high excitation powers. Here, an experimental transmission spectrum through a strongly-coupled system is shown and fitted with the same quantum optical model as in Fig. \ref{figure:4-5} (red line). The green and blue decompositions will be discussed later.

However, the L3 photonic crystal has a rich mode structure, as shown in Fig. \ref{figure:4-7}c, that is not necessarily well-approximated by a single mode operator~$a(t)$. Instead, a better approximation is to consider an additional scattering channel that is due to a roughly constant background density of photonic states; this scattering pathway is referred to as the continuum channel. The discrete and continuum channels can interfere with one another to generate a lineshape that is closely related to a Fano resonance, and this is modeled with the operator $A(t)\rightarrow a(t)+\alpha(t)$ instead of just $a(t)$ where $\alpha(t)= \alpha\mathcal{E}(t)$, $\mathcal{E}(t)$ is the Gaussian pulse shape, and $\alpha$ is a c-number \citep{Fischer2016-dj,Fischer2016-nk}. The operator $\alpha(t)$ represents the laser light reflected into the cross-polarized output channel via the continuum modes of the photonic crystal. The lineshape can be changed between Lorentzian-like or Fano-like in cross-polarized reflectivity by altering the focal spot size and the excitation/detection polarizations, an effect which has been theoretically verified in L3 photonic crystal cavities through a rigorous scattering-matrix formalism \citep{Vasco2013-xt}.

While the difference between these two transmission profiles may initially appear small from a comparison between these two lineshapes, the effects on nonclassical light generation are dramatic and manifest in the grey boxed regions that represent the frequency of the emitter-like polariton. In fact, the mixing action of combining the reflected laser light with light scattered by the strongly-coupled system is a type of homodyne measurement, which has the power to emphasize the incoherent or nonclassical portion of the scattered light over the coherent or classical portion. Because the mixing occurs at the level of the photonic crystal, this effect is named a self-homodyne interference (SHI) \citep{Fischer2016-dj}. To further explore this point, the transmission profiles in Figs. \ref{figure:4-7}b and~\ref{figure:4-7}d were decomposed into their incoherent and coherent portions of emission. The coherent portions (blue) are primarily due to the classically scattered light from a subset of almost harmonically spaced dressed states or the continuum modes and hence look predominantly like the Lorentzian or Fano lineshapes. This light is due to the mean of the electric field, i.e. ${I_\textrm{\scriptsize coh}\propto\langle a^\dagger \rangle\langle a \rangle}$. Meanwhile, the incoherent portions of the emissions (green) are the result of the nonlinearity in the Jaynes--Cummings system and hence from the nonclassically scattered light. This light is due to the fluctuations of the electric field, i.e. $I_\textrm{\scriptsize inc}\propto\langle a^\dagger a \rangle - \langle a^\dagger \rangle\langle a \rangle$.

Now, we revisit Figs. \ref{figure:4-7}b and~\ref{figure:4-7}d and compare the coherent and incoherent portions in the insets. Under the Lorentzian-like conditions (Fig.~\ref{figure:4-7}b), the coherent portion of the transmitted light (blue) dominates the incoherent portion (green). However, under the Fano-like conditions (Fig.~\ref{figure:4-7}d), the incoherent portion of the transmitted light dominates and over 90\,\% of the coherently scattered portion is suppressed at the frequency of the emitter-like polariton. In this way, the effect of the quantum nonlinearity in the Jaynes--Cummings ladder is emphasized, with great potential to allow highly dissipative systems to still exhibit robust signatures of nonclassical light generation.

%%%%%%%%%%%%%%%%%%%%%%%%%%%%%%%%%%%%%%%%%%%%%%%%%%%%%%%%%%%%%%%%

\subsubsection{Effects on Emission Spectra}

\begin{figure}
\textsf{
  \includegraphics[width=11.43cm]{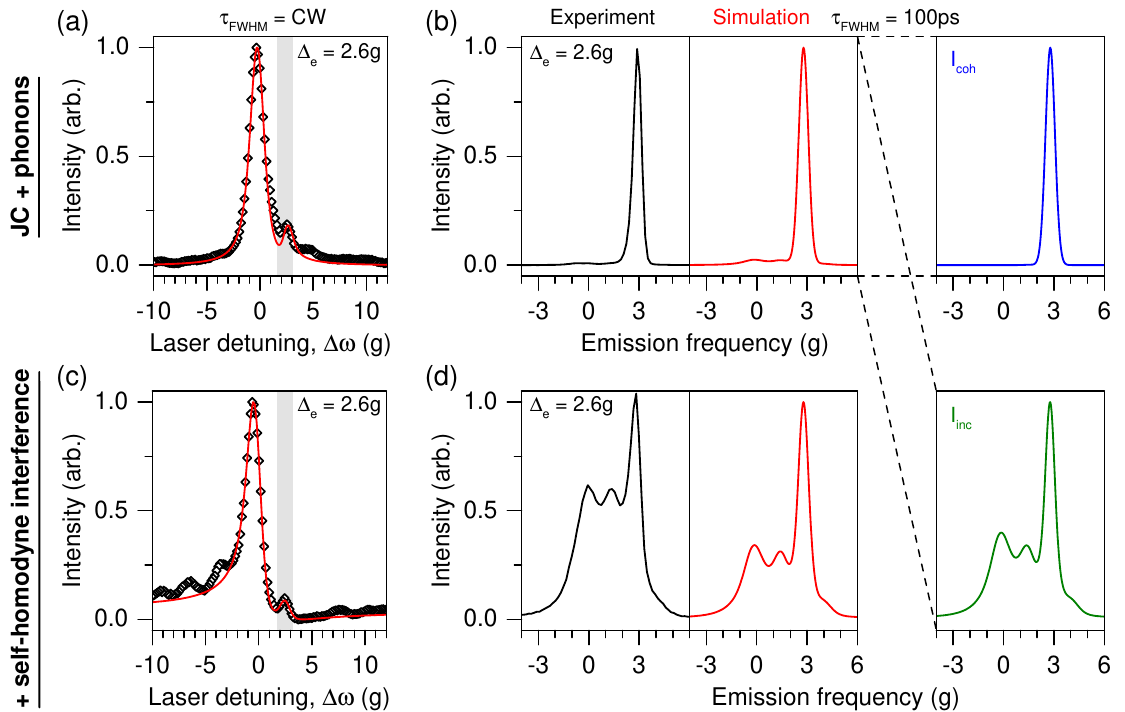}
  \caption{Effect of self-homodyne interference on resonance fluorescence from a dissipative strongly-coupled system. (a) Transmission profile under excitation conditions that produce a Lorentzian-like profile. (b) Experimental and simulated resonance fluorescence under drive on the emitter-like polariton [grey box in (a)] by a high-power $\tau_\textrm{\tiny FWHM}=100$\,ps pulse. (c) Transmission profile under excitation conditions that produces a Fano-like profile. (d) Experimental and simulated resonance fluorescence under drive on the emitter-like polariton [grey box in (c)] by a high-power $\tau_\textrm{\tiny FWHM}=100$\,ps pulse, including self-homodyne interference. (a-d) Quantum optical fits in red, and decomposition into coherent ($I_\textrm{\scriptsize coh}$) and incoherent ($I_\textrm{\scriptsize inc}$) components in blue and green, respectively. \textit{Data and simulations from \citet{Fischer2016-dj}.} All simulated spectra were convolved with the response function of the spectrometer (linewidth $\Gamma_\textrm{\tiny FWHM}/2\pi = 4.5$\,GHz).
}
\label{figure:4-8}}
\end{figure}

As our first example of the dramatic influence of self-homodyne interference on emphasizing nonclassical light generation, we review data from \citet{Fischer2016-dj}. In Fig. \ref{figure:4-8}, we consider the effect of self-homodyne interference on the spectra of pulsed resonance fluorescence from a solid-state strongly-coupled system. The ideal spectrum of pulsed resonance fluorescence for an operator $A(t)$ is calculated with
\begin{equation}
S(\omega)=\iint_{\mathbb{R}^2} \mathop{\textrm{d} t} \mathop{\textrm{d} \tau} \textrm{e}^{-\textrm{\footnotesize i}\omega\tau} \langle A^\dag(t+\tau) A(t) \rangle\textrm{.} \label{eq:spec_2d}
\end{equation}
The spectrum for resonance fluorescence from a Jaynes--Cummings system (even with phonons) is given when $A(t)\rightarrow a(t)$ and the spectrum with self-homodyne interference is given when $A(t)\rightarrow a(t)+\alpha(t)$. The two-time correlations are again computed with a time-dependent version of the quantum regression theorem, using the system Liouvillian
\begin{equation}
\mathcal{L}_\textrm{\scriptsize sc}(t)=\tilde{\mathcal{L}}_\textrm{\scriptsize JC}+ \tilde{\mathcal{L}}_\textrm{\scriptsize phonon} + \tilde{\mathcal{L}}_\textrm{\scriptsize drive}(t)\textrm{.}
\end{equation}
Note: although we do not include blinking in this subsection, it has very minimal effects on the results since the dot is detuned from the bare cavity.

In the experiment, the focal conditions were first tuned to the Lorentzian-like conditions (Fig.~\ref{figure:4-8}a) and we then excited the emitter-like polariton (denoted by the grey box) with a high-power $\tau_\textrm{\tiny FWHM}=100$\,ps pulse. The spectrum of resonance fluorescence was measured on a spectrometer and is shown in Fig. \ref{figure:4-8}b. Notably, the spectrum simply shows a single peak with a linewidth determined by the laser pulse (black). Simulating the spectrum with a Jaynes--Cummings model including phonon effects, i.e. with the Liouvillian $\mathcal{L}_\textrm{\scriptsize sc}(t)$, the singly peaked spectrum can be reproduced (red). Like the transmission spectra, the fluorescence spectrum can also be decomposed into the coherent and incoherent potions by making the replacement of $\langle A^\dag(t+\tau) A(t) \rangle \rightarrow \langle a^\dag(t+\tau) \rangle \langle a(t) \rangle$ and $\langle A^\dag(t+\tau)A(t) \rangle \rightarrow \langle a^\dag(t+\tau)a(t) \rangle - \langle a^\dag(t+\tau) \rangle \langle a(t) \rangle$, respectively, in equation (\ref{eq:spec_2d}). These decompositions are shown in the blue and green lineshapes, and they show that the coherently scattered light completely dominates the emission spectrum at high powers. \textit{Again, we emphasize that this is a standard feature of highly dissipative but strongly-coupled Jaynes--Cummings systems, that well-known cavity QED effects can be unobservable!} On the other hand, the incoherent portion shows a very interesting quadruplet structure that is closely related to the Mollow triplet, a hallmark of quantum-mechanically scattered light. This quadruplet structure was discussed thoroughly in \citet{Fischer2016-dj} and its origins will not be discussed here; suffice to say that it is a structure arising from the nonclassical light emission of the cavity QED system.

In order to experimentally observe the quadruplet, self-homodyne interference was critical. We next tuned to the Fano-like conditions (Fig.~\ref{figure:4-8}b) and again excited the emitter-like polariton (denoted by the grey box) with the same high-power $\tau_\textrm{\tiny FWHM}=100$\,ps pulse. Now, the optimally tuned self-homodyne interference experimentally reveals the interesting quadruplet structure, which is completely unobservable otherwise. The interference was also included theoretically in simulation to reveal the quadruplet (Fig.~\ref{figure:4-8}). We make two practical notes here regarding the modeling of these experiments. First, an entire power-series of spectra was necessary to accurately fit excitation powers [see Fig. 3 in \citet{Fischer2016-dj}]. Second, the experimental data shows a better defined quadruplet, with more energy in the sidebands than the model. We expect this difference is due to an inaccuracy of the effective phonon master equation at large driving powers.

%%%%%%%%%%%%%%%%%%%%%%%%%%%%%%%%%%%%%%%%%%%%%%%%%%%%%%%%%%%%%%%%

\subsubsection{Effects on Nonclassical Light Generation}

As our second application of self-homodyne interference, we consider its effect on nonclassical light generation. As we saw in the Section 3.6.1, an optimally tuned self-homodyne interference (SHI) has the ability to remove un-wanted coherently scattered light from the system. On a similar principle, we would expect the interferometric technique to be capable of enhancing nonclassical light generation by isolation of the quantum-mechanical signal. In prior work, we explored the precise mechanism for how SHI is capable of enhancing both single- and multi-photon emission in dissipative Jaynes--Cummings systems, but under continuous-wave excitation \citep{Fischer2016-nk}. In this section, we will examine the enhancement under pulsed excitation and in doing so will almost perfectly fit the experimental data from \citet{Muller2015-il} and \citet{Dory2016-ca}.

First, we discuss our complete model that we believe captures nearly all experimental effects relevant for photon blockade and photon tunneling. It is again based off of the system Liouvillian
\begin{equation}
\mathcal{L}_\textrm{\scriptsize sc}(t)=\tilde{\mathcal{L}}_\textrm{\scriptsize JC}+ \tilde{\mathcal{L}}_\textrm{\scriptsize phonon} + \tilde{\mathcal{L}}_\textrm{\scriptsize drive}(t)\textrm{,}
\end{equation}
but now we calculate $g^{(2)}[0]$ with a self-homodyne interference. Specifically, that means to calculate
\begin{equation} \label{eq:2}
g^{(2)}[0] = \frac{\int_0^T \int_0^T \mathop{\textrm{d} t} \mathop{\textrm{d} t'} \, \langle \mathcal{T}_-[A^{\dagger}(t) A^{\dagger}(t')] \mathcal{T}_+[A(t') A(t)] \rangle}{\left(\int_0^T \mathop{\textrm{d} t} \, \langle A^{\dagger}(t) A(t) \rangle\right)^2}\equiv\frac{G^{(2)}_{AA}[0]}{\langle N_A  \rangle^2}\textrm{,}
\end{equation}
where $A(t)\rightarrow a(t)+\alpha(t)$ and again $\alpha(t)= \alpha\mathcal{E}(t)$. To incorporate blinking, we simply change equation~(\ref{eq:22}) into
\begin{equation} \label{eq:2}
g^{(2)}[0] = \frac{\left(1-f_\textrm{\scriptsize blink}\right)G^{(2)}_{AA}[0] + f_\textrm{\scriptsize blink}\langle N_\textrm{\scriptsize blink}  \rangle^2}{\left(1-f_\textrm{\scriptsize blink}\right)\langle N_\textrm{\scriptsize A}  \rangle^2+f_\textrm{\scriptsize blink}\langle N_\textrm{\scriptsize blink}  \rangle^2}\textrm{.}
\end{equation}
Now, $\langle N_\textrm{\scriptsize blink}  \rangle=\int_0^T \mathop{\textrm{d} t} \, \langle A^{\dagger}(t) A(t) \rangle$ and is calculated using the Liouvillian $\mathcal{L}(t)=\tilde{\mathcal{L}}_\textrm{\scriptsize blink}+  \tilde{\mathcal{L}}_\textrm{\scriptsize drive}(t)$; again, $A(t)\rightarrow a(t)+\alpha(t)$ .

\begin{figure}
\textsf{
  \includegraphics[width=11.43cm]{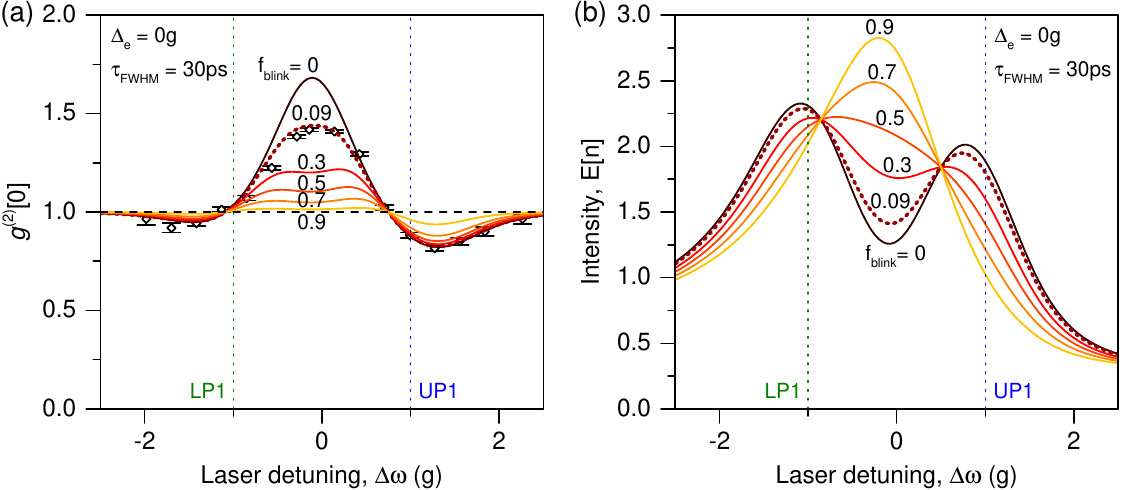}
  \caption{Effects of strong emitter blinking on a resonant strongly-coupled system. (a) Second-order coherence ${g}^{(2)}[0]$ as a function of laser detuning, \textit{experimental data from \citet{Muller2015-il}.} The ideal fit to ${g}^{(2)}[0]$ is given by the brown dashed line. Horizontal black dashed line represents statistics of the incident
laser pulses. (b) Pulsed transmission spectrum, with quantum simulation only. In both sub-figures, the statistics or intensities are given for different blinking fractions.
  }
\label{figure:4-9}}
\end{figure}

With the necessary theoretical machinery established, we are now finally ready to fully model the ${g}^{(2)}[0]$ versus laser detuning scans. First, we revisit the data from Fig. \ref{figure:4-6}, but with self-homodyne interference, in Fig. \ref{figure:4-9}a. The optimal fit is given by the dashed brown line, which almost perfectly matches the data with the addition of an optimized SHI. The exact same trends are visible for blinking with or without the interference, where the tunneling regions are affected first with increasing blinking while the blockade regions are relatively unaffected. Now, we additionally show the simulated transmission plots for the incident pulses (Fig.~\ref{figure:4-9}b). We emphasize here that when the system blinks the SHI still occurs, and hence the system blinks with a Fano lineshape.

Next, we consider detuned blockade and tunneling with an emitter detuning of $\Delta_\textrm{\scriptsize e}=3.2g$. Comparing the resonant and detuned cases in Fig. \ref{figure:4-10}, one can again observe the general trend of enhanced photon blockade with increasing detuning, as was discussed in Section~\ref{sec:det}. (We provide a brief technical note that, just as with the resonance fluorescence experiments in the Section 3.6.1, before performing any ${g}^{(2)}[0]$ scan the photon blockade region was optimized. At that point, a transmission experiment would reveal a Fano-like lineshape, incorporating SHI; these interference conditions were held constant over the course of the ${g}^{(2)}[0]$ scan.) Back to both subfigures, the quantum-optical model is broken down into several different lines, to separate out the effects of blinking, electron-phonon interaction, and self-homodyne interference. The primary effects are as follows:
\begin{figure}
\textsf{
  \includegraphics[width=11.43cm]{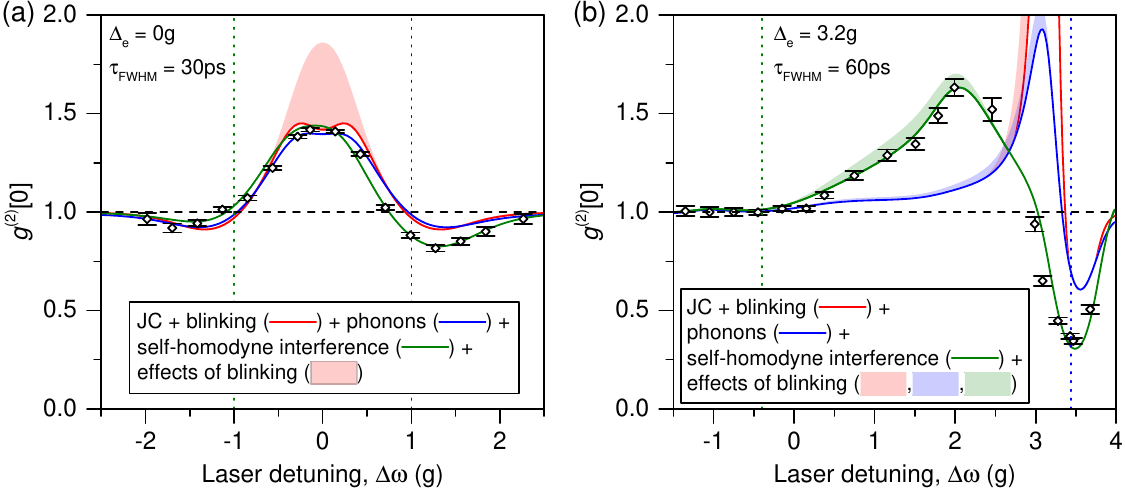}
  \caption{Perfectly modeled photon blockade and tunneling under pulsed excitation. Second-order coherence ${g}^{(2)}[0]$ as a function of laser detuning on an (a) resonant and (b) detuned system. The legends discuss the complete decomposition of the quantum-optical model into its constituent effects. Green and blue dashed lines represent the LP1 and UP1, respectively. Horizontal black dashed lines represent statistics of the incident laser pulses. A laser power of 5\,nW was used. \textit{Experimental data in (a) and (b) from \citet{Muller2015-il} and \citet{Dory2016-ca}, respectively.}
}
\label{figure:4-10}}
\end{figure}
\begin{enumerate}
\item Blinking has the strongest effect in the zero dot detuning case and minimal effect in the detuned case. For zero detuning, the effect of blinking is not just to decrease the maximum achievable value of ${g}^{(2)}[0]$ in the tunneling region, but also to broaden the width of the tunneling region. We note that interestingly, none of the other tunable parameters are able to control this effect---without blinking, the resonant tunneling region could never be fitted properly. Additionally, the blinking fraction was taken from the fit in Section~\ref{sec:blink}.
\item Electron-phonon interaction has almost negligible effects on photon blockade for either the resonant or detuned system, because the effect of the dissipation is only to change the frequency and coherence of any emitted single photons \citep{Muller2016-fs}. In the tunneling region, phonon-induced transfers have the largest effect on the detuned system since the ${g}^{(2)}[0]$ values in the tunneling region are much larger due to the decrease in photon transmission, as discussed in Section~\ref{sec:det}.
\item Self-homodyne interference has the strongest effect on the detuned system, though it's certainly important in the resonant case as well. For both system configurations, SHI enables much lower values of ${g}^{(2)}[0]$ in the blockade regions, thus significantly improving the quality of single-photon emission. Because SHI occurs on one side of the cavity profile or the other, in the resonant case the LP1 blockade is worsened while the UP1 blockade is improved. This evidence of enhanced photon blockade fully suggests that in some previous experiments with strongly-coupled systems where the simulations were surprisingly unable to perform as well as the experiments [e.g. those in \citet{Muller2015-il}, \citet{Reinhard2011-ye}, \citet{Muller2015-om}, and \citet{Kim2014-yh}], researchers may have unknowingly utilized SHI.
\end{enumerate}

To further discuss the effect of SHI, we note that the detuned tunneling region is dramatically different with SHI, where the peak of the ${g}^{(2)}[0]$ scan no longer occurs at the point of minimal photon transmission. Because the unwanted coherently scattered light is removed with SHI, the tunneling region is now a much stronger indicator of the multi-photon processes occurring in the detuned strongly-coupled system. Additionally, we note that driving the quantum dot term directly (as opposed to through the cavity) was proposed as one method to increase the single-photon purity of photon blockade and could cause asymmetries like the one observed in the resonant case \citep{Tang2015-mw}; however, this process only decreases the amplitude in the tunneling region and does not shift its maximum point of ${g}^{(2)}[0]$. Since no fitting parameter other than SHI was able to enhance the tunneling region, it is clear that SHI plays an important role in nonclassical light generation.

\begin{figure}
\textsf{
  \includegraphics[width=11.43cm]{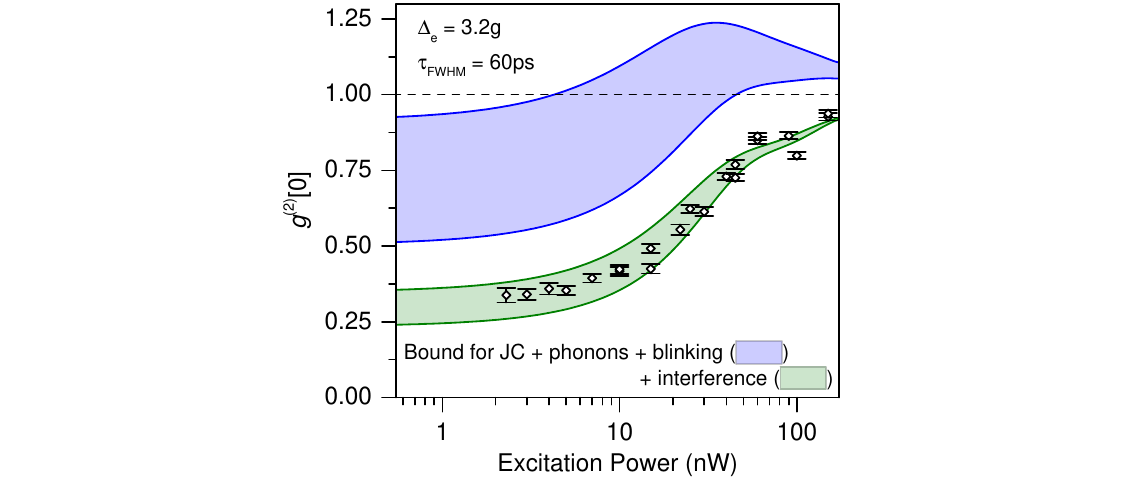}
  \caption{Detuned photon blockade as a function of excitation power. Data and fit parameters identical to those in Fig. \ref{figure:4-10}. Horizontal black dashed line represents statistics of the incident laser pulses.
}
\label{figure:4-11}}
\end{figure}

Next, we explore detuned photon blockade as a function of excitation power, both by presenting new experimental data and with quantum-optical fits (Fig.~\ref{figure:4-11}). This power-dependent data and fitting is important in verifying that we have not over-fit our experimental data and in reaffirming the strength of the self-homodyne interference to improve photon blockade. In performing the power scan of ${g}^{(2)}[0]$, the experimental uncertainty in the precise laser detuning is given by the spectrometer linewidth of $\Gamma_\textrm{\tiny FWHM}\approx g/3$ (though there is little uncertainly in the laser detuning for the ${g}^{(2)}[0]$ versus laser detuning scans due to the high \textit{relative} precision of the experimental pulse shaper). Therefore, we have simulated both the minimum ${g}^{(2)}[0]$ values in photon blockade and the values under a system that was imprecisely tuned by $\Gamma_\textrm{\tiny FWHM}/2$. This procedure pictorially shows the potential uncertainty in the correct laser detuning for the simulated values. Using the optimal fit from Fig. \ref{figure:4-10}b, the green simulated blockade region matches almost perfectly with the experimental values. Meanwhile if SHI is excluded, then the blockade values are not just worse at the minimum, but much more sensitive to any possible imprecision in laser detuning. Because both the blockade and tunneling regions are sensitive to the excitation power, the strong fit with experiment  helps confirm our complete model of an experimental strongly-coupled system.

With such a complex model, one must be wary of over-fitting the data. However, each of the elements has been independently verified and fitted through a large series of a experimental data to extract the Jaynes--Cummings dissipation, electron-phonon effects, blinking, and self-homodyne interference. Taken as a whole, \textit{each of the non-ideal effects almost independently tunes different aspects of the emission statistics and hence we believe we have identified the appropriate number of model parameters.}

\subsection{Outlook for Single-Emitter Cavity QED}

In summary, we proposed a complete model for photon blockade and tunneling in III-V quantum dot cavity QED systems. We found that the pure Jaynes--Cummings model was incapable of accurately modeling either the spectra or photon statistics in transmission or emission from a strongly-coupled system based on InGaAa quantum dots. However, by including dissipation, dot-cavity detuning, pulsed dynamics, effects of phonons, blinking, and a new effect called self-homodyne interference we were able to almost perfectly model the nonclassical light generation from our cavity QED system.

By incorporating frequency filtering of the emission, we recently showed highly indistinguishable photon generation and evidence for two-photon generation from such a strongly-coupled system \citep{Dory2016-ca,Muller2016-fs}. Looking towards future experiments, by adapting the self-homodyne interference technique to on-chip photonic crystal waveguide devices, we expect that this work could easily become a standard feature of optical solid-state platforms in enhancing nonclassical light generation \citep{Fischer2016-nk}. We believe this technique should also enable the first direct observation of a solid-state system's higher-order Jaynes--Cummings structure and the efficient generation of N-photon states.

Moving towards more interesting and complex level structures will allow for a much richer set of dynamics and possibilities for nonclassical light generation. For instance, charged III-V quantum dots in magnetic field \citep{Carter2013-bm}, III-V quantum dot molecules \citep{Vora2015-af}, and group-IV color-centers \citep{Riedrich-Moller2014-tb} may allow for the exploration of cavity QED with a single, multi-\textit{level} quantum emitter coupled to a cavity \citep{Bajcsy2013-ss}. These posses untapped level structures for improving photon blockade and studying multiphoton transitions, and they may allow experimentalists to more readily probe quantum nonlinearities in a solid-state environment. One of their most promising applications is to realize arbitrary single-photon generation in a solid-state nanocavity \citep{Sweeney2014-vz,Santori2009-qe}. In this scheme, the cavity mediates the generation or annihilation of an arbitrarily shaped single-photon through a Raman transition. Such devices form the backbone of spin-photon interfaces in some theoretically proposed quantum networks \citep{Cirac1997-kw}. Unlike the already-demonstrated spin-photon interfaces that rely only on weak cavity coupling, it is possible for a flying photonic qubit to be perfectly absorbed by a cavity QED device operating in the Raman single-photon regime.

Fundamentally, however, single-emitter systems with multiple levels are also limited in their potential for nonclassical light emission. Instead, we expect the future of cavity QED, utilizing rich level structures for studying quantum light and computation, lies in multi-\textit{emitter} cavity QED. In the next section, we cover an emerging field whereby multiple color-centers couple to a single cavity mode.

%%%%%%%%%%%%%%%%%%%%%%%%%%%%%%%%%%%%%%%%%%%%%%%%%%%%%%%%%%%%%%%%
% Section 4
%%%%%%%%%%%%%%%%%%%%%%%%%%%%%%%%%%%%%%%%%%%%%%%%%%%%%%%%%%%%%%%%

\section{Overview of Multi-Emitter Cavity QED}

\begin{figure}
\textsf{
\centering
\includegraphics[width=11.43cm]{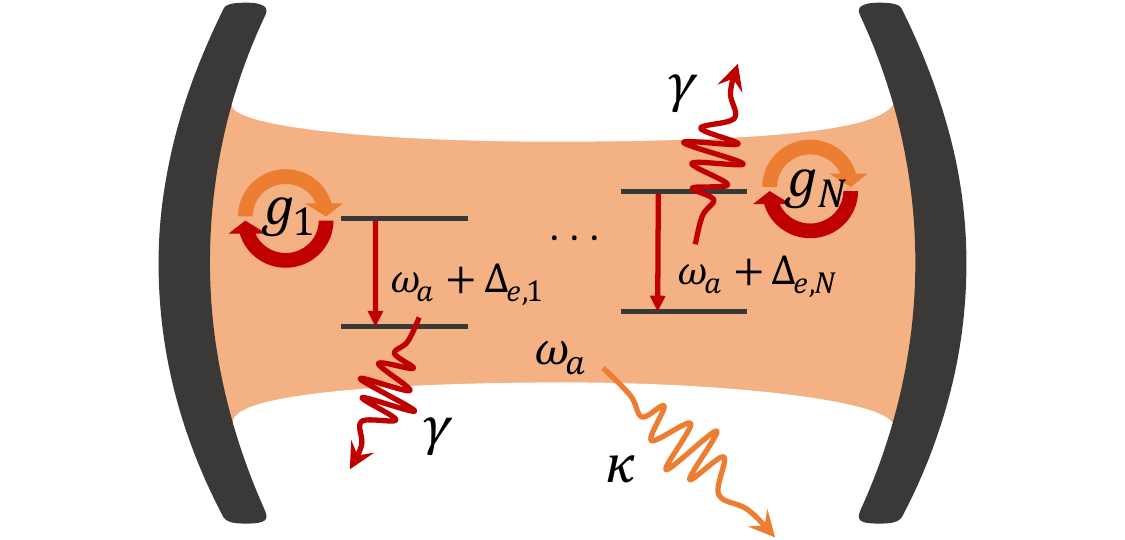}
\caption{\label{figure:5-1} Illustration of $N$ non-identical emitters coupled to a cavity mode.}
}
\end{figure}

While the previous sections illustrate the limit of what has been experimentally demonstrated so far, here we propose a direction for the future cavity QED experiments. The nonlinearity in a cavity QED system can be significantly increased through the collective coupling of multiple emitters to the resonant mode (Fig.~\ref{figure:5-1}). In this picture, $N$ emitters are effectively described as a single emitter strongly coupled to the cavity with an increased interaction rate by a factor of $\sqrt{N}$. Such coupling is achievable in systems where the inhomogeneous broadening is comparable to the collective coupling rate \citep{diniz2011strongly}. Collective coupling is possible even when the inhomogeneous broadening exceeds the cavity linewidth, and the polariton width is dominantly defined by cavity and emitter linewidths, due to an effect called \emph{cavity protection}. This phenomenon was previously investigated with rare-earth ions in a solid-state cavity for a very large ensemble of emitters approximating a continuum \citep{zhong2016chip}. In contrast, we are interested in a regime of several emitters coupled to a nano-optical cavity and giving rise to a discretized energy ladder suitable for demonstrations of advanced photon blockade effects. Similar multi-emitter cavity QED systems have been demonstrated in atomic \citep{thompson1992observation, neuzner2016interference} and superconducting circuit systems \citep{fink2009dressed}. With the developments in techniques of substrate growth and processing, the implementation of these systems is expected  with color-centers in solid-state nanocavities as well, which will have an impact on the development of GHz-speed optical switches and high-quality integrated sources of single photons.

%%%%%%%%%%%%%%%%%%%%%%%%%%%%%%%%%%%%%%%%%%%%%%%%%%%%%%%%%%%%%%%%

\subsection{The Tavis--Cummings Model}

We first discuss the original Tavis--Cummings model for atoms, which have negligible inhomogeneous broadening, and then extend to the model to include the inhomogeneous broadening of solid-state quantum emitters. The Tavis--Cummings model was developed to describe an ensemble of atoms that strongly couple to a cavity mode \citep{tavis1968exact}, and gives rise to the Hamiltonian
\begin{equation}
H_\textrm{\scriptsize TC} = \omega_a a^\dag a + \sum_{n=1}^N{\left[ \omega_{a} \sigma_n^\dag \sigma_n + g_n \left(\sigma_n^\dag a + a^\dag \sigma_n\right) \right]}.
\end{equation}

Here, a single cavity mode individually couples to each of the $N$ emitters that otherwise do not interact with one another. As before, $a$ and $\omega_a$ represent the cavity operator and frequency, while the atoms are characterized by their dipole operator $\sigma_n$ and cavity coupling rate $g_n$. The variables $g_n$ account for non-equal positioning of emitters relative to the cavity field intensity.

Coherent interactions in the system give rise to a new set of eigenfrequencies that form a dressed ladder of states. The first rung contains $N+1$ states, corresponding to an additional excitation of the cavity mode or one of the emitters. The second rung contains $\frac{(N+1)(N+2)}{2}$ states, representing either an excitation of one of the previously non-excited emitters or a new photon in the cavity mode.

\begin{figure}
\centering
\textsf{
\includegraphics[width=9cm]{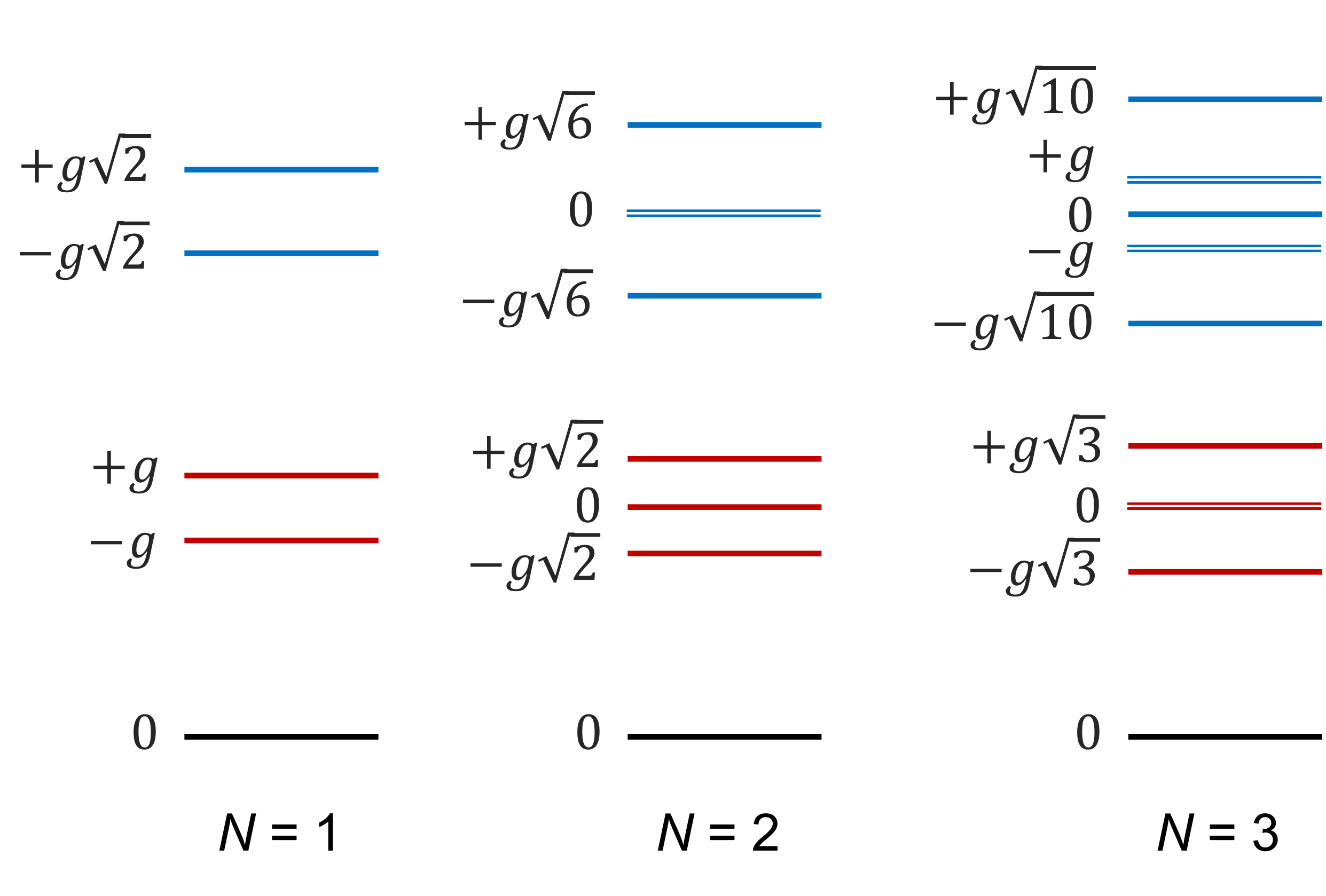}
\caption{\label{figure:5-2} Dressed ladder of states for $N=1,2,3$ identical emitters coupled to a single cavity mode.}
}
\end{figure}

For a system with equally coupled atoms ($g_n=g$), we illustrate these levels in Fig. \ref{figure:5-2}. In particular, the levels in the first rung represent two polaritonic states and $N-1$ degenerate states. The degenerate states are at the cavity frequency and their eigenvectors have no cavity component; therefore, the states do not couple to the environment and are referred to as the subradiant states. For the first rung, the eigenenergies are given by
\begin{eqnarray}
E_{1}& = &\omega_c - g\sqrt{N},\\
E_{2,...,N}& = &\omega_a,\\
E_{N+1}& = &\omega_a + g\sqrt{N}.
\end{eqnarray}
The splitting between the polaritonic states $E_1$ and $E_{N+1}$ is a result of the collective coupling with an effective coupling rate of $G_N=g\sqrt{N}$. When the emitters are unequally coupled (i.e. $g_n\neq g_m$), then the collective coupling rate can be calculated as
\begin{equation}
G_N=\sqrt{\sum_{n=1}^{N}g_n^2}.
\end{equation}

\textit{However, the degeneracy between the other $N-1$ states occurs only for identical emitters and is lifted with the introduction of any non-identical quantum emitters.} Compared to atoms, color-centers are non-identical quantum emitters, and therefore we need to expand the original Hamiltonian to include a set of emitter frequencies that capture the inhomogeneous broadening in the ensemble, i.e. with
\begin{equation}
H_\textrm{\scriptsize TC}' = \omega_a a^\dag a + \sum_{n=1}^N{\left[ \left(\omega_{a}+\Delta_{\textrm{\scriptsize e},n}\right) \sigma_n^\dag \sigma_n + g_n \left(\sigma_n^\dag a + a^\dag \sigma_n\right) \right]}.
\end{equation}
Specifically, the inhomogeneous broadening is represented by the emitter detunings $\Delta_{\textrm{\scriptsize e},n}$. In the following sections, we will explore how the inhomogeneous broadening, resulting in $\Delta_{\textrm{\scriptsize e},n}\neq \Delta_{\textrm{\scriptsize e},m}$, affects the ability to observe collective oscillations, subradiant states, and quantum coherent phenomena.

%%%%%%%%%%%%%%%%%%%%%%%%%%%%%%%%%%%%%%%%%%%%%%%%%%%%%%%%%%%%%%%%

\subsection{Strong-Coupling Cavity QED With an Ensemble of Color-Centers}

In our CQED model of an ensemble of color-centers in a cavity, the following factors are considered:

1. The loss of photons from the cavity to the environment. The loss is defined by the quality factor of the mode $Q = \omega_a/\kappa$, and is often practically imposed by fabrication limitations.

2. The loss of photons from the emitters to the environment. This mechanism is governed by the ratio of color-center emission that is directed into the cavity mode and is affected by the density of states. A high density of photonic states results in a low photon loss rate, which can be encouraged by high quality factor and small mode volume cavities, as well as by the good positioning of the emitters within the cavity.

3. The distribution of emission frequencies in the ensemble. This factor is usually influenced by the local strain in the lattice and can be especially pronounced in nanoparticles or heteroepitaxial layers, compared to bulk substrates.

4. The variable positioning of individual emitters in the cavity. Some control over the depth of the color-centers in the substrates can be gained though selective doping during substrate growth or by defining the ion energy during irradiation of the pre-grown sample. Laterally, the use of a focused ion beam or masked apertures can provide a degree of localization.

5. The ratio of emission into the zero-phonon line (ZPL). This is an intrinsic property of each color-center, and it is the property that motivates the search for new systems with high Debye-Waller factor.

We now discuss how these effects are incorporated into our quantum-optical model. The cavity and emitter losses can be modeled through the Liouville's equation, as presented in Section~\ref{sec:osc}. Now, the super-operator that describes the system dynamics has the form
\begin{equation}
\mathcal{L}_{\textrm{\scriptsize TC}}\rho(t)=\textrm{i} \left[\rho(t), H_{\textrm{\scriptsize TC}}' \right]+ \frac{\kappa}{2}\mathcal{D}[a]\rho(t)+\sum_{n=1}^N\frac{\Gamma}{2}\mathcal{D}[\sigma_n]\rho(t),
\end{equation}
where $\kappa$ represents the cavity energy decay rate, and $\Gamma$ represents the individual emitter linewidth which is assumed to be constant within the ensemble.

The emission frequencies are sampled from a Gaussian distribution centered at $\omega_a+\Delta_\textrm{\scriptsize e}$ with standard deviation of $\frac{\delta}{2}$, where $\delta$ is the inhomogeneous linewidth of the ensemble. It is worth noting that the shape of individual color-center emission is a Lorentzian (homogeneous broadening), however, the distribution of central frequencies in the ensemble is a Gaussian (inhomogeneous broadening).

The coupling strength is limited by the maximal value $g_{\textrm{\scriptsize max}}$ imposed by the system parameters: ZPL extraction ratio (Debye-Waller factor) $\rho_\textrm{\scriptsize ZPL}$, emitter lifetime $\tau$, index of refraction $n$, cavity mode volume $V$ and operating frequency~$\omega$ \citep{mccutcheon2008design}
\begin{equation}
g_{\textrm{\scriptsize max}}=\sqrt{\frac{3\pi c^3 \rho_{\textrm{\scriptsize ZPL}}}{2\tau \omega^2n^3V}}.
\end{equation}
The coupling rate is reduced by imperfect positioning of emitters $r_n$ relative to the resonant field maximum $E_{\textrm{\scriptsize max}}$ and the angle between the dipole and field orientation $\phi_n$
\begin{equation}
g_n = g_{\textrm{\scriptsize max}} \left| \frac{E(r_n)}{E_{\textrm{\scriptsize max}}}\textrm{cos}(\phi_n) \right|.
\end{equation}
The distribution of the intensity in range $[0, g_{\textrm{\scriptsize max}}]$ is also specific to the system, mainly to the spatial positioning of emitters and the resonant field intensity distribution. Here, we will sample the coupling $g_n$ from a uniform distribution.

\begin{figure}
\centering
\textsf{
\includegraphics[width=11.43cm]{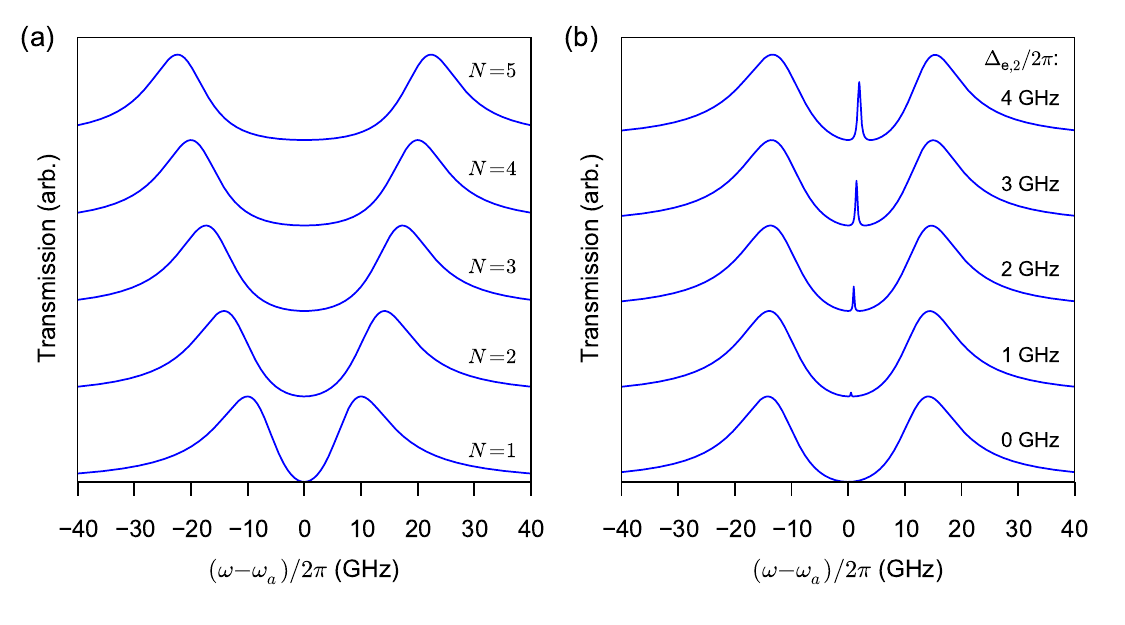}
\caption{\label{figure:5-3} (a) Transmission spectra of $N$ identical and equally coupled emitters. (b) Transmission spectra showing the emerging subradiant peak as the second of $N=2$ emitters becomes off-resonant.}
}
\end{figure}

We choose a realistic set of parameters $$\{\kappa/2\pi, \Gamma/2\pi, g_{\textrm{\scriptsize max}}/2\pi\} = \{25\,\textrm{GHz}, 0.1\,\textrm{GHz}, 10\,\textrm{GHz}\}$$ to model the dynamics of the multi-emitter cavity system. In order to numerically investigate the transmission spectra of the multi-emitter cavity system, we use the incoherent \textit{cavity} pumping technique discussed in  Section~\ref{sec:osc}. From these simulations, in Fig. \ref{figure:5-3}a we plot the increasing polaritonic spacing with an increased number of identical and equally-coupled emitters. The trend is consistent with the lossless case analysis where the collective coupling is given as $G_N = g\sqrt{N}$. As discussed previously, non-identicallity of emitters lifts the degeneracy of the new states in the ladder and makes them visible in the spectrum. With increasing emitter-emitter detuning, the subradiant states centered between the emitter frequencies become more visible. This is illustrated with the emergence of an intermediate peak in the transmission spectrum for $N=2$ non-identical emitters in Fig. \ref{figure:5-3}b.

\begin{figure}
\centering
\textsf{
\includegraphics[width=11.43cm]{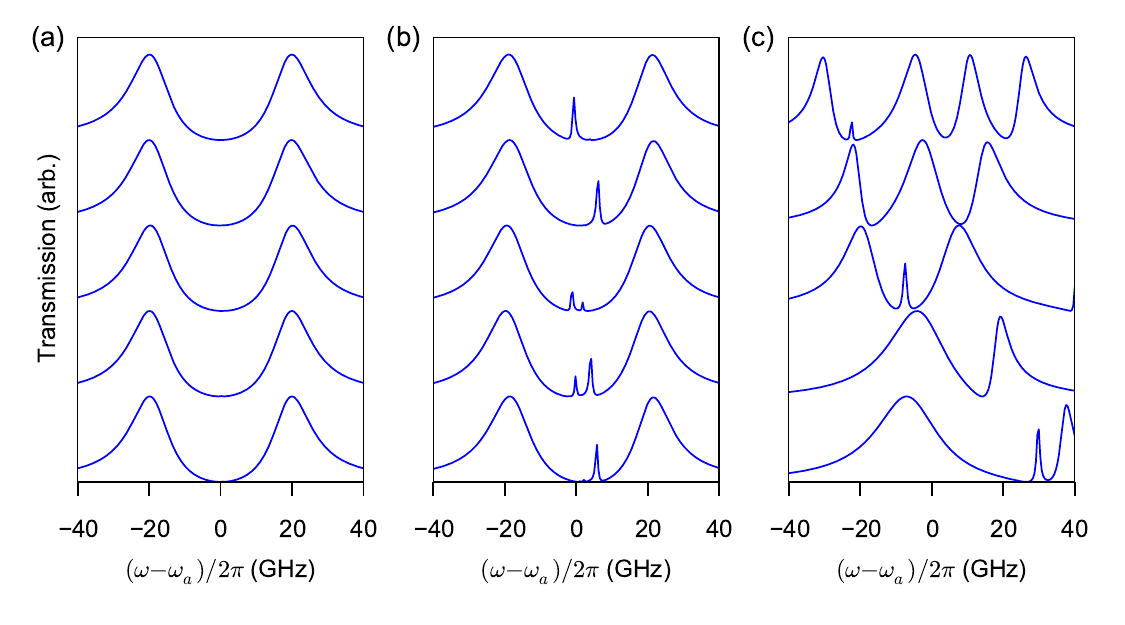}
\caption{\label{figure:5-4} Randomly generated transmission spectra for $N=4$, $\{\kappa/2\pi, \Gamma/2\pi, g_{\textrm{\scriptsize max}}/2\pi\} = \{25\,\textsf{GHz}, 0.1\,\textsf{GHz}, 10\,\textsf{GHz}\}$ and $\delta/2\pi =$ (a) 1\,GHz, (b) 10\,GHz, and (c) 100\,GHz.}
}
\end{figure}

Next, we analyze the influence of the inhomogeneous broadening to the transmission spectrum and present one of our most important findings, that the inhomogenous broadening will not obscure the observation of collective many-body physics in systems comprising color-centers with small inhomogeneous broadening. First, we investigate this phenomenon in Fig. \ref{figure:5-4}, which shows randomly generated spectra with $N=4$ emitters for variable inhomogeneous broadening with $$\delta/2\pi \in \{1\,\textrm{GHz}, 10\,\textrm{GHz}, 100\,\textrm{GHz}\}.$$ The first two sets of spectra show a small or a moderate perturbation to the system with identical emitters, featuring two polariton peaks and several subradiant peaks, suggesting that silicon-vacancy related complexes could be used in multi-emitter cavity QED systems. In the case of large (100\,GHz) inhomogeneous broadening, however, the spectrum is significantly perturbed and the identification of the polariton peaks is difficult or impossible (Fig. \ref{figure:5-4}c). Therefore, a multi-emitter systems with such broadening may not show the signs of collective strong coupling.

\begin{figure}
\centering
\textsf{
\includegraphics[width=11.43cm]{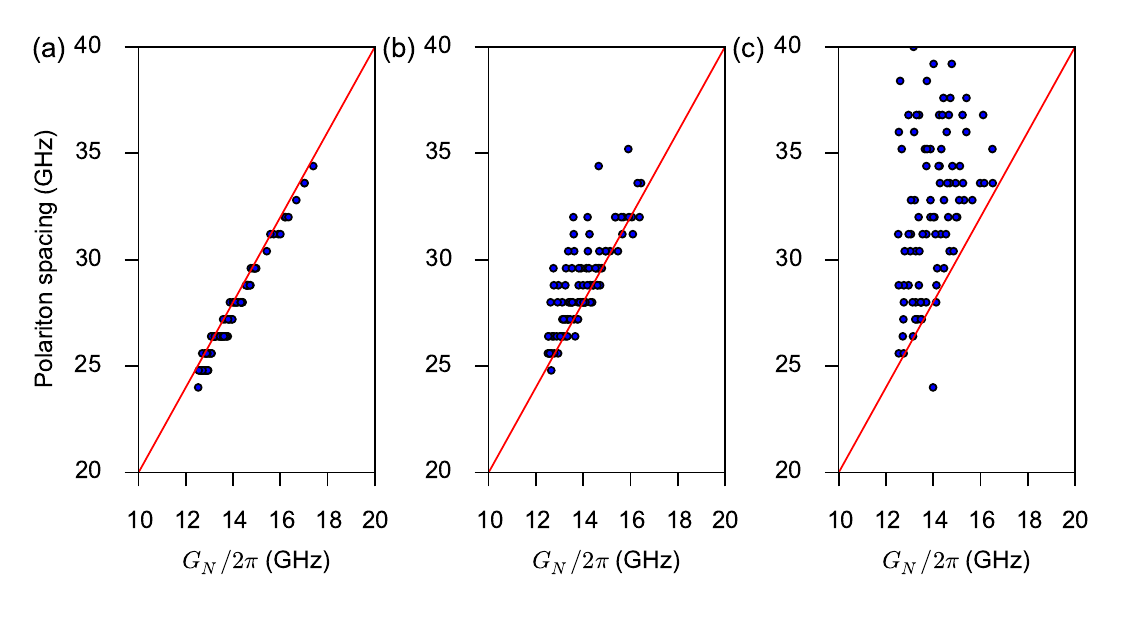}
\caption{\label{figure:5-5} Polariton spacing of randomly generated strongly coupled systems vs. expected collective coupling rate for $N=4$, $\{\kappa/2\pi, \Gamma/2\pi, g_{\textrm{\scriptsize max}}/2\pi\} = \{25\,\textsf{GHz}, 0.1\,\textsf{GHz}, 10\,\textsf{GHz}\}$ and $\delta/2\pi =$ (a) 1\,GHz, (b) 10\,GHz and (c) 20\,GHz. The red line corresponds to the system with identical emitters.}
}
\end{figure}

To further analyze the character of the collective strong coupling, we generate 100 systems ($G_N>\kappa/2$) for variable inhomogeneous broadening with $$\delta/2\pi \in \{1\,\textrm{GHz}, 10\,\textrm{GHz}, 20\,\textrm{GHz}\}$$ and plot the spacing between the emitters against the expected collective coupling $G_N$. We use the same formula as in the case of identical emitters: $G_N=\sqrt{\sum_{n=1}^{N}g_n^2}$, so that we benchmark the results with inhomogeneous broadening against the ideal multi-emitter cavity systems. As seen in Fig. \ref{figure:5-5} the effective coupling rate is fit well by $2G_N$, implying that the collective coupling rate of non-identical emitters is comparable to identical emitters for small inhomogeneous broadening. Crucially, because the numerically investigated polariton splittings closely match the $G_N$ for the cases of diamond SiV ($\delta \sim 1$\,GHz) and SiC silicon vacancy ($\delta \sim 20$\,GHz) systems, we expect these emitters to exhibit multi-emitter cavity QED phenomenon in spite of their inhomogeneous broadening.

%%%%%%%%%%%%%%%%%%%%%%%%%%%%%%%%%%%%%%%%%%%%%%%%%%%%%%%%%%%%%%%%

\subsection{Effective Hamiltonian Approach to Multi-Emitter Cavity QED} \label{sec:effH}

The quantum master equation provides a full numerical treatment of the system dynamics in transmission, however, the diagonalization of the density matrix makes this approach computationally challenging. This practically limits the number of emitters in the system to $N \lesssim 5$. The effective Hamiltonian approach introduced in the Section~\ref{sec:dissipative} can provide a significant speedup. For example, in the modeling of the transmission spectra, computation is effectively reduced to the diagonalization of a $(N + 1) \times (N + 1)$ matrix, from the initial $2^{(N +1)} \times 2^{(N+1)}$ size.

\begin{figure}
\centering
\textsf{
\includegraphics[width=11.43cm]{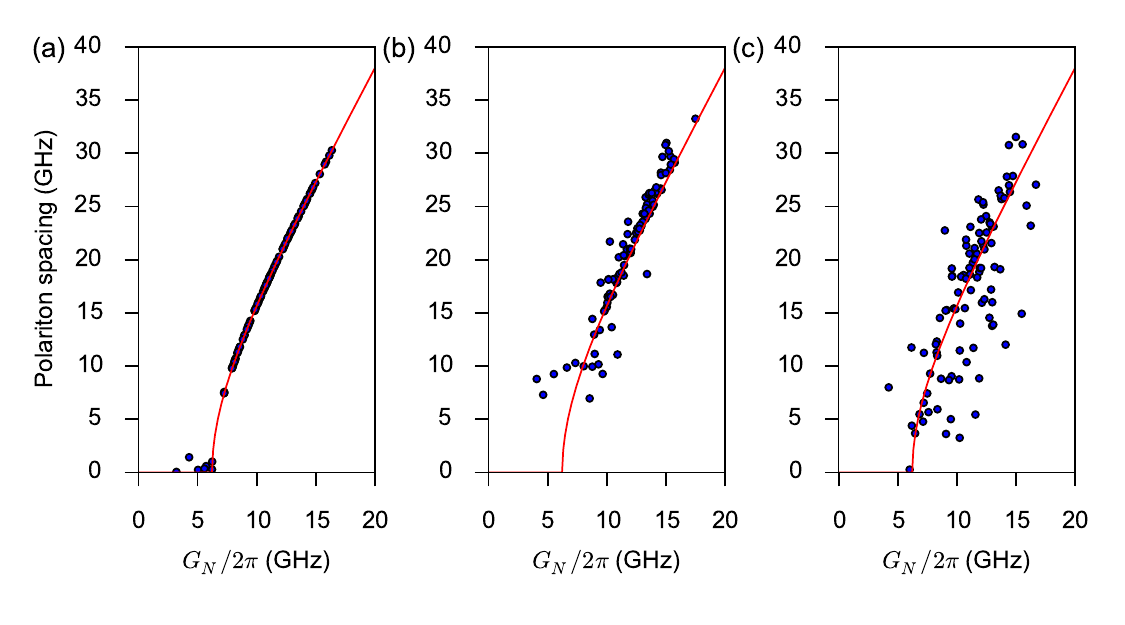}
\caption{\label{figure:5-6} Polariton spacing calculated using the effective Hamiltonian diagonalization vs. expected collective coupling rate for $N=4$, $\{\kappa/2\pi, \Gamma/2\pi, g_{\textrm{\scriptsize max}}/2\pi\} = \{25\,\textsf{GHz}, 0.1\,\textsf{GHz}, 10\,\textsf{GHz}\}$ and $\delta/2\pi =$ (a) 1\,GHz, (b) 10\,GHz and (c) 20\,GHz. The red line corresponds to the system with identical emitters.}
}
\end{figure}

Extended from the single emitter case presented in Section~\ref{sec:dissipative}, the effective Hamiltonian for multi-emitter system is
\begin{equation}
H_\textrm{\scriptsize EFF}=H_\textrm{\scriptsize TC}'-\textrm{i} \frac{\kappa}{2}a^{\dag}a - \textrm{i} \sum_{n=1}^N \frac{\Gamma}{2}\sigma_n^{\dag}\sigma_n.
\end{equation}
Its eigensystem can be represented as a set of eigenenergies and eigenvectors $\left\{E_n^\textrm{\scriptsize EFF}, \psi_n^\textrm{\scriptsize EFF}\right\}$. Bearing in mind that $\textrm{Re}\,\{E_n^\textrm{\scriptsize EFF}\}$ gives the frequency of an energy level and $2\,\textrm{Im}\{E_n^\textrm{\scriptsize EFF}\}$ represent its linewidth, we can reconstruct some of the spectral information obtained through the quantum master equation approach. We again generate 100 random systems for $$\delta/2\pi \in \{1\,\textrm{GHz}, 10\,\textrm{GHz}, 20\,\textrm{GHz}\}$$ and plot the frequency difference between the two eigenstates with the highest linewidths against the expected strong coupling rate $G_N$ (Fig. \ref{figure:5-6}). Here, the eigenstates with the highest linewidths represent the two states of strongest cavity emission since we operate in the bad-cavity limit. The correlation between the simulated polariton splitting for non-identical emitters to the one expected for identical emitters is excellent. The latter can be analytically calculated by approximating the multi-emitter system as a single-emitter system with effective coupling $G_N$. Using the derivation presented in Section~\ref{sec:dissipative}, the polariton splitting is expressed as
\begin{equation}
E_{N+1} - E_1 = 2\sqrt{G_N^2-\left( \frac{\kappa-\Gamma}{4} \right)^2}.
\end{equation}

While here we confirm that the non-identical multi-emitter systems behave similarly to identical ones in the effective Hamiltonian approximation, additional steps are needed to derive a close approximation to the transmission spectrum. To do this, we first introduce a driving term to the effective Hamiltonian
\begin{equation}
H_\textrm{\scriptsize drive}(\omega_\textrm{\scriptsize L})=\mathcal{E} (a e^{\textrm{\footnotesize i} \omega_\textrm{\tiny L} t}+a^{\dag}e^{-\textrm{\footnotesize i}\omega_\textrm{\tiny L} t}),
\end{equation}
where $\omega_\textrm{\scriptsize L}$ is the laser frequency and $\mathcal{E}$ is proportional to the laser field intensity. Transforming the Hamiltonian into the rotating-wave frame (as discussed in Section~\ref{sec:osc}) and diagonalizing the effective Hamiltonian, we obtain the eigensystem $\{E_n^\textrm{\scriptsize EFF}(\omega_\textrm{\scriptsize L}), \psi_n^\textrm{\scriptsize EFF}(\omega_\textrm{\scriptsize L})\}$. Now, we assume that each energy level emits light with Lorentzian intensity distribution centered at $\textrm{Re} \,\{E_n^\textrm{\scriptsize EFF}(\omega)\}$ and with linewidth of $2\textrm{Im}\,\{E_n^\textrm{\scriptsize EFF}(\omega)\}$. To combine the individual levels into a transmission spectrum, we weight each of the eigenstates by their cavity occupation terms $\langle \psi_n^\textrm{\scriptsize EFF}(\omega_\textrm{\scriptsize L}) | a^{\dag}a | \psi_n^\textrm{\scriptsize EFF}(\omega_\textrm{\scriptsize L}) \rangle$. From there, we derive our first effective Hamiltonian approximation to the spectrum $S^\textrm{\scriptsize EFF}(\omega)$ as
\begin{equation}\label{eq:47}
S^\textrm{\scriptsize EFF}_I(\omega)=\sum_{n=1}^{N} \langle \psi_n^\textrm{\scriptsize EFF}(\omega) | a^{\dag}a | \psi_n^\textrm{\scriptsize EFF}(\omega) \rangle L\left(\omega;\, \textrm{Re} \,\{E_n^\textrm{\scriptsize EFF}(\omega)\}, \textrm{Im}\,\{E_n^\textrm{\scriptsize EFF}(\omega)\}\right),
\end{equation}
where $L(\omega;\, \omega_0, \eta)$ represents Lorentzian distribution defined as
\begin{equation}
L(\omega; \, \omega_0, \eta) = \frac{1}{\pi \eta \left[1+\left(\frac{\omega-\omega_0}{\eta}\right)^2\right]}.
\end{equation}

\begin{figure}
\centering
\textsf{
\includegraphics[width=11.43cm]{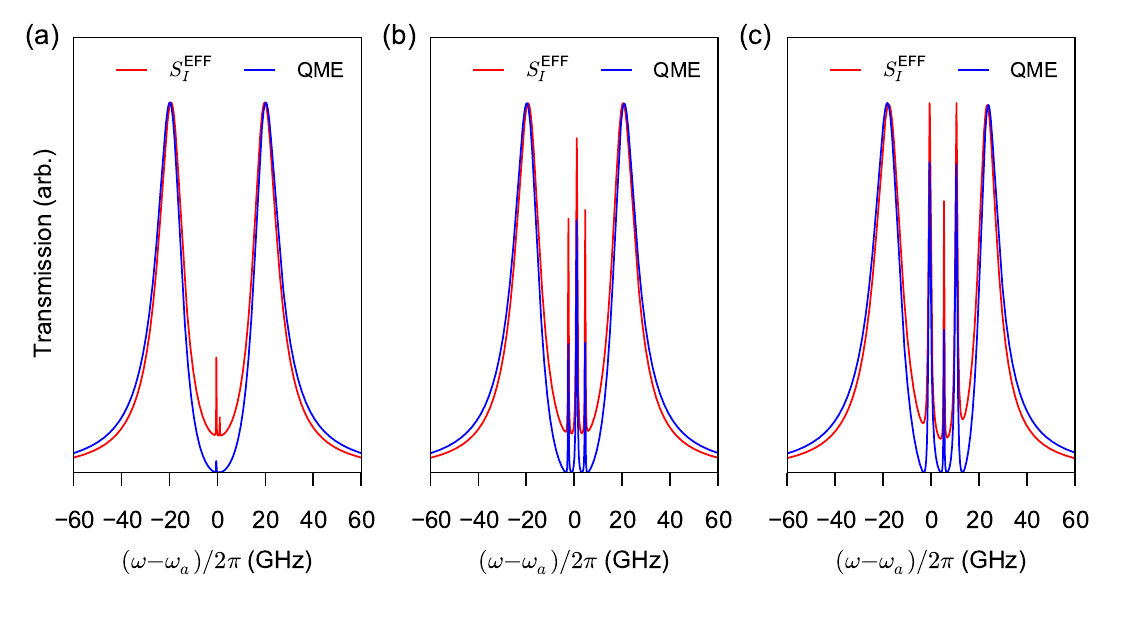}
\caption{\label{figure:5-7} A comparison between the transmission spectra calculated using quantum master equation (QME) and the first effective Hamiltonian approximation ($S^\textrm{\scriptsize EFF}_{I}$) for a $N=4$ multi-emitter system $\{\kappa/2\pi, \Gamma/2\pi, g_{\textrm{\scriptsize max}}/2\pi\} = \{25\,\textsf{GHz}, 0.1\,\textsf{GHz}, 10\,\textsf{GHz}\}$ and emitter frequencies sampled for  $\delta/2\pi =$ (a) 1\,GHz, (b) 10\,GHz, and (c) 20\,GHz.}
}
\end{figure}

In Fig. \ref{figure:5-7} we compare the spectra obtained through the effective Hamiltonian approach to the ones calculated by the quantum master equation. The qualitative agreement is very good, while the quantitative match between peak locations and intensities is close, but not complete, because it does not capture the interference effects between the polaritons. To improve on this, we derive another approximation which models the light field interference more reliably.

\begin{figure}
\centering
\textsf{
\includegraphics[width=11.43cm]{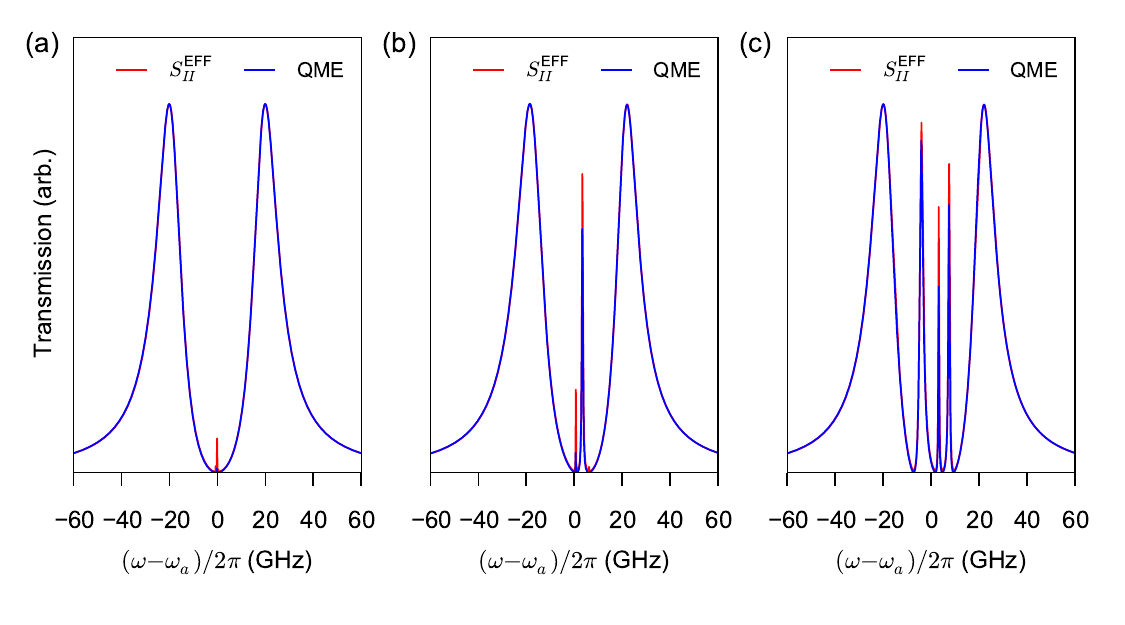}
\caption{\label{figure:5-8} A comparison between the transmission spectra calculated using quantum master equation (QME) and the second effective Hamiltonian approximation ($S^\textrm{\scriptsize EFF}_{II}$) for a $N=4$ multi-emitter system $\{\kappa/2\pi, \Gamma/2\pi, g_{\textrm{\scriptsize max}}/2\pi\} = \{25\,\textsf{GHz}, 0.1\,\textsf{GHz}, 10\,\textsf{GHz}\}$ and emitter frequencies sampled for $\delta/2\pi =$ (a) 1\,GHz, (b) 10\,GHz, and (c) 20\,GHz.}
}
\end{figure}

Because we consider the transmission spectra when the emitters are all weakly excited (as discussed in Section~\ref{sec:osc}), they may be considered to primarily emit coherent radiation \citep{Steck2007-qk}. Thus, we can ignore their incoherent portions of emission and use the classical limit
\begin{equation}
\langle a^{\dag}a \rangle \approx \langle a ^{\dag}\rangle \langle a \rangle.
\end{equation}
Now, we can take the fields ($\propto \langle a \rangle $) rather than intensities ($\propto \langle a^{\dag}a \rangle$) to combine the transmission profiles in the spectrum, thereby incorporating the interference effects between the different polaritons and subradiant states. Then, in our second effective Hamiltonian approximation, the transmission spectrum is calculated as
\begin{equation}
S^\textrm{\scriptsize EFF}_{II}(\omega)=\left| \sum_{n=1}^{N} \langle \psi_n^\textrm{\scriptsize EFF}(\omega) | a | \psi_n^\textrm{\scriptsize EFF}(\omega) \rangle \sqrt{L\left(\omega;\, \textrm{Re} \,\{E_n^\textrm{\scriptsize EFF}(\omega)\}, \textrm{Im}\,\{E_n^\textrm{\scriptsize EFF}(\omega)\}\right)} \right|^2
\end{equation}
where now the terms inside the summation represent field contributions that are interfered before the final modulus squared is taken to calculate the intensity. This result is formally very similar to that obtained by diagonalizing a set of coupled Heisenberg-Langevin equations and using the approximation that the emitters are all in their ground states \citep{Waks2006-cc}. While these two methods arrive at similar answers, we believe the intuitive connection to the complex eigenstates of the non-Hermitian effective Hamiltonian provides additional insight. Figure \ref{figure:5-8} shows that this approximation provides a close fit to the quantum master equation results.

\begin{figure}
\centering
\textsf{
\includegraphics[width=11.43cm]{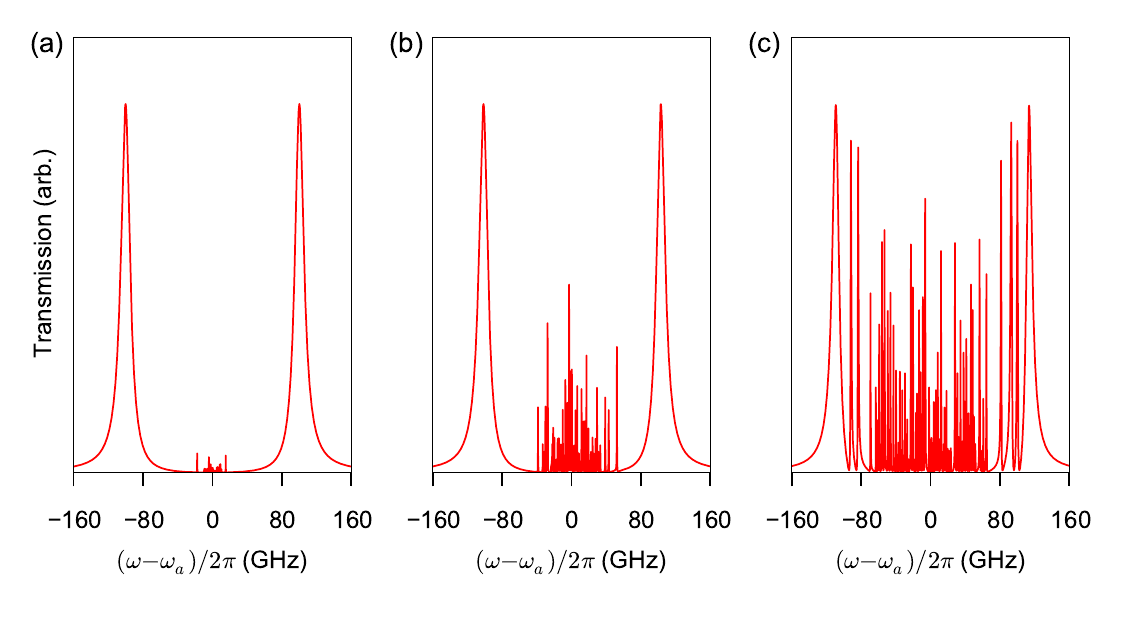}
\caption{\label{figure:5-9} Transmission spectra calculated by the second effective Hamiltonian approximation for a $N=100$ multi-emitter system $\{\kappa/2\pi, \Gamma/2\pi, g_{\textrm{\scriptsize max}}/2\pi\} = \{25\,\textsf{GHz}, 0.1\,\textsf{GHz}, 10\,\textsf{GHz}\}$ and emitter frequencies sampled for $\delta/2\pi =$ (a) 10\,GHz, (b) 40\,GHz, and (c) 80\,GHz.}
}
\end{figure}

With these approximations at hand, we can conclude that the effective Hamiltonian approach can provide an excellent insight into transmission properties of a cavity QED system. This is especially valuable for systems with a large number of emitters where the solving of the quantum master equation would require unrealistic computational resources. We analyze one such system based on $N = 100$ emitters for $\delta/2\pi \in \{10\,\textrm{GHz}, 40\,\textrm{GHz}, 80\,\textrm{GHz}\}.$ Figure~\ref{figure:5-9} shows the extension of the cavity protection effects to systems with larger inhomogeneous broadening, granted by the increase in the collective coupling amounting from an increased number of emitters in the system. Despite the large inhomogeneous broadening, collective oscillations as the highest-energy and lowest-energy states are still readily apparent. These collective excitations manifest as two polaritons, each with approximately half the cavity linewidth, that are separated by $2G_N$.

In the next section, we will analyze new opportunities for nonclassical light generation in multi-emitter cavity QED systems.

%%%%%%%%%%%%%%%%%%%%%%%%%%%%%%%%%%%%%%%%%%%%%%%%%%%%%%%%%%%%%%%%
% Section 5
%%%%%%%%%%%%%%%%%%%%%%%%%%%%%%%%%%%%%%%%%%%%%%%%%%%%%%%%%%%%%%%%

\section{Nonclassical Light Generation With Multi-Emitter Cavity QED Systems} \label{sec:meCQED}

As discussed in Section~\ref{sec:seCQED}, nonclassical light generation in cavity QED systems finds its origin in the discretized and anharmonic character of the energy states. The differences in the dressed ladder of states for an increasing number of emitters are presented in Fig. \ref{figure:5-2} and compared to the Jaynes--Cummings ladder. These more complicated level structures for multi-emitter systems bring in new opportunities for $n$-photon emission.

%%%%%%%%%%%%%%%%%%%%%%%%%%%%%%%%%%%%%%%%%%%%%%%%%%%%%%%%%%%%%%%%

\subsection{Resonant Photon Blockade}

In previous sections, we only considered photon blockade operated in a pulsed regime. Here, we consider it under continuous-wave (CW) excitation. Emerging technologies for photodetection such as superconducting single-photon detectors are beginning to provide the timing resolution required for the study of CW photon correlations. Because these correlations do not involve averaging over an entire pulse, they can act as a more sensitive probe of the underlying system dynamics.

For the CW case, the second-order coherence $g^{(2)}(0)$ is defined as 
\begin{equation}
g^{(2)}(0) = \frac{\langle a^\dag a^\dag a a \rangle}{\langle a^\dag a \rangle^2},
\end{equation}
and it is related to the statistics of simultaneously emitting two (or more) photons \citep{Glauber1963-io}. Just like for the pulsed statistic $g^{(2)}[0]$, when the CW statistic $g^{(2)}(0)$ is less than one (ideally zero), the light has a sub-Poissonian nonclassical character that indicates its potential for single-photon emission. However, to confirm the single-photon character of the emitted light in the CW case, higher-order coherences need also to be taken into account \citep{carreno2016criterion}.

\begin{figure}
\centering
\textsf{
\includegraphics[width=11.43cm]{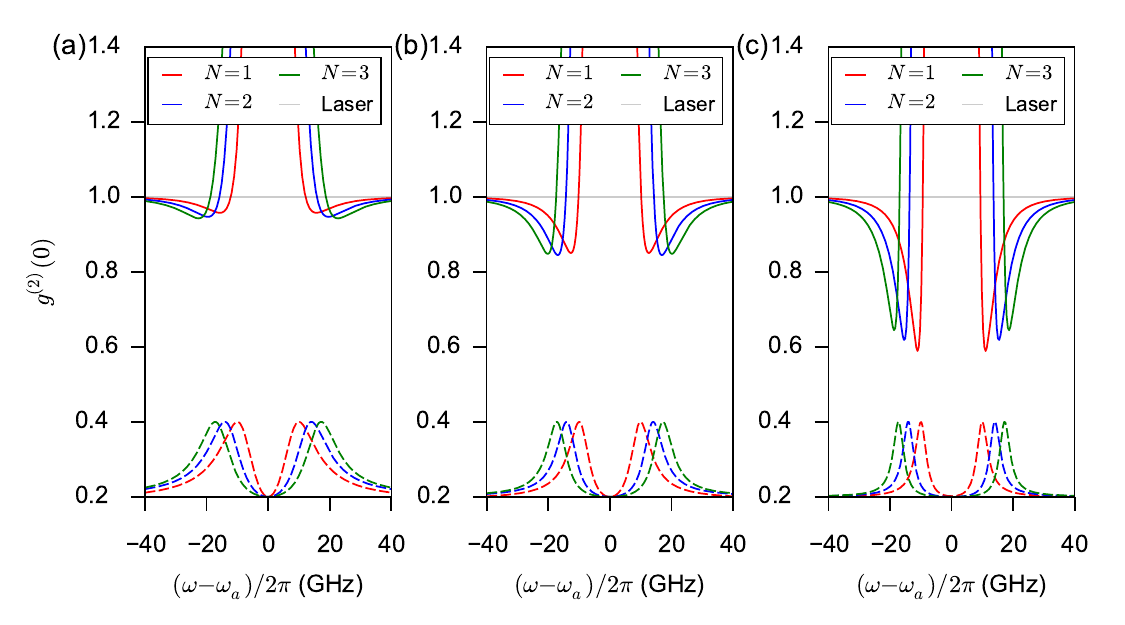}
\caption{\label{figure:6-1} Photon blockade trends with an increasing $N$ for parameters $\{\Delta_{\textrm{\scriptsize e}, n}/2\pi, \Gamma/2\pi, g_{n}/2\pi\} = \{0\,\textsf{GHz}, 0.1\,\textsf{GHz}, 10\,\textsf{GHz}\}$ and $\kappa/2\pi =$ (a) 25\,GHz ($G_N<\kappa$), (b) 15\,GHz ($G_N\sim \kappa$), and (c) 8\,GHz ($G_N>\kappa$). Dashed lines illustrate the transmission spectra.}
}
\end{figure}

Now, we use this tool to study transmission through the collective polaritonic excitations of multi-emitter cavity QED systems. Considering the transmission through polaritonic states, we naively expect that the photon blockade effect should become even more pronounced with a $\sqrt{N}$ increase in coupling strength. However, this seems to be the case only while $G_N<\kappa$---we attribute this somewhat unexpected finding to the addition of energy states in the second-rung (see Fig. \ref{figure:5-2}) with an increasing number of emitters. While a detailed investigation is needed for full understanding of this phenomenon, we suggest that the photon blockade effects are greatly influenced by the dipolar coupling mechanisms between the two energy rungs. We present these effects on a system of identical emitters, resonantly and equally coupled to the cavity. Figure \ref{figure:6-1} compares second-order coherence $g^{(2)}(0)$ values with an increasing $N$ for three systems whose parameters capture the change in the trend around $G_N\approx \kappa$. As with the single-emitter photon blockade, the points of best single-photon emission are located around the frequencies of the polariton transmission peaks. When $G_N<\kappa$, the photon blockade improves with the increasing $N$, while for $G_N>\kappa$ the behavior reverses. Hence, we have identified a potential fundamental limit in multi-emitter photon blockade under resonant conditions, in that high quality factor cavities may not necessarily yield the best photon blockade in a multi-emitter cavity QED system. Notably, this limit is relaxed for detuned photon blockade.

\begin{figure}
\centering
\textsf{
\includegraphics[width=9cm]{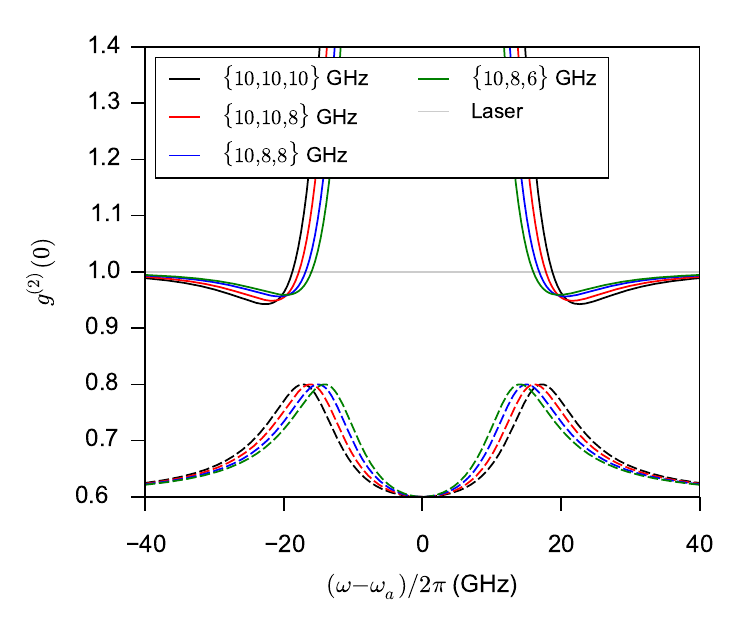}
\caption{\label{figure:6-2} Photon blockade trends for $N=3$, $\{\kappa/2\pi, \Gamma/2\pi, \Delta_{\textrm{\scriptsize e}, n}/2\pi\} = \{25\,\textsf{GHz}, 0.1\,\textsf{GHz}, 0\,\textsf{GHz}\}$ and the variable coupling parameters $\{g_1/2\pi, g_2/2\pi, g_3/2\pi\}$ noted in the legend.}
}
\end{figure}

Next, we analyze the effect of unequal coupling to the photon blockade, and compare the second-order coherence values for systems with variable coupling of three emitters (Fig. \ref{figure:6-2}). Starting with a system of identically coupled emitters, we gradually decrease coupling rates of the second and the third emitter, which corresponds to the decrease of the collective coupling rate $G_N$. We observe that the quality of the photon blockade decreases ($g^{(2)}_{\textrm{\scriptsize min}}(0)$ increases) with the lowering of the collective coupling rate.

%%%%%%%%%%%%%%%%%%%%%%%%%%%%%%%%%%%%%%%%%%%%%%%%%%%%%%%%%%%%%%%%

\subsection{Detuned Photon Blockade in the Multi-Emitter System}

\begin{figure}
\centering
\textsf{
\includegraphics[width=11.43cm]{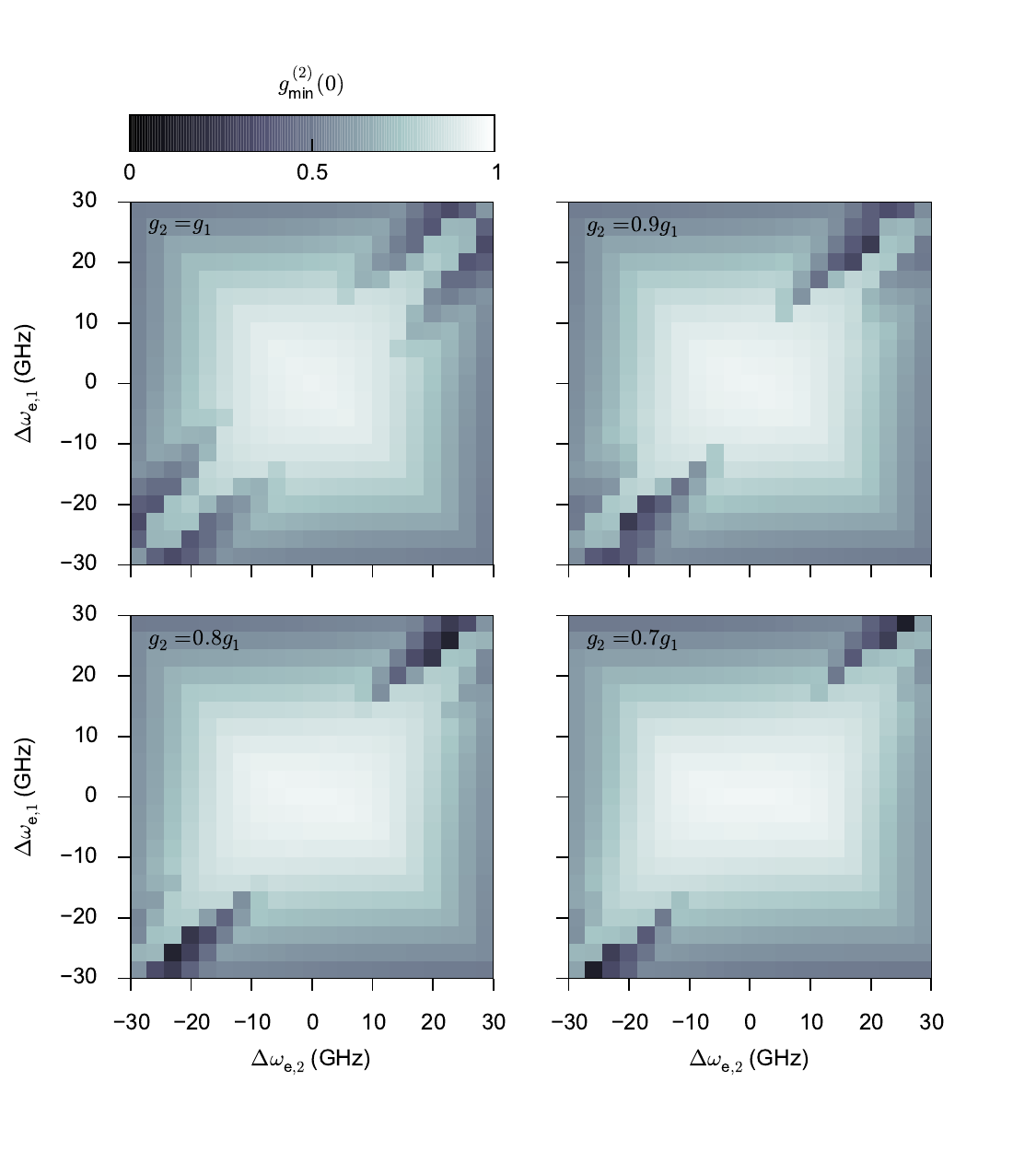}
\caption{\label{figure:6-3} Zero-time second-order coherence minima for a detuned system with $N=2$ emitters and $\{\kappa/2\pi, \Gamma/2\pi, g_1/2\pi\} = \{25\,\textsf{GHz}, 0.1\,\textsf{GHz}, 10\,\textsf{GHz}\}$; $g_2$ is quoted in each panel.}
}
\end{figure}

As in the single emitter case, we expect an enhanced photon blockade for a system where multiple emitters are detuned from the cavity. Here, we focus on a two-emitter system and analyze the minima of $g^{(2)}(0)$ as a function of the emitters' detuning from the cavity. In Fig. \ref{figure:6-3} we see that the photon blockade improves as the emitters detune from the cavity. For unequally coupled emitters, the trend becomes less pronounced for the lowered $G_N=\sqrt{g_1^2+g_2^2}$. Peculiarly, we also see an effect that additionally lowers $g_{\textrm{\scriptsize min}}^{(2)}(0)$ and occurs when both emitters are highly detuned from the cavity and a little detuned from one another. This behavior becomes even more pronounced for unequally coupled emitters providing a lower second-order coherence value, and shifting the $g_{\textrm{\scriptsize min}}^{(2)}(0)$ further away with the emitters' detuning from the cavity. The effect is asymmetric and favors systems where the most stongly coupled emitter is also the highest detuned one.

\begin{figure}
\centering
\textsf{
\includegraphics[width=11.43cm]{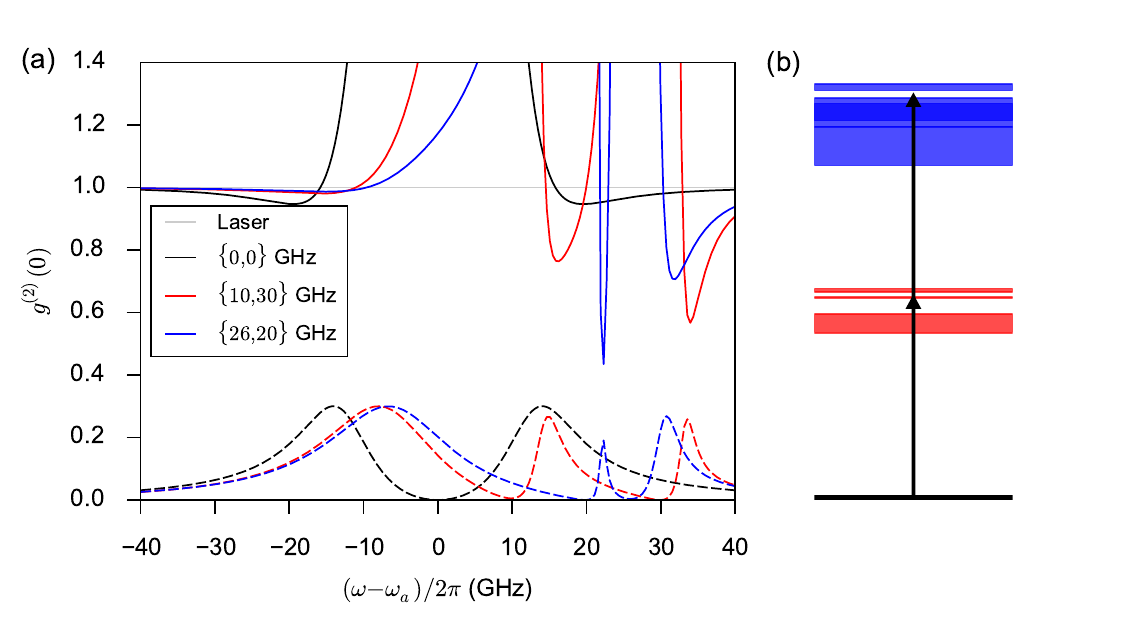}
\caption{\label{figure:6-4} (a) Zero-time second-order coherence for $N=2$, $\{\kappa/2\pi, \Gamma/2\pi, g_{1,2}/2\pi\} = \{25\,\textsf{GHz}, 0.1\,\textsf{GHz}, 10\,\textsf{GHz}\}$  and the different emitter detunings $\{\Delta_{\textrm{\scriptsize e},1}/2\pi, \Delta_{\textrm{\scriptsize e},2}/2\pi\}$ noted in the legend; plotted with transmission spectra (dashed lines). (b) Dressed ladder of states calculated with our effective Hamiltonian approach for $\{\Delta_{\textrm{\scriptsize e},1}/2\pi, \Delta_{\textrm{\scriptsize e},2}/2\pi\}=\{26\,\textsf{GHz}, 20\,\textsf{GHz}\}$, plotted with their linewidths; black vertical arrows illustrate first and second excitation at the point of the best photon blockade. Energy spacing between the rungs is significantly reduced for clarity.}
}
\end{figure}

Let us look more closely into what causes this additional second-order coherence reduction. Figure \ref{figure:6-4}a shows the $g^{(2)}(0)$ dependence on the transmission wavelength. We see that the minima of the function occur around the frequencies of the transmission peaks. When the emitters are identical there are only two local minima corresponding to transmission near the polariton peaks. For the non-identical emitters the second-order coherence has three local minima, with the middle one corresponding to the frequency of the emerging peak. For large detuning, transmission through this intermediate peak gains an advantageous second-order coherence value, which corresponds exactly to the peculiar minima in $g^{(2)}(0)$ seen in Fig. \ref{figure:6-3}. The origin of this effect has been traced back to the specifics of the dressed ladder of states \citep{radulaski2016nonclassical}. Time evolutions of the second-order coherence function [for details see \citet{radulaski2016nonclassical}] show that the dynamics of the single-photon emission from either of the photon blockade frequencies evolves at the 100\,ps scale, which represents a significant speedup in the single-photon emission relative to the expected 10\,ns lifetime of individual color-centers.

%%%%%%%%%%%%%%%%%%%%%%%%%%%%%%%%%%%%%%%%%%%%%%%%%%%%%%%%%%%%%%%%

\subsection{Effective Hamiltonian Approach to Photon Blockade}

The effective Hamiltonian approach can provide an insight into the photon blockade properties of the system. The frequency spacing between the eigenstates in the first and the second rung is calculated by the $H_\textrm{\scriptsize EFF}$ diagonalization previously discussed in Section~\ref{sec:effH}. Figure \ref{figure:6-4}b illustrates the energy levels of the system with 26 and 20\,GHz emitter detunings, plotted in red in Fig. \ref{figure:6-4}a. The levels of the first and the second rung are depicted in red and blue, respectively. Each state's width corresponds to its full width at half maximum. The spacing between the rungs is not true to size. The black vertical arrows indicate the frequencies of the best photon blockade (22.6\,GHz), owing their performance to the highest frequency clearance from a second photon absorption. Due to the variable linewidth of the states, it is important to consider the widths of the individual levels in the analysis of their frequency overlap. When there is less frequency overlap between subsequent jumps up our extended Tavis--Cummings ladder, the quality of the photon blockade increases.

This approach can be used in more complex systems, where the calculation of quantum master equation is too lengthy, to indicate potentially advantageous parameters for high quality single-photon emission.

\subsection{Effective Hamiltonian Approach to N-Photon Generation}

Multi-emitter cavity QED systems can generate light in the so-called \emph{un-conventional photon blockade} regime. Here, the second-order coherence has sub-Poissonian character, while the third-order coherence has super-Poissonian statistics, therefore promoting the generation of three-photon states.

\begin{figure}
\centering
\textsf{
\includegraphics[width=11.43cm]{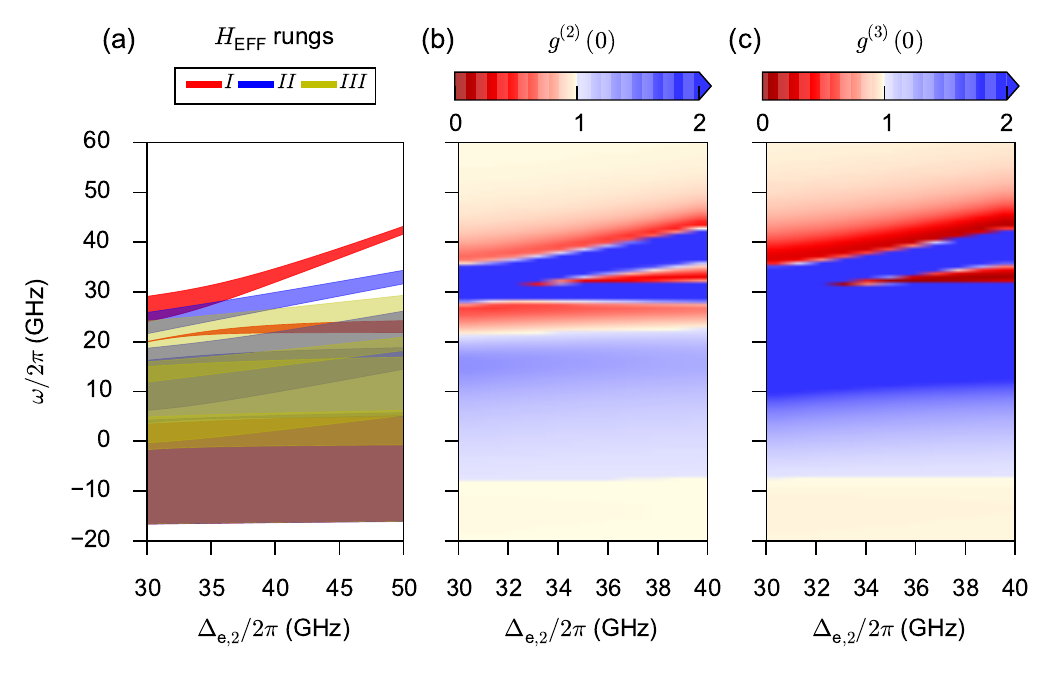}
\caption{\label{figure:6-5} (a) Frequency overlap $\left(E_n-n\omega_a\right)/n$ where $E_n$ are eigenenergies of $n$-th rung, between three rungs of the dressed ladder for $N=2$ emitters and $\{\kappa/2\pi, \Gamma/2\pi, g_{1,2}/2\pi, \Delta_{\textrm{\scriptsize e},1}/2\pi\} = \{25\,\textsf{GHz}, 0.1\,\textsf{GHz}, 10\,\textsf{GHz}, 30\,\textsf{GHz}\}$. (b) Second and (c) third-order coherence functions for the same system. \textit{Data from \citet{radulaski2016nonclassical}.}}
}
\end{figure}

Presenting data from \citet{radulaski2016nonclassical}, we show how the effective Hamiltonian approach can be used to identify such regimes. We focus on the system with $N=2$ emitters, $$\{\kappa/2\pi, \Gamma/2\pi, g_{1,2}/2\pi, \Delta_{\textrm{\scriptsize e},1}/2\pi\} = \{ 25\,\textrm{GHz}, 0.1\,\textrm{GHz}, 10\,\textrm{GHz}, 30\,\textrm{GHz} \}$$ and variable detuning of the second emitter $\Delta_{\textrm{\scriptsize e},2}/2\pi \in [30\,\textrm{GHz}, 40\,\textrm{GHz}]$. Figure \ref{figure:6-5}a shows the frequency overlap of the first three rungs of the dressed ladder calculated using $H_\textrm{\scriptsize EFF}$, while Figs. \ref{figure:6-5}b-c show the second- and the third-order coherences calculated using the quantum-optical master equation. For the region corresponding to $\omega / 2\pi = 35$\,GHz, we see that the eigenstate of the third rung has no frequency overlap with states of the first two rungs and should promote three photon emission. This is confirmed by the coherence functions which show a suppression of the two photon emission, but an enhancement in the three photon emission as discussed in \citet{radulaski2016nonclassical}. These results show that effective Hamiltonian approach can be used to identify parameters advantageous for higher order $n$-photon generation.

%%%%%%%%%%%%%%%%%%%%%%%%%%%%%%%%%%%%%%%%%%%%%%%%%%%%%%%%%%%%%%%%

\subsection{Outlook for Multi-Emitter Cavity QED}

Multi-emitter cavity QED at the level of several emitters is close to its first demonstrations in a solid-state platform. Recent developments in low-strain material growth, nanofabrication in bulk substrates, and controlled implantation of color-centers are setting the path for achieving high quality factor, small mode volume cavities with preferentially positioned nearly-identical quasi-emitters.

We have set forth an extended Tavis--Cummings model that captures the core dynamics of a multi-emitter cavity QED system. The initial theoretical findings on improved photon blockade and photon bundle generation are expected to be expanded and unveil rich physics, even with the inhomogeneous broadening of emerging color centers. Robust regimes of operation have been predicted, with promising applications in reliable quantum optical networks. Experimental data will set this theory to a test, and based on the experience with the single emitter systems, we envision additional inclusion of factors in the quantum model explicitly, such as dephasing and phonon interaction.

The systems with lossy cavities will require a higher number of emitters to reach the strong coupling regime. Our new effective Hamiltonian approach to estimating transmission spectra could provide insights into the dynamics of such systems by finding the best fit to the transmission spectra, identifying individual emitter frequencies, and establishing regimes of operation. From there, the same framework can be used to evaluate frequency overlap between the rungs of excited states and target operating frequencies for nonclassical light generation.

%%%%%%%%%%%%%%%%%%%%%%%%%%%%%%%%%%%%%%%%%%%%%%%%%%%%%%%%%%%%%%%%
% Section 6
%%%%%%%%%%%%%%%%%%%%%%%%%%%%%%%%%%%%%%%%%%%%%%%%%%%%%%%%%%%%%%%%

\section{Conclusions}

In this chapter we presented the state-of-the-art in the generation of nonclassical states of light using semiconductor cavity QED platforms, focusing in particular on the photon blockade effects that enable the generation of indistinguishable photon streams with high purity and efficiency. InGaAs quantum dots in optical nanocavities have been the leading platform for such experiments for many years, and have enabled exciting fundamental science demonstrations (including the first demonstration of photon blockade in solid-state), with direct applications to nonclassical light generation. However, the limitations of these quantum dots, particularly in terms of their random positioning and large inhomogeneous broadening, are impeding their employment in systems requiring more than one emitter. This is in turn has generated much interest in systems based on color-centers in group-IV semiconductors---diamond and SiC, which feature very small inhomogenous broadening and even room temperature operation. This facilitates interference of photons emitted from different quantum emitters, and it enables the implementation of multi-emitter cavity QED systems that feature richer dressed-states ladder structures and offers opportunities for studying new regimes of photon blockade \citep{radulaski2016nonclassical}. However, the remaining piece of the puzzle is the demonstration of a strongly coupled cavity QED platform based on a few color-centers coupled to a cavity, which should be within reach with today's technologies \citep{burek2012free, radulaski2016nonclassical}. Once this milestone is achieved, not only will it open the door to implementation of new quantum light sources, but also will create opportunities beyond nonclassical light, including quantum many-body physics simulation \citep{greentree2006quantum, hartmann2008quantum, carusotto2013quantum}.

%%%%%%%%%%%%%%%%%%%%%%%%%%%%%%%%%%%%%%%%%%%%%%%%%%%%%%%%%%%%%%%%
% Section 7
%%%%%%%%%%%%%%%%%%%%%%%%%%%%%%%%%%%%%%%%%%%%%%%%%%%%%%%%%%%%%%%%

\section{Acknowledgments}

The authors thank Kai M\"uller for productive and helpful discussions in framing the context of this work, and for contributing experimental data to Figs. \ref{figure:4-5}, \ref{figure:4-6}, and \ref{figure:4-11}. The authors are grateful to Tomas Sarmiento and Jingyuan Linda Zhang for providing data to Figs. \ref{figure:1-1} and \ref{figure:2-1}.

This material is based upon work supported by the Air Force Office of Scientific Research under award number FA9550-17-1-0002 and the National Science Foundation Division of Materials Research, Grant Numbers 1406028 and 1503759. KAF acknowledges support from the Lu Stanford Graduate Fellowship and the National Defense Science and Engineering Graduate Fellowship.

\section*{References}

\bibliography{mybibfile}
\end{document}